\documentclass{aa}  

\usepackage{graphicx}

\usepackage{txfonts}
\usepackage[font=small]{caption}
\usepackage{geometry}
\usepackage{array}
\usepackage{makecell}
\usepackage{dblfloatfix}
\usepackage{float}
\usepackage{caption}
\usepackage{multirow}
\usepackage{subcaption}
\usepackage{hyperref}
\hypersetup{
    colorlinks=true,
    linkcolor=red,
    urlcolor=cyan,
    citecolor=blue
    }
\usepackage[capitalize]{cleveref}
\usepackage{tablefootnote}
\usepackage{longtable}
\usepackage{footmisc}
\usepackage{gensymb}
\usepackage{xcolor}
\newcolumntype{P}[1]{>{\centering\arraybackslash}p{#1}}

\begin{document} 

   \title{\texttt{TUVOpipe}: A pipeline to search for UV transients with \textit{Swift}-UVOT\thanks{A reproduction package for this paper is available at \url{https://doi.org/10.5281/zenodo.5946940/}.}}

      \author{David Modiano\inst{1} \and Rudy Wijnands\inst{1} \and Aastha Parikh\inst{1} \and Jari van Opijnen\inst{1} \and Sill Verberne\inst{1,2} \and Marieke van Etten\inst{1}}

   \institute{Anton Pannekoek Institute for Astronomy, University of Amsterdam, Postbus 94249, 1090 GE Amsterdam, The Netherlands\\
              \email{d.modiano@uva.nl}
         \and
             Leiden Observatory, Leiden University, PO Box 9513, 2300 RA, Leiden, The Netherlands
             }
 
  \abstract
   {Despite the prevalence of transient-searching facilities operating across most wavelengths, the ultraviolet (UV) transient sky remains to be systematically studied. Therefore, we recently initiated the Transient Ultraviolet Objects (TUVO) project, with which we search for serendipitous UV transients in data obtained using currently available UV instruments with a strong focus on the UV and Optical (UVOT) telescope aboard the \textit{Neil Gehrels Swift Observatory} (an overview of the project is described in a companion paper). Here, we describe the pipeline (named \texttt{TUVOpipe}) we constructed in order to find such transients in the UVOT data, using difference image analysis. The pipeline is run daily on all new public UVOT data (which are available 6-8 hours after the observations are performed), so we discover transients in near real time. Transients that last $>$0.5 days are therefore still active when discovered, allowing for follow-up observations to be performed. From 01 October 2020 to the time of submission, we used the \texttt{TUVOpipe} to process 75\,183 individual UVOT images, and we currently detect an average rate of $\sim$100 transient candidates per day. Of these daily candidates, on average $\sim$30\% are real transients (separated by human vetting from the remaining `bogus' transients which were not discarded automatically within the pipeline). Most of the real transients correspond to known variable stars, though we also detect a significant number of known active galactic nuclei and accreting white dwarfs. The \texttt{TUVOpipe} can additionally run in archival mode, whereby all the archival UVOT data of a given field is scoured for `historical' transients; in this mode, we also mostly find variable stars. However, some of the transients we find (in particular in the real-time mode) represent previously unreported new transients or undiscovered outbursts of previously known transients, predominantly outbursts from cataclysmic variables. In this paper, we describe the operation of (both modes of) \texttt{TUVOpipe} and some of the initial results we have obtained so far.}


   \maketitle

\section{Introduction}  \label{introduction}

In the last decade, there have been many advances in the field of time-domain astronomy. An increasing range of facilities have been performing large-scale surveys with the aim of detecting and studying transients and variable sources (collectively referred to as `transients' hereafter). These facilities have operated across the electromagnetic (EM) spectrum, ranging from (very-high energy) gamma rays down to low-frequency radio waves (see \citealp{Sagiv_2014} for a review of many of the existing transient facilities). Recently, this wealth of knowledge has been further expanded with the development of multi-messenger transient-searching facilities such as LIGO-Virgo (gravitational waves; \citealp{Abbott_2016}) and IceCube (neutrinos; \citealp{IceCube_2018}). \

A noticeable exception, however, exists in the ultraviolet (UV) regime. Although UV observations are often used for follow-up studies of transients discovered at other wavelengths (e.g. \citealp{Cenko_2012}; see also \citealp{Middleton_2017} for an overview of multi-wavelength astronomy including the utility of the UV) and through other messengers (e.g. \citealp{Abbott_2017}), the UV has not been utilised for systematic large-scale searches for serendipitous transients. Some relatively small projects have carried out searches for UV transients, but with very limited scopes (e.g. \citealp{Welsh_2005,Wheatley_2008,Gezari_2013} using data obtained with \textit{GALEX}). The primary reason for the lack of such UV transient studies is that the Earth's atmosphere is opaque to most UV radiation, prohibiting ground-based facilities\footnote{Although most UV is blocked by the Earth's atmosphere, a significant fraction of the near-UV can still be observed from the ground, and some new facilities are currently undertaking transient searches in bluer bands (i.e. the U band) than typically used in transient surveys (e.g. \textit{MeerLICHT}; \citealp{Bloemen_2016}, and in the near future, \textit{BlackGEM}; \citealp{Groot_2019}).}. UV telescopes mounted on satellites are therefore the primary option for UV astronomy. However, there are no large UV surveying facilities currently in operation, though some are planned or proposed (see, e.g. \textit{ULTRASAT}, \citealp{Sagiv_2014}; \textit{CASTOR}, \citealp{Patrick_2012}; \textit{Dorado}, \citealp{Singer_2021}; \textit{UVEX}, \citealp{Kulkarni_2021}). See \cite{Kulkarni_2021} for a review of past and future missions with UV-transient searching capabilities.\ 

This gap in our surveying capability is particularly noteworthy given that the UV can provide valuable information about many types of interesting high-energy sources, many of which are known to exhibit strong UV emission (and may even peak in the UV at early times during their outbursts). Examples of such phenomena include UV bright flares from active stars and interacting binaries, outbursts from accreting white dwarfs, namely novae and dwarf novae (DNe), outbursts from accreting neutron stars and black holes (e.g. X-ray binaries; XRBs), supernovae (SNe), tidal disruption events (TDEs), variability of active galactic nuclei (AGNs), and kilonovae (see \citealp{Sagiv_2014} for an overview of the types of transients expected to show strong UV emission). Discovering such sources in the UV can provide important information about the early UV emission as well as how they evolve in time. Therefore, such studies could significantly help to understand the physics of the UV emission processes in these transients (which in many cases is still poorly understood) and potentially uncover previously unknown behaviour. Perhaps most importantly, since systematic, large-scale, blind UV transient searches have not been undertaken, such studies would have the potential to discover completely new types of sources.\

Despite the lack of dedicated UV transient facilities, currently operational UV telescopes mounted on satellites can be utilised to perform effective searches for serendipitous UV transients. These facilities have relatively small fields of view (FoV; typically up to a few arcminutes to at most a few tens of arcminutes), reducing the number of possible transients discovered per field studied. Nonetheless, one can take advantage of the repeating observations of the same fields that these telescopes frequently carry out in order to perform relatively large-scale transient surveys of the UV sky.\

To this end, we initiated the Transient UV Objects (TUVO) project (see Wijnands et al., in prep., for an overview of the TUVO project), with which we aim to study the UV transient sky. The instrument we primarily use in the TUVO project is the Ultraviolet and Optical Telescope (UVOT; see \citealp{Roming_2005,Breeveld_2010,Breeveld_2011}) aboard the \textit{Neil Gehrels Swift Observatory} (\textit{Swift}; see Sect. \ref{swift} for a description of the observatory and instrument). The reasons why, in the TUVO project, we so far have focused on the UVOT is because of the high flexibility and rapid pointing capabilities of \textit{Swift} (which allow a large number of fields to be observed multiple times) in combination with the accessibility of the data: all data are public and accessible within a few hours of the observations being performed. Furthermore, there are $\sim$17 years of archival UVOT data and up to a few hundred new observations each day. Since the UVOT is primarily used to follow up on individual previously discovered sources, this huge amount of data has hardly been explored to search for serendipitous UV transients. \ 

To undertake such a study, we built \texttt{TUVOpipe}\footnote{Note the dropped `S' with respect to the name of the pipeline in \cite{Modiano_2020}, which referred to \textit{Swift}. This designation of the facility was originally added to the name of the pipeline because different pipelines were initially envisaged for the different currently active UV instruments. However, we have now decided that if additional UV instruments are implemented in the TUVO project in the future, the processing of their data will be incorporated directly into the current pipeline by making any necessary modifications to the code (for details, see Wijnands et al., in prep.).}, a  pipeline that processes all newly available UVOT data every day  in order to search for UV transients. As the main goal of the TUVO project is to discover and study currently active transients, \texttt{TUVOpipe} primarily runs on the most recent \textit{Swift} data (called the 'Quick-Look' or 'QL' data; see Sect. \ref{uvot_data_structure}). We denote this default mode of \texttt{TUVOpipe} as the real-time mode. A secondary mode of the pipeline is available and can be used to process all archival observations of a given field, allowing `historical' unknown transients to be discovered (i.e. transients that were active during archival images and were likely never studied).

\section{Swift and UVOT}  \label{swift_uvot}

\subsection{Swift}  \label{swift}

\textit{Swift} was launched in 2004 with the primary science goal of understanding the origin of gamma-ray bursts (GRBs; see \citealp{Gehrels_2004} for a detailed description of \textit{Swift}). The facility houses three instruments: a gamma-ray detector called the Burst Alert Telescope (BAT; 15-150 keV; \citealp{BAT_2005}); the X-ray Telescope (XRT; 0.2-10 keV; \citealp{XRT_2005}); and the Ultraviolet and Optical Telescope (UVOT; 1700-8000\AA; see Sect. \ref{uvot} for more details). When a GRB is detected by the BAT, \textit{Swift} quickly repoints itself so that the XRT and UVOT point at the position of this GRB in order to obtain immediate X-ray, UV, and optical follow-up observations of the source. The rapid pointing capability is one of the unique features of \textit{Swift}. When GRBs are not being observed, \textit{Swift} carries out observations based on a schedule consisting of targets from a wide variety of Guest Investigator and Target of Opportunity observing programmes.\

\begin{figure*}
    \centering
    \includegraphics[width=.85\textwidth]{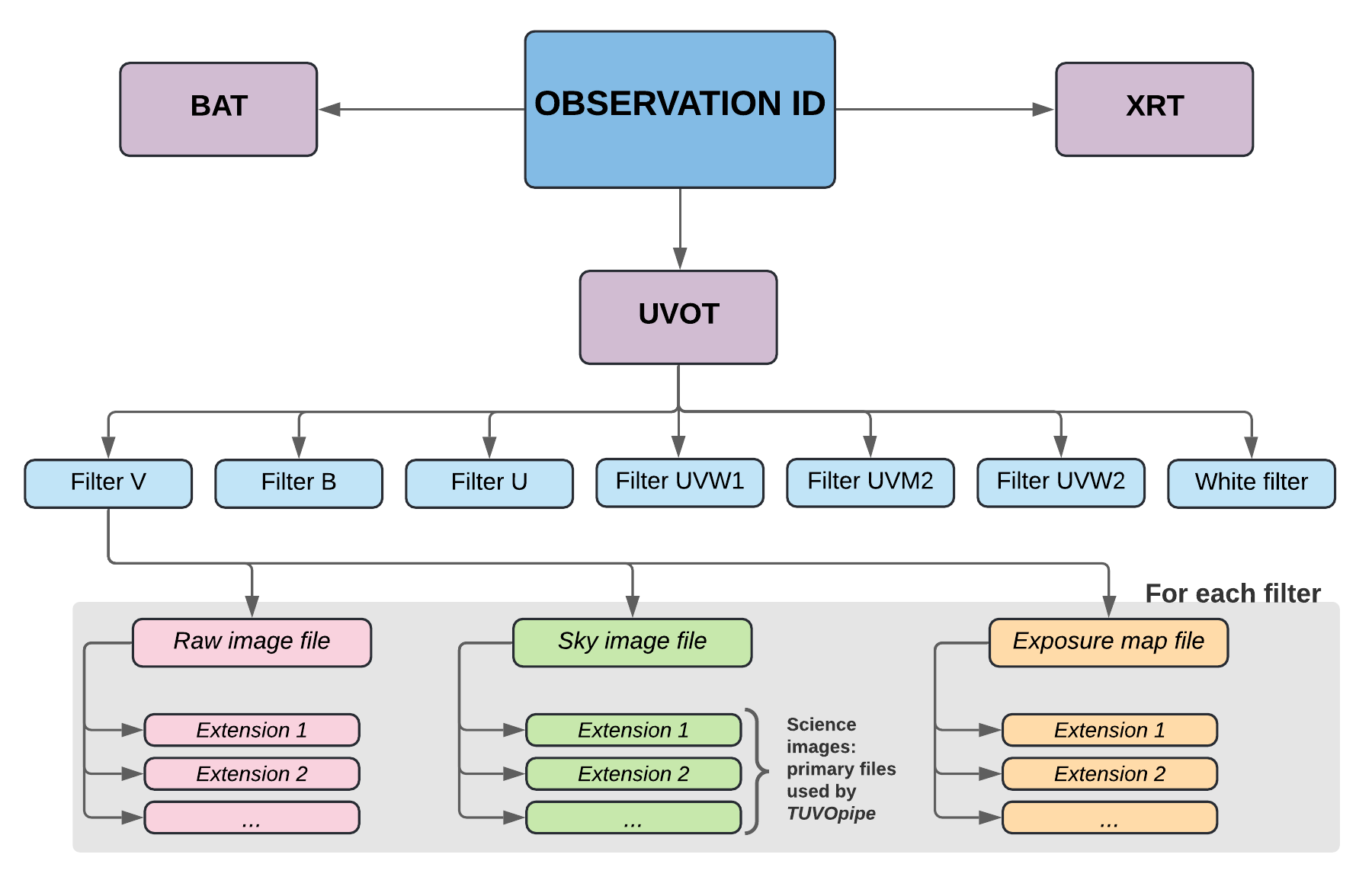}
    \caption{Data structure of Swift ObsIDs, which contain data obtained with the BAT, the XRT, and the UVOT. For the UVOT, the data are then further separated by filter, for all filters with which data were obtained for the given ObsID. We note that the UVOT is also equipped with a blocking filter, not shown in the figure. Within each filter, the data are separated into three primary files: the raw image file, the reduced sky image file, and the exposure map file (see Sect. \ref{uvot_data_structure}). Each of these three files is composed of a number of extensions, which can be extracted and analysed independently of each other. Each extension is an image created from a single UVOT exposure in a particular filter (i.e. a snapshot). Extensions of the different file types correspond to each other (e.g. extension 1 of the raw image file is used to create extension 1 of the sky image file).}
    \label{fig:uvot_data_structure}
\end{figure*}

Each \textit{Swift} observation is identified with a unique Observation ID (ObsID). The data of each ObsID consist of all BAT, XRT, and UVOT data obtained during that specific observation. Each observation is further split into the individual exposures undertaken for that observation (i.e. Swift may observe a target several times in one day, and all exposures are then part of the same ObsID; see Sect. \ref{uvot_data_structure}  and Fig. \ref{fig:uvot_data_structure} for a full description of the structure of UVOT data files).\ 
A few times per day, all of the newly obtained \textit{Swift} data are down-linked to the primary ground station in Malindi, Kenya, after which they are processed and uploaded to the Quick-Look (QL) page\footnote{\url{https://swift.gsfc.nasa.gov/sdc/ql/}}. This means that preliminary \textit{Swift} data from all three instruments are typically accessible to the public just 4-6 hours after they are taken. We note that `preliminary' here does not mean raw; all public data have already been processed automatically by the \textit{Swift} reduction pipelines. However, the data on the QL page are updated as additional exposures are carried out; these are then added to their corresponding ObsID with every down-link of data. With every such down-link, all QL data are also reprocessed with updated housekeeping information and tagged with a new version number. When the data corresponding to an ObsID have been on the QL page for $\sim$1 week, they are fully updated with all exposures that the ObsID will comprise and are processed with the most up-to-date housekeeping information. At this point, the data are transferred to the High Energy Astrophysics Science Archive Research Centre (HEASARC)\footnote{\url{https://heasarc.gsfc.nasa.gov/cgi-bin/W3Browse/swift.pl}} for archiving and are not modified further (unless a calibration update is deemed significant enough to warrant a `grand reprocessing' of the whole archive\footnote{See \url{https://www.swift.ac.uk/archive/reprocessing.php} for details.}.

\subsection{UVOT characteristics} \label{uvot}

The UVOT is the primary instrument used by the TUVO project (see \citealp{Roming_2005,Breeveld_2010,Breeveld_2011}, also see the UVOT instrument page\footnote{\url{https://swift.gsfc.nasa.gov/about_swift/uvot_desc.html}}). In short, the UVOT is a diffraction-limited, 30cm Ritchey-Chretien reflector with a 17'x17' FoV and a photon-counting CCD detector, constructed with the primary science goal of detecting and characterising the optical and UV afterglows of GRBs. The UVOT is equipped with six filters covering the UV and optical wavelength range 1700-6500 \AA~(see Table \ref{tab:uvot_tab} for details of the central wavelengths and full width at half maximum, FWHM, of the filters), plus a white filter covering the 1700-8000 \AA\ range. In addition, a blocked filter, a magnifier, and two grisms are available for use during UVOT observations, though data obtained using these filters are not used in the TUVO project; we only process data that were obtained using the seven primary UV and optical filters. 

\begin{table}[]
    \centering
    \begin{tabular}{c|c|c}
        \textbf{Filter} & \textbf{Central Wavelength ($\AA$)} & \textbf{FWHM ($\AA$)} \\ \hline
        V & 5468 & 769 \\
        B & 4392 & 975 \\
        U & 3465 & 785 \\
        UVW1 & 2600 & 693 \\
        UVM2 & 2246 & 498 \\
        UVW2 & 1928 & 657 \\
    \end{tabular}
    \caption{Central wavelengths and FWHM of the six primary UVOT filters. Table adapted from \citet{Breeveld_2011}.}
    \label{tab:uvot_tab}
\end{table}

Most \textit{Swift} observations contain UVOT exposures\footnote{The rare exceptions are typically fields that contain very bright stars that would damage the detector, and hence the blocked filter is used.}. UVOT observations are by default performed  using the `filter of the day' \citep{Page_2014}, which involves cycling through the U, UVW1, UVM2, and UVW2 filters daily. This filter strategy is not used if the principal investigator (PI) of an observing programme specifies a different filter or set of filters. As with all \textit{Swift} ObsIDs, any UVOT ObsID is composed of one or multiple snapshots, which are individual exposures typically taken throughout the course of one day.\

\subsection{UVOT data structure}  \label{uvot_data_structure}

The UVOT data within an ObsID are first separated by filter (see Fig. \ref{uvot_data_structure}). The primary data components of a UVOT observation for a particular filter are a raw image, a sky image, and an exposure map\footnote{Exposure maps are arrays with each pixel representing the effective exposure time of the corresponding pixel in the sky image (see the instrument software guide at \url{https://swift.gsfc.nasa.gov/analysis/UVOT_swguide_v2_2.pdf} for details).} (see Fig. \ref{fig:uvot_data_structure} for a visual representation of the structure of UVOT data). Raw images are known as `Level 1' images; these data are processed by the standard UVOT reduction pipeline\footnote{\url{https://swift.gsfc.nasa.gov/quicklook/swift_process_overview.html}} in order to produce the sky images (the `Level 2' images). Level 2 images have been flat-fielded, cleared of bad pixels, and astrometrically solved to a precision of a few arcseconds \citep{Poole_2008,Breeveld_2010}. The standard UVOT reduction pipeline also attempts to astrometrically solve images to  $\leq0.5$" (\citealp{Poole_2008,Breeveld_2010}). However, this advanced solution may fail on certain fields; for example, when the UV sky is very different from the optical (UVOT images are astrometrically solved by matching source positions to sources in the USNO-B1 catalogue; \citealp{Monet_2003}) or when fields are very crowded \citep{Breeveld_2010}. In such cases, the misalignment between UVOT images of the same field may be up to several arcseconds.\ 

Primary UVOT data files (sky images, raw images, and exposure maps) are in the Flexible Image Transport System (FITS; \citealp{FITS_2010}) format, and each file type is comprised of up to several extensions (see Fig. \ref{fig:uvot_data_structure}). Each individual extension is a single exposure, or snapshot (see Sect. \ref{uvot}). Extensions of the three different file types correspond to each other: for example, extension 1 of the raw image file is the raw image used to produce extension 1 of the sky image file. QL ObsIDs are updated with each subsequent data down-link: additional snapshots taken since the previous down-link may be added to the ObsID, all images are reprocessed with the standard UVOT reduction pipeline and updated housekeeping information, and each image is tagged with a new version number (reflecting the number of times it has been reprocessed). All data within an ObsID were typically obtained by \textit{Swift} over the course of one day, so the extensions within each file represent exposures separated by a few hours to a day. In \texttt{TUVOpipe}, we made use of the Level 2 sky images and their exposure maps.

\section{TUVOpipe}  \label{tuvopipe}
To automatically process the UVOT data and search for serendipitous UV transients, we used our purposely built pipeline, \texttt{TUVOpipe}. In this section, we describe all aspects of the operation of \texttt{TUVOpipe}in detail. The pipeline can run in two different modes: real-time or archival. The real-time mode is of prime interest to the TUVO project since it discovers (in the QL data) sources that are currently active and can therefore be studied in detail with follow-up observations. Therefore, in this paper when discussing \texttt{TUVOpipe,} we generally refer to the real-time mode. The functionality of the two modes is, however, very similar; any discrepancies are briefly discussed in the relevant sections.\

\texttt{TUVOpipe} functions in five parts, each with a different purpose. Part I (PI) downloads all required data; Part II (PII) searches for candidate transients using difference imaging and performs some tests to filter out false positive detections; Part III (PIII) creates light curves of all remaining candidates; Part IV (PIV) produces preliminary classifications of each candidate through light-curve analysis and source matching with external astronomical catalogues; and Part V (PV) creates long-term light curves of the most interesting candidates by using all archival UVOT observations of the field. By default, parts PI-PIV are linked, and thus each of these parts is initiated upon completion of the previous part. PV is only run when a candidate transient of sufficient interest is found. However, the scripts can also be used independently; for example, when undertaking tests for improvements or bug fixes.\

\texttt{TUVOpipe} is written entirely in \texttt{python} (v3.7)\footnote{Python Software Foundation, \url{https://www.python.org/}} and makes use of several software packages. The primary software components integrated in the pipeline include several \texttt{astropy}\footnote{\url{https://www.astropy.org/}}\citep{astropy:2018} packages, the image subtraction software \texttt{hotpants} (which is an abbreviation of High Order Transform PSF and Template Subtraction; \citealp{Becker_2015}; see Sect. \ref{image_subtraction}), and HEASARC's software suite HEASoft\footnote{\url{http://heasarc.gsfc.nasa.gov/ftools}} (v6.28). The latter requires the calibration database CALDB\footnote{\url{https://heasarc.gsfc.nasa.gov/docs/heasarc/caldb/swift/}} (we used version 20201215). A few additional software packages are used at various stages of the pipeline; these are described and referenced in the relevant sections in our paper.\

Currently, \texttt{TUVOpipe} is run daily by members of the TUVO project. In Figs.~\ref{fig:flow_pi}, \ref{fig:flow_pii}, \ref{fig:flow_piii}, \ref{fig:flow_piv} and \ref{fig:flow_pv}, we show the workflow of the various parts of the pipeline. The details of \texttt{TUVOpipe} are discussed in the following sections.

\subsection{User inputs}  \label{user_inputs}

\texttt{TUVOpipe} runs with minimal user input: only a parent directory on a user's local machine needs to be specified via a small input text file. There are many adjustable parameters within the scripts that can be modified for testing and improvements, though they are now generally considered optimal and are therefore typically not modified from their default values. We note that `optimal' concerns the best parameters for the majority of UVOT fields. The pipeline may perform better with different parameters for some specific fields - for example, very crowded fields such as globular clusters (see \citealp{Modiano_2020}) - but so far it has not been possible to implement this programmatically within the code. This is both because it is difficult for the code to automatically recognise fields as certain types (e.g. those with diffuse emission) and because it is difficult to determine sets of parameters that would work optimally for all fields of a given type.\ 
One item the user can manually specify is which UVOT filters should be processed by the pipeline. However, by default all filters are selected, causing the pipeline to run on each filter independently, since only images taken in the same filter can be compared to each other to search for transients. For the archival mode of \texttt{TUVOpipe}, three additional user inputs are required (compared to the real-time mode), namely the right ascension (RA) and declination (Dec) of the field of interest and the search radius (see Sect. \ref{pi} for details).

\subsection{Part I}  \label{pi}

\begin{figure}[!t]
    \centering
    \includegraphics[width=.49\textwidth]{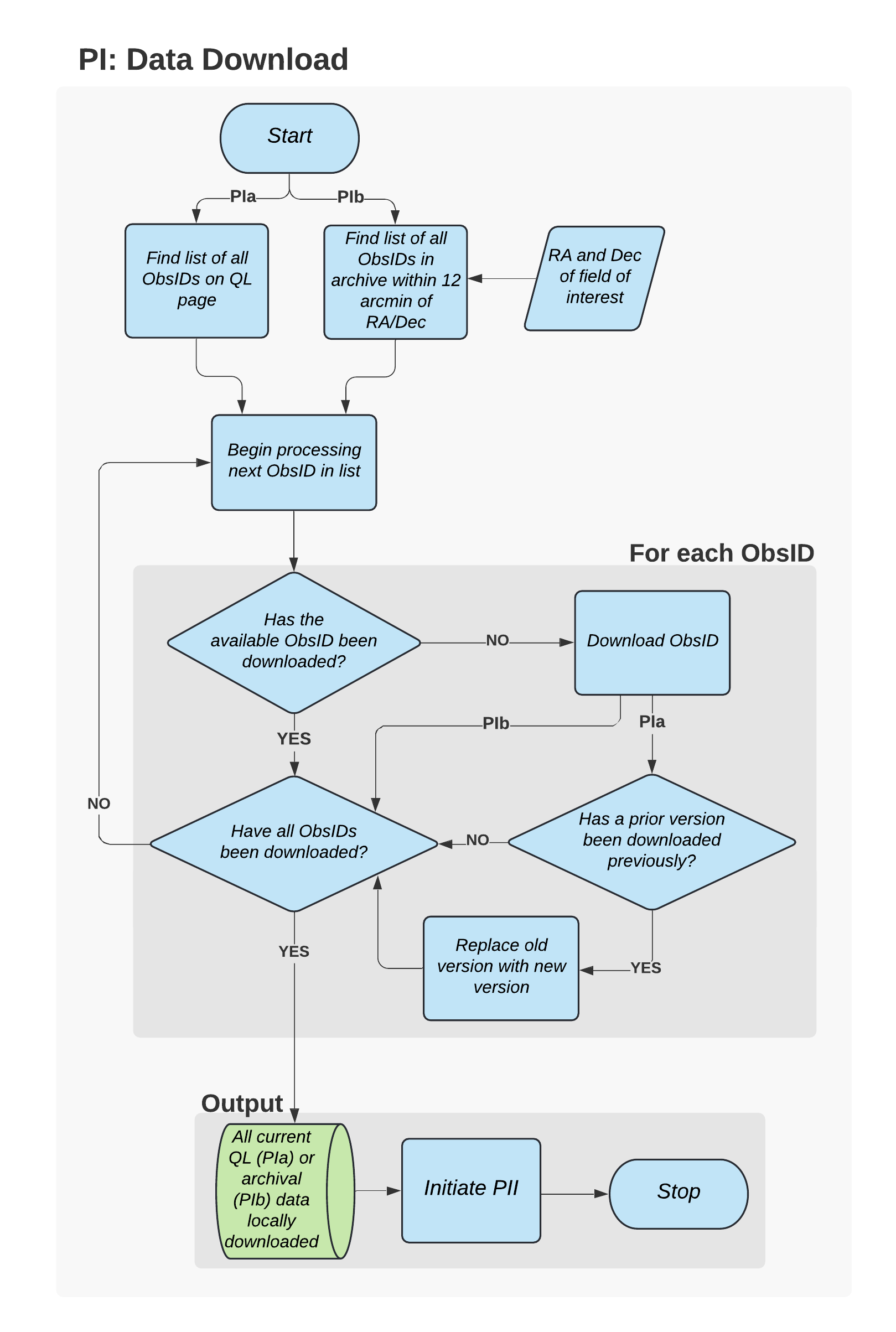}
    \caption{Workflow of PI (discussed in detail in Sect. \ref{pi}. This part of \texttt{TUVOpipe} downloads all UVOT data used to search for transients. In the archival mode, the RA and Dec of the field of interest is required as input. In the real-time mode, no input is required, since it always downloads all data available on the QL page at the time when PI is initiated. In the archival mode, the ObsID version check is bypassed, since ObsID versions of data in the archive are not further updated (except in rare cases when a reprocessing of the entire archive is carried out; see Sect. \ref{swift}). Upon completion of PI, Part II (PII) of the pipeline (see Sect. \ref{pii}) is automatically initiated.}
    \label{fig:flow_pi}
\end{figure}

Part I (PI) of \texttt{TUVOpipe} downloads the UVOT data to be processed. Since the data to be downloaded differ in real-time and archival mode, PI consists of two versions: PIa for the real-time mode and PIb for the archival mode. The downloading processes for the two versions are slightly different (see Fig. \ref{fig:flow_pi} for a chart displaying the workflow of PI in both modes). \

In PIa, first a list is created of all the fields and ObsIDs available on the current \textit{Swift} QL page. As discussed, new data are transferred from the QL page to the archive after around one week, so the list created in PIa includes all data taken up to around one week before the time when PIa is initiated. This typically consists of up to several hundred observations of many different fields. PI then loops through each ObsID in the list to check if a download is required. If there are UVOT observations for a given ObsID, then a download of the Level 2 sky images and corresponding exposure maps is carried out if either the ObsID has not been downloaded before (i.e. it is not present in the user's local archive), or if the ObsID has already been downloaded but a newer version of the data is available (e.g. if in a previous run of PIa a given ObsID was downloaded, and before the next run of PIa, a further image was taken by UVOT and added to that ObsID). In the latter case, PIa detects the newer version and performs the download, replacing the old version with the new version. After looping through all available observations in this way, PIa is completed. Since we run \texttt{TUVOpipe} once per day, with each run one day's worth of new observations are downloaded, as well as any updated versions of older observations. Since the data remain on the QL page for around one week, \texttt{TUVOpipe} is guaranteed to process all available data unless it runs less frequently than once per week.\

In the case of PIb, the user provides the RA and Dec of the field of interest. The user also selects a search radius, which defines the distance from the specified RA and Dec out to which UVOT data will be downloaded - in other words, all UVOT images with pointings within the search radius from the input coordinates. Typically, a radius of 12' is chosen, meaning that all archival UVOT images with pointings $<$12' from the specified RA and Dec are downloaded for processing. This value is chosen because the FoV of the UVOT is 17'x17', so even images whose centres are offset with respect to each other's by 12' will still have considerable overlapping areas. This is important because to feasibly perform transient searches, images must have large overlapping areas. This is necessary both in order to successfully align images (see Sect. \ref{pii_dataprep}) and to avoid very inefficient searches, since searching for transients in images with very small overlapping areas would yield very few transients. PIb then creates a list of all such available observations from the \textit{Swift}-UVOT archive and downloads all corresponding ObsIDs similarly to the operation of PIa. To avoid unnecessary downloads, checks for previously downloaded ObsIDs in the user's local directory are carried out.\

\subsection{Part II} \label{pii}

\begin{figure*}
    \centering
    \includegraphics[width=0.95\textwidth]{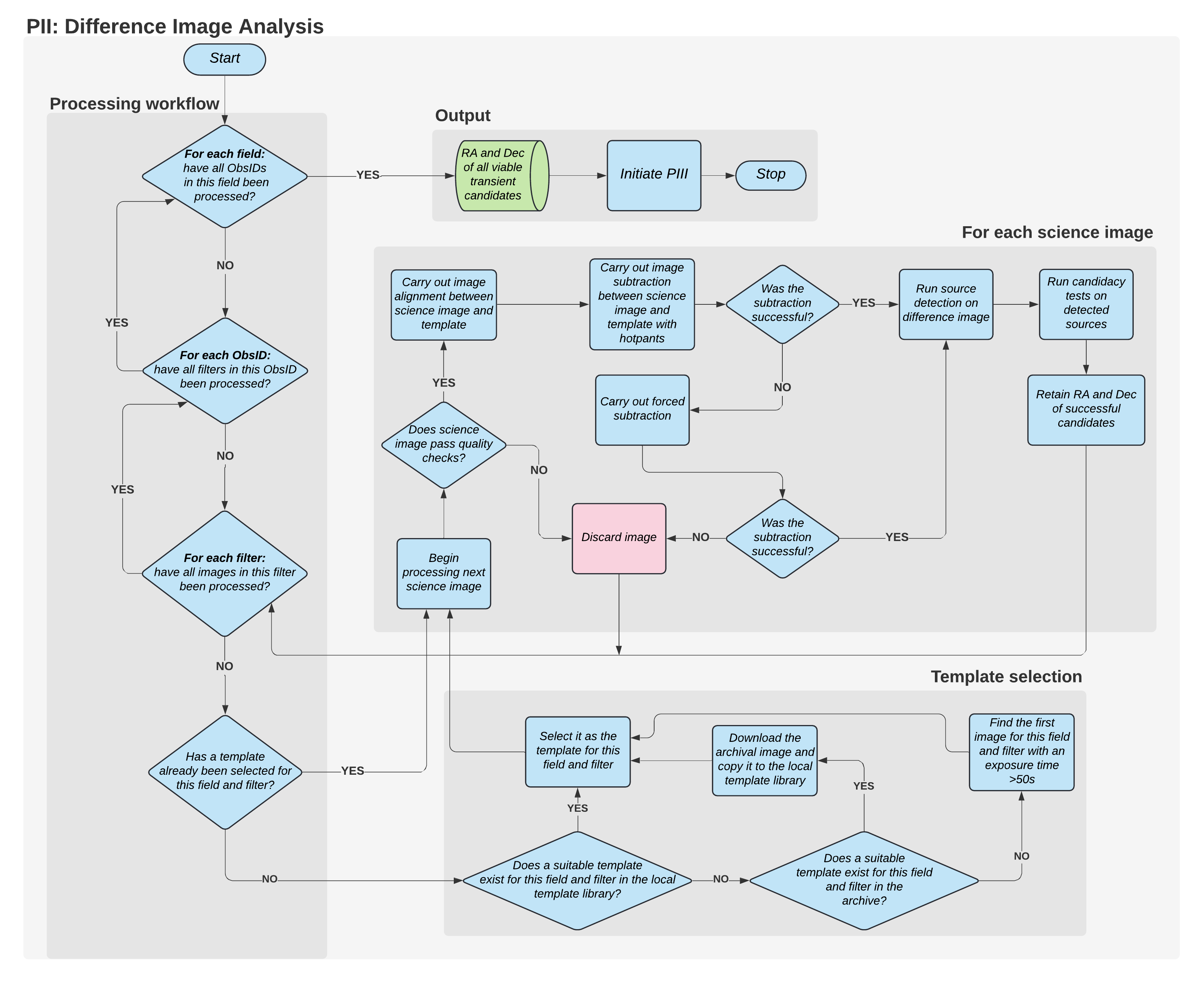}
    \caption{Workflow of PII (discussed in detail in Sect. \ref{pii}). This is the part of \texttt{TUVOpipe} that searches for transient candidates in UVOT images through difference image analysis. The operation of PII is identical for the real-time and archival modes, since in archival mode it simply stops after the field of interest is completed. When PII terminates, it has produced lists of candidate transients for each field and filter. These are passed to Part III of the pipeline (see Sect. \ref{piii}), which is automatically initiated upon completion of PII.}
    \label{fig:flow_pii}
\end{figure*}

Once all new UVOT data have been downloaded, the search for candidate transients is carried out through Part II (PII) of the pipeline (see Fig. \ref{fig:flow_pii} for a chart showing the workflow of PII). PII consists of several levels of processing, taking UVOT sky images as input and ultimately producing lists of viable candidate transients for each field and filter. The workflow of PII is such that all observations of all fields present in the user's local archive are processed in sequence automatically. Each field is checked to determine if it has already been processed by PII. If so, PII moves to the next field. When a field is encountered that has not been processed by PII (i.e. the pipeline determines that there are unprocessed ObsIDs within the field), all ObsIDs of the field are checked. When an ObsID that has not been processed is encountered, PII begins processing that ObsID. All PII processing steps are described in this section and outlined in Fig. \ref{fig:flow_pii}.

\subsubsection{Data preparation}  \label{pii_dataprep}

For each ObsID that requires processing, PII carries out some initial data processing steps necessary for the image subtraction  (see Sect. \ref{image_subtraction}) to perform optimally. For each ObsID, all extensions are extracted using the HEASoft \texttt{fextract}\footnote{\url{https://heasarc.gsfc.nasa.gov/lheasoft/ftools/fhelp/fextract.html}} tool. This is done because the extensions may not always be perfectly aligned to each other, so we later perform the image alignment on all extensions. Additionally, by processing each extension separately we obtain a larger number of data points for the light curves of each transient we detect (see Sect. \ref{piii}).\ 

Transient searches in \texttt{TUVOpipe} require at least two images of the same field: a template and a science image. The field for each image is defined by the field name assigned to it by \textit{Swift}, which refers to the target of the observations. The science and template images must be taken in the same filter; all filters are therefore processed independently in \texttt{TUVOpipe}. PII begins by processing each field independently. In any field directory, one image is selected as the template, and all other images are defined as science images. The template is the image with respect to which all science images will be aligned and subtracted (i.e. it is the image to which all the science images will be compared when searching for transients). For each field, PII thus begins by checking if a template image has already previously been defined within a given set in an ObsID. If it has, then it processes that ObsID. If no template image has been selected, it attempts to select a suitable template image. First, a template image library, located in the user's local machine and containing many UVOT archival images, is searched. We search for template images that are in the required filter, have $>200s$ exposure time (higher quality templates are preferred for image subtraction, so we try to avoid using low-exposure images as templates), and have a pointing within 12' of the first image in the ObsID being processed. If no such template is found in the local template library, PII undertakes the same search but in the entire UVOT archive. Archival images are preferred as templates because if a transient is variable on timescales larger than a few days to a week, then comparing only QL images (that at most span a time of approximately a week) may not reveal any variable behaviour. By selecting an archival template taken weeks to years previously, sources that vary on a larger range of timescales can be identified. If a suitable template is found in the archive, it is downloaded and selected as the template for the set. It is also copied to the local template image library. We note, therefore, that the local template image library contains only copies of UVOT archival images; it is simply a way of reducing processing time as it avoids downloads. If no suitable template image is found in the archive, the first science image of the field and filter being processed that passes the exposure time and image quality checks (basic checks performed for every image we process, described below) is selected as the template.\

\begin{figure}
    \centering
    \includegraphics[width=0.45\textwidth]{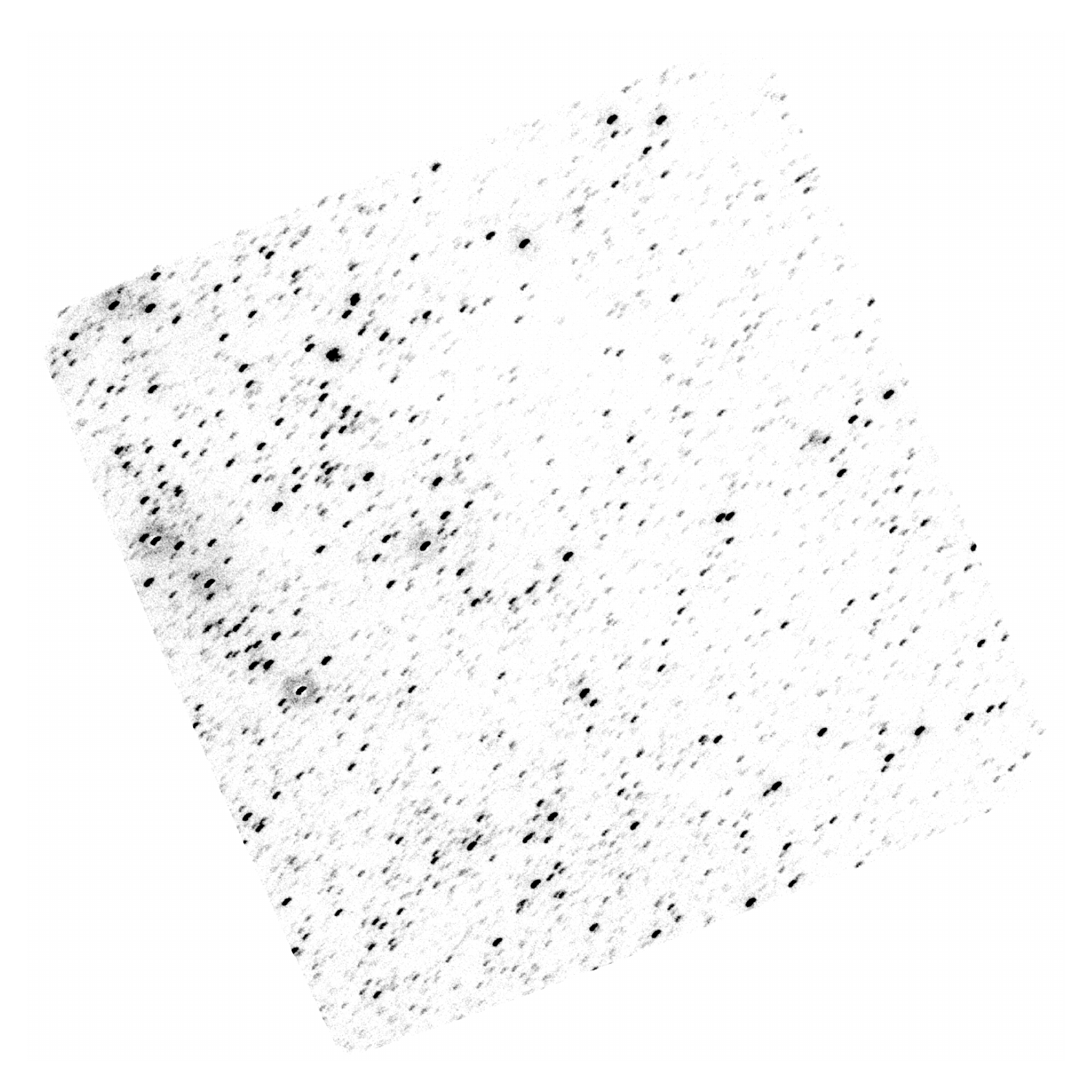}
    \caption{Example of a UW2 image (ObsID 00048753209, extension 2, field ASASSN-20ni) that did not pass the image quality check and was therefore rejected by PII and not processed further in \texttt{TUVOpipe}. The measured mean elongation for sources in this image is 1.6 (i.e. greater than the cut-off of 1.4).}
    \label{fig:elongation}
\end{figure}

After selecting a template, PII begins processing all science images. For each science image, PII carries out an exposure time check and an image quality check. If the image has an exposure time $\leq$50s then it is flagged and not processed in PII. We implemented this check because when using lower exposures times, we found that the image subtraction performs poorly (typically, the difference images produced contain many artefacts). We note that although these science images are not processed in PII, they are not discarded, as they are still used later in the pipeline to create light curves (see Sect. \ref{piii}). Next, an image quality check is carried out in order to improve the performance of PII by rejecting science images that are not usable to search for transients. The HEASoft tool \texttt{uvotdetect}\footnote{\url{https://heasarc.gsfc.nasa.gov/ftools/caldb/help/uvotdetect.html}}, the standard tool used for source detection in UVOT images, is run on the image. For each detected source a measure of its elongation is obtained: if the total mean source elongation in the image is larger than 1.4 (a user-defined value chosen empirically), then the image is rejected and not processed further at any stage of \texttt{TUVOpipe}. This check prevents low-quality images (e.g. those in which stars may be elongated due to either a telescope tracking error or the exposure commencing before the telescope had finished slewing) from being processed. In Fig. \ref{fig:elongation}, we show an example of such an image. They represent $\lesssim$1\% of UVOT data that we process, both by number of images and by exposure time. \ 

\begin{figure}
    \includegraphics[width=0.45\textwidth]{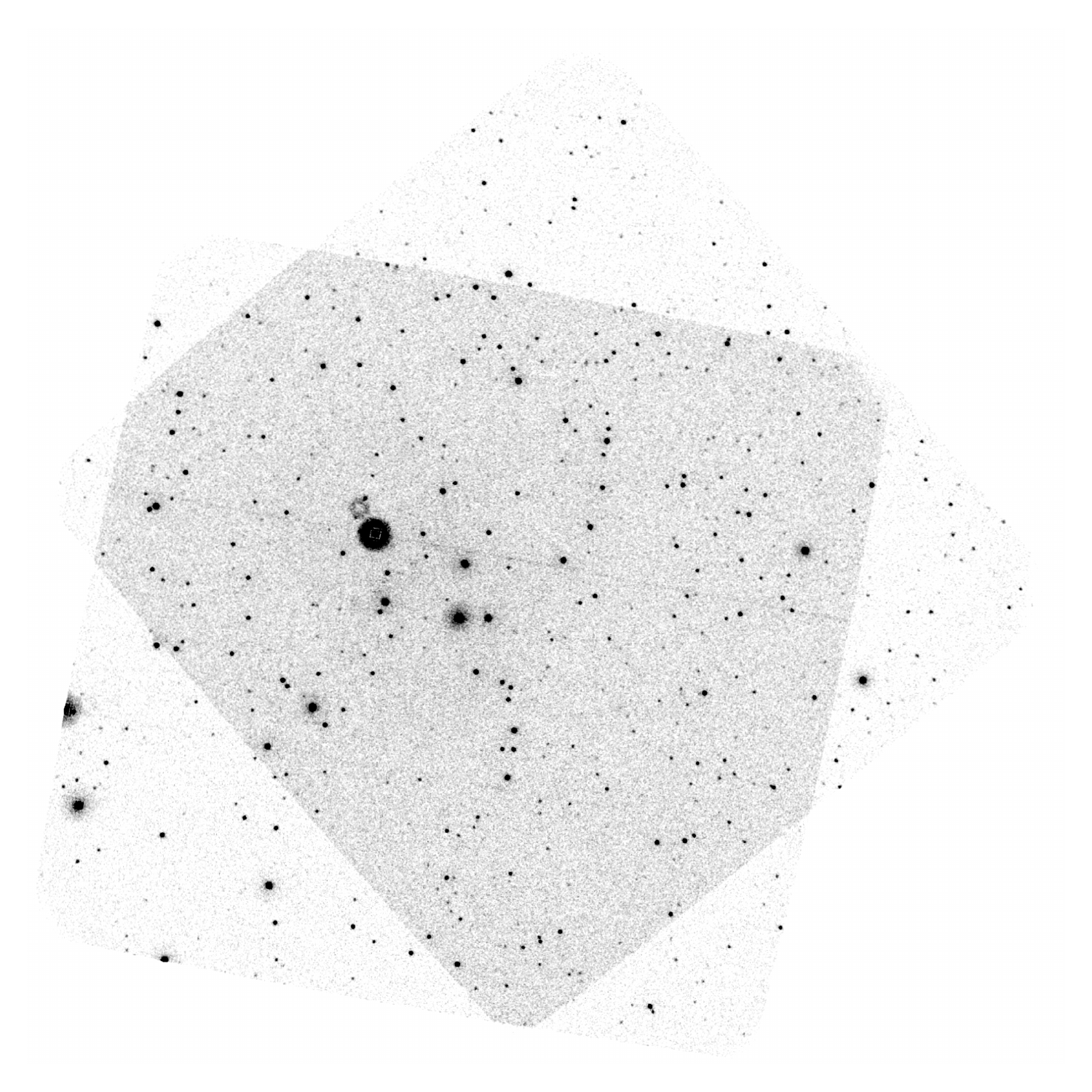}
    \caption{Example of translational and rotational offset between two UVOT images of the same field, taken around one month apart. The image is composed of two stacked UVOT UVW1 exposures of the field of Nova Per 2020. The ObsIDs and extensions of the images are 00013909005[1] and 00013909010[1]. The position angle of the telescope was $\sim$40$^{\circ}$ offset between the two images ($320^{\circ}$ for ObsID 00013909005; $280^{\circ}$ for ObsID 00013909010). The displayed images clearly show the result of this rotational offset, as well as a translational offset due to the pointing accuracy of \textit{Swift}. Searches for transients can be performed within the overlapping area. To prepare images for alignment and image subtraction in PII, the images are cropped (and saved to two new images) so as to contain only the overlapping area.}
    \label{fig:crop}
\end{figure}

For each science image that passes these initial checks, the science and template images are cropped such that two new images are created that cover the maximum overlapping area between the original science and template images. This is necessary because the image alignment routines used require input images to be of the same shape and size, and due to the pointing accuracy of the UVOT (up to a few arcminutes\footnote{\url{https://swift.gsfc.nasa.gov/proposals/tech_appd/swiftta_v14/node23.html}}) and the orientation of the images in subsequent images (usually the images are obtained at different roll angles of the satellite), this is not usually the case. To illustrate this effect, we stacked two UVOT images of the same field and filter taken at different times, using the \textit{HEASoft} tools \texttt{fappend}\footnote{\url{https://heasarc.gsfc.nasa.gov/lheasoft/ftools/fhelp/fappend.txt}} and \texttt{uvotimsum}\footnote{\url{https://heasarc.gsfc.nasa.gov/lheasoft/ftools/headas/uvotimsum.html}} (see Fig. \ref{fig:crop} for an illustration of this effect).\ 

Since image subtraction requires precisely aligned images, the next step in PII is to align each (cropped) science image to the (cropped) template. Image alignment is carried out using the image registration tool \texttt{register\_translation} of the \texttt{scikit-image}\footnote{\url{https://scikit-image.org/}} package, which uses fast Fourier transforms and cross-correlation to perform image translation with sub-pixel accuracy (see \citealp{Guizar_2008}). The tool returns the X and Y pixel translation  required to align the science image to the template image. This translation is then applied to both the cropped science image and the original, uncropped science image. The former will be used for image subtraction (since only the parts of the image that overlap between the science and template are useful to search for transients); the latter will be used to create light curves of our candidate transient sources, where using full images is optimal in order to maximise the number of data points per light curve (see Sect. \ref{piii}).\ 

We note that the template chosen by PII is not necessarily the image with the highest astrometric precision. Solving UV images astrometrically can be difficult, in particular when the UV field looks significantly different compared to optical images of the same field. Since at the PII stage we are only interested in detecting transients, having the most precise astrometric coordinates for each candidate is not vital; thus, we simply ensure that images are correctly aligned relatively to each other. This means that the error on the position for each transient we detect (e.g. all the coordinates given in the example transients we show in Sect. \ref{results}) can be up to a few arcseconds (see Sect. \ref{uvot_data_structure}; see also \citealp{Poole_2008,Breeveld_2010}).\

On some occasions, the image alignment process fails. It is not always clear what the cause of a failed alignment is, though certain kinds of sky fields are more likely to be problematic for alignment, such as very empty (meaning very few stars detected), or conversely, very crowded fields. Although the tool we use always returns some alignment (i.e. it does not explicitly state when alignment has failed), we performed extensive tests to identify incorrectly aligned images and found that these occur most frequently when the returned translation necessary to align the images is found to be greater than 8 pixels. Therefore, if the returned translation required is above this value, we do not align the images at all, and we continue to process them using the coordinates automatically assigned by the UVOT reduction pipeline. This is because UVOT images are often already well aligned before we process them, but on occasion the alignment tool may still fail and attempt a $>8$ pixel shift. In these cases, by refusing the translation but continuing to process the images, we may still obtain useful data. In our pipeline, these images are flagged as unaligned images, and corresponding data points in the light curves we create are labelled as such (see Sect. \ref{piii}). We note that on occasion, the alignment tool may also not correctly align images even when the returned translation is less than 8 pixels. Therefore, poor alignments are not always guaranteed to be flagged automatically by our code, though this happens rarely and is easily recognised by users when inspecting the image stamps in the final output of the pipeline (see Fig. \ref{fig:PIV_output}). \

\begin{figure*}[!t]
    \centering
    \includegraphics[width=0.85\textwidth]{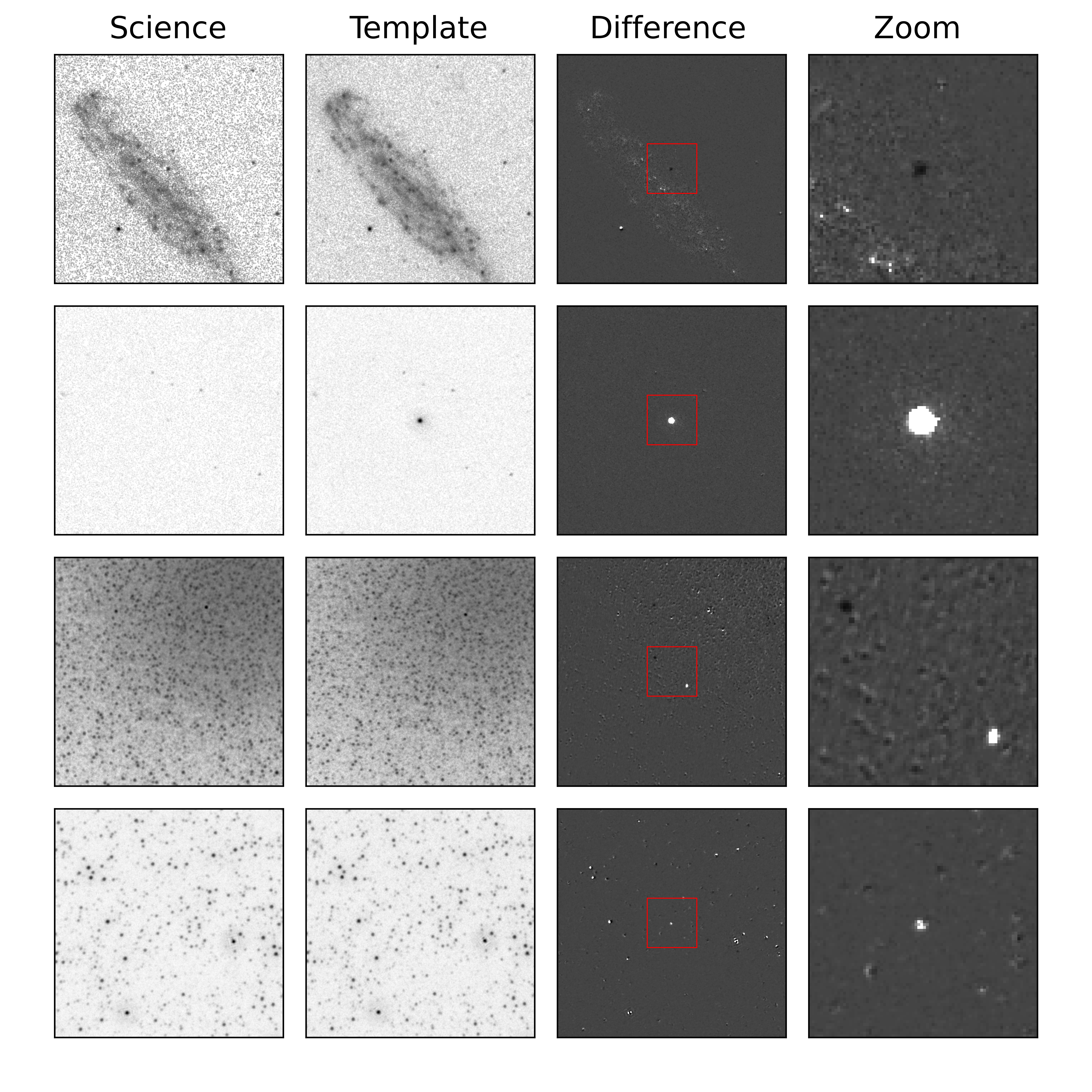}
    \caption{Examples of transient candidates found through the image subtraction process. Each row displays examples for a particular field and filter in which transients were found. From left to right, the columns show the science image, the template image, the difference image, and a zoomed-in view of the difference image to display one or more transients. The red square in the difference images (third column) shows the region displayed in the fourth column. The examples displayed were chosen to show different types of fields observed by UVOT and thus processed by TUVO. From top to bottom, the rows show a field with diffuse emission (where the transient is the supernova SN2021pit), a very empty field (where the transient is the X-ray binary SWIFT J1357.2-0933), a very crowded field with diffuse emission (where the transient in the top left of the zoom stamp is the star Cl* NGC 5139 NJL 47 and the transient in the bottom right is the star Cl* NGC 5139 NJL 14), and a crowded field (where the transient shown is the RR Lyrae OGLE LMC-SC3 324213). In the displayed stamps, positive flux is shown as black, so `negative' transients, that is, those that were brighter in the template image than in the science image, appear white in the difference images; and `positive' transients, representing sources that are brighter in the science image than in the template image, are black in the difference images. The field names, ObsIDs, and extensions of the science and template images shown (from the top left) are SN2001EL-HOST/00049929023[1]; NGC1448/00031031001[5]; SwiftJ1357.2-0933/00031918123[1]; SwiftJ1357.2-0933/00031918002[1]; OmegaCentauri1/00037520001[2]; OmegaCentauri1/00037520004[1]; LMCFIELD6/00094074015[1]; and LMCFIELD6/00094074007[1]).}
    \label{fig:image_subtraction_main}
\end{figure*}

\subsubsection{Image subtraction} \label{image_subtraction}

Once a science image is fully prepared (see Sect. \ref{pii_dataprep}), image subtraction is carried out for the template with respect to it. This is done using using the \texttt{hotpants} \citep{Becker_2015} software. This routine was constructed based on the image subtraction algorithm developed by \citet{Alard_1998} and \citet{Alard_2000}. It works by first constructing two matching stamps (stamp-pairs) from the science and template images, and then convolving the point spread functions (PSFs) of the pairs (on the assumption that the stamps are small enough that PSF variations within it for a given image are negligible). The PSF matching\footnote{We note that, typically, PSF matching is crucial for image subtraction using data from ground-based facilities because sky conditions can significantly affect the PSF of different images; space-based telescopes such as UVOT do not suffer from these sky effects. However, the PSF of sources in UVOT images can also change, for example due to temperature changes in the telescope during observations, or due to the location on the CCD in which a particular source falls (i.e. the PSF of a source that falls on the centre of the UVOT CCD in one image may be different to the PSF of that same source if it is located near the corner of the CCD in another image).\label{psf_fn_label}} is then modelled across the entire image by a combination of the functions obtained for each stamp pair. By default, the science-image PSF is convolved to match that of the template. In addition to the PSF convolution, \texttt{hotpants} carries out flux normalisation (to account for exposure time differences in the science and template images), before finally subtracting the two PSF-convolved and flux-normalised images pixel-by-pixel, producing a difference image. We run the image subtraction in both directions, that is, first subtracting the template image from the science image and then vice-versa. This is because the source detection software we use (see Sect. \ref{pii_sextractor}) only picks up positive sources in the difference images, so to detect transients that were brighter in the science than the template image and also vice-versa, we need two difference images, with the second created by flipping the direction of the subtraction used in the first (we denote this the `flipped difference image').\ 

In some cases, such as when processing very empty fields, \texttt{hotpants} cannot find suitable sources with which to match the PSF of the stars in the science image to those in the template image, and the subtraction fails. Within \texttt{TUVOpipe}, this occurs on $\sim10\%$ of subtractions. In these cases, PII performs a `manual' subtraction by simply normalising the flux in the two images by their exposure time and then carrying out a pixel-by-pixel subtraction. This produces difference images that are less accurate than when using \texttt{hotpants} (because no PSF matching was performed), but in these difference images transients can still be detected. In Fig. \ref{fig:image_subtraction_main} we show examples of difference images produced.

\subsubsection{Source extraction and candidacy tests} \label{pii_sextractor}

\begin{figure*}[!t]
    \centering
    \includegraphics[trim={2.5cm 0 0.5cm 0},clip,width=.95\textwidth]{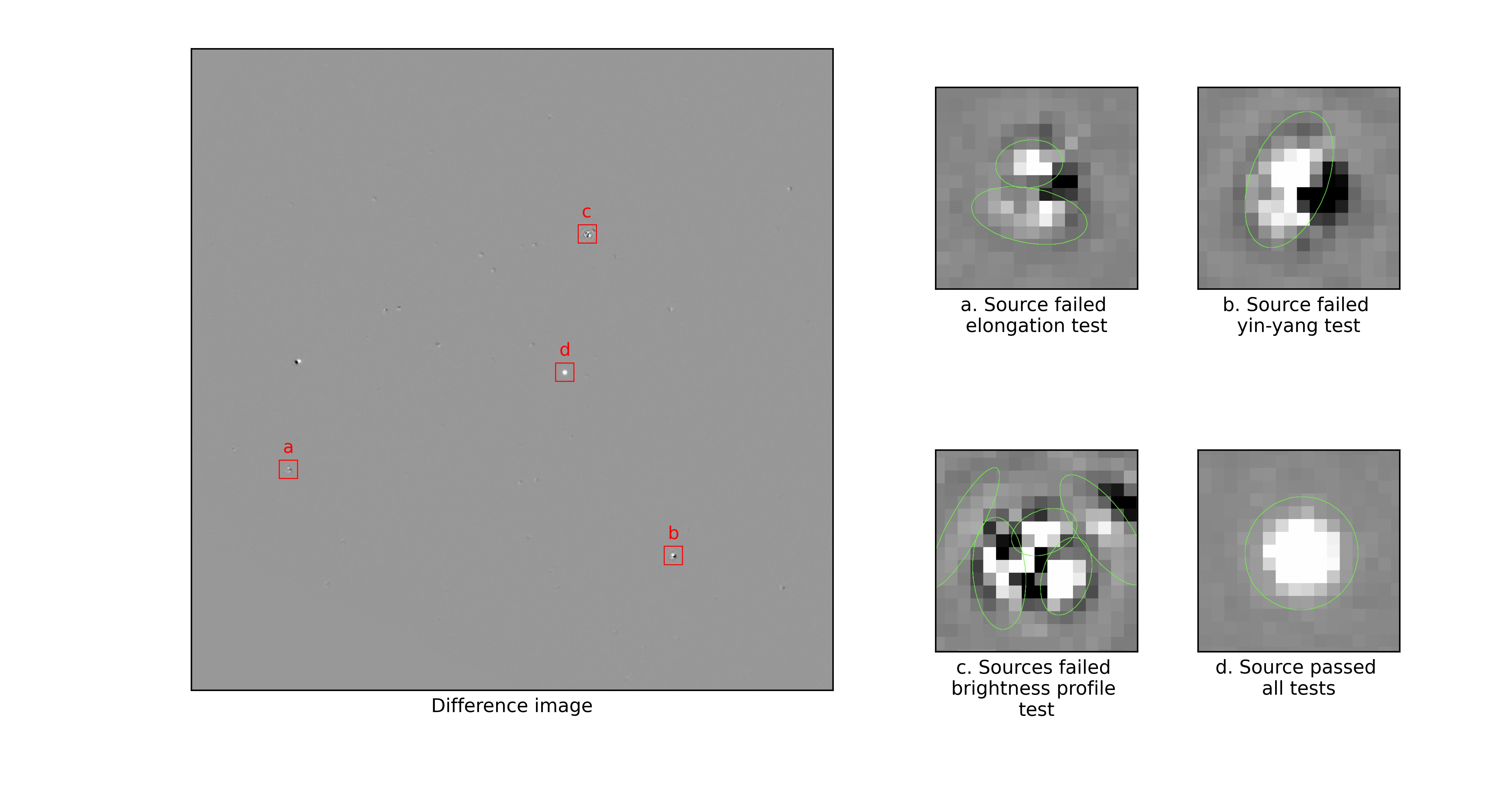}
    \caption{Example of a difference image and the outcome of our candidacy tests for some examples of the detected sources. Left panel: Typical difference image produced by \texttt{TUVOpipe} (science image: ObsID 00034748064, extension 2, \textit{u} filter; template image: ObsID 00035028027, extension 1, \textit{u} filter). In this difference image, 38 sources were detected. Only one passed all three tests (Sect. \ref{pii_sextractor}) and was retained as a candidate transient (source d - this is a real transient, the BL Lac prototype system, which is often monitored by \textit{Swift}). The labels indicate the example sources displayed in the right panel. Right panel: examples of candidate transients detected in the difference image that did not pass one of the tests (a-c), and the only source that passed all tests (d). The elliptical regions are the source regions extracted by the source detection software (in this case \texttt{uvotdetect}).}
    \label{fig:candidacy_tests}
\end{figure*}

Once a difference image is produced, PII searches it for candidate transients. Real variable sources will appear in the difference image as point sources (either positive or negative, depending on whether the source was brighter in the template or in the science image; see Fig. \ref{fig:image_subtraction_main}). A first set of candidates is detected by running source detection software on the difference images. This is a standard strategy used in transient searches which utilise image subtraction methods (e.g. \textit{ZTF}, \citealp{Masci_2019}; \textit{MeerLICHT}, \citealp{Hosenie_2021}, the Deeper Wider Faster Survey with the Dark Energy Camera, \citealp{Andreoni_2017, Andreoni_2019}; see also \citealp{Hu_2021}). \

We run both \texttt{uvotdetect} (the standard UVOT source detection tool) and \texttt{sextractor}\footnote{\url{https://sextractor.readthedocs.io/en/latest/}} \citep{sextractor_1996} to detect sources in the difference images. We use both methods, because after testing the software on many different fields and with different settings, and inspecting the results by eye, we found that some sources were only detected by one of the two methods. The clearest sources in the difference images are picked up by both methods, and virtually all sources are picked up by at least one of the methods (with \texttt{sextractor} typically being more effective than \texttt{uvotdetect}, but the latter still occasionally detecting sources that are missed by the former). We run both detection methods on both the normal and flipped difference images (see Fig. \ref{fig:image_subtraction_main} for examples of template, science, and difference image sequences and the appearance of transients in the difference images). Once both source detection methods have been run on both difference images, the detected sources are compiled into a single transient candidate list (i.e. removing multiple detections of the same source) for that image.\

In practice, image subtraction is not a perfect process, due to various possible factors. First of all, there may be artefacts in the input data, such as stripes, optical ghosts, saturated stars, or imperfect mod-8 noise correction (i.e. related to bright or saturated stars; see Sect. 4 of \citealp{Breeveld_2010} for details about the mod-8 noise) in the science and template images, or imperfect alignments. These can result in artefacts being carried over in some form to the difference images. However, the majority of the issues exhibited in difference images are due to complexities in the image subtraction process itself (in our case \texttt{hotpants}). To create the convolution kernel to match PSFs, \texttt{hotpants} uses Gaussian functions, which are not exact representations of the image PSFs (see \citealp{Cao_2016, Hu_2021} for descriptions of this caveat in \texttt{hotpants} and see \citealp{Breeveld_2010} for a detailed description of the PSFs of UVOT images). More specifically, for example, asymmetric variations in the PSFs of the two input images may cause problems in the subtraction, since \texttt{hotpants} models the PSFs with circularly symmetric Gaussian functions \citep{Hu_2021}. The image subtraction algorithm may also encounter significant issues when dealing with diffuse emission. Collectively, these complications ultimately lead to imperfect difference images, and hence a large number of candidate sources detected in the difference images are `bogus' transients (false positives). The next stage of PII is therefore to reject candidates that are not likely to be true transient sources. A series of tests is performed on each detected source in order to produce a list of reliable transient candidates (see Fig. \ref{fig:candidacy_tests} for an example of a difference image with all detected sources shown and examples of the results of these tests). \ 

To begin with, for each image the overall performance of the image subtraction is checked: if the number of sources detected in the difference image is $>$90\% of the number of sources in the science image, then it is considered a bad subtraction. We note that here for consistency we run the test once for each method; in other words, we compare the number of sources detected by \texttt{uvotdetect} in the science and difference images, and then again for \texttt{sextractor}. If either of the two methods fails the test, then it is considered a bad subtraction. Typical causes of bad subtractions are images that were not correctly aligned; if the template and science images are misaligned, every source in both images will appear in the difference image. Very crowded fields are also frequent causes of bad subtractions; crowded fields can cause severe problems for image subtraction even on well-aligned images, and many stars may not be subtracted properly, leaving residuals in the difference images that are picked up as sources. Badly subtracted difference images are rejected and not processed further, though their corresponding science images are kept for creating light curves if any transients were detected in that field (see Sect. \ref{piii}). These unsuitable subtractions occur on 20-30\% of images\footnote{This problem is found to be very challenging to solve, though we are currently working on methods to reduce the fraction of badly subtracted images.}.\

If the difference image is accepted, PII proceeds to run three tests on each individual detected source to determine whether it is a viable transient candidate. If a source fails any test, it is discarded; if it passes all tests, it is considered a viable candidate transient and is retained. The three tests carried out are described in the following paragraphs, in order of when they are processed within PII.\

The first is the elongation test. If the source has an elongation higher than a specific value then it fails the test (see Fig. \ref{fig:candidacy_tests}, right panel, a). By default, we use a value of 2.0 (i.e. higher than the elongation cut for low-quality science images of 1.4; see Sect. \ref{pii_dataprep}; this number can be adjusted by users if necessary) because during the image subtraction process, the convolution of the PSF between the template and science images can cause real transient sources to deviate slightly from point sources in the difference images. Many poorly-subtracted sources in the differences images have brightness profiles consisting of multiple positive and negative brightness components. Most cases of detections of elongated sources in the difference images are due to the source detection software identifying these components as real sources (see Fig. \ref{fig:candidacy_tests}, right panel, a).

Secondly, we have the yin-yang test. One of the most common artefacts in difference images is when sources appear as distinct dipoles - the so-called yin-yang pattern (see e.g. \citealp{Hu_2021}). This is when a source appears composed of both positive and negative contributions (see Fig. \ref{fig:candidacy_tests}, right panel, b). These patterns are often detected in the difference images (for sources that are not truly variable) and are artefacts introduced by the image subtraction method. Typically this occurs when there are asymmetric differences in the PSFs of the input images, which are not well accounted for by \texttt{hotpants} \citep{Hu_2021}, or when the images were not properly aligned. To reject these artefacts, a box (twice the size of the semi-major axis of the source region detected by the source detection software) is created centred at each source's location; for a source to pass the yin-yang test, the total negative counts in the box must not be comparable (less than a factor of 2.5, tested empirically on many sources) to the total positive counts.

The brightness profile test provides a useful way of determining whether a source is real by examining characteristics of its PSF; for example, by fitting a Gaussian to the source profile. Although a Gaussian is not a perfect match to the UVOT PSF (see \citealp{Breeveld_2010}), extensive testing on our sources has revealed that it is still adequate for our purposes. Sources in the difference images that exhibit irregular brightness profiles that do not resemble real sources could be the result of subtracting saturated stars, blended sources, or \texttt{hotpants} failing to perfectly cross-convolve the PSFs of the two images. The final test therefore uses the same small image stamp created during the yin-yang test and fits a two-dimensional Gaussian to the brightness profile of the source within the stamp. The source passes the test if the full width at half maximum of the fitted Gaussian is between 2.8 and 3.7 pixels and the $\chi^{2}$ squared value of the fit is less than $10^{3}$. The values for these cut parameters have been determined by performing tests across many fields with different characteristics, including empty fields and very crowded fields. These thresholds we chose for passing the test are relatively loose in order to account for the fact that real source profiles are not expected to exhibit perfect Gaussian shapes (see Fig. \ref{fig:candidacy_tests}, right panel, c, for an example of sources that did not pass this test).\

The described process in PII so far (from extraction of the image from the ObsID to the candidacy tests per source) is then repeated for every image in a given field, after which PII moves to the next field. Therefore, when PII has finished, a list has been produced of all the viable transient candidates for each field and for each UVOT filter. These sources are passed to Part III of \texttt{TUVOpipe} for further processing.\

There are no major differences in the functionality of PII between the archival and the real-time modes of the pipeline, since for the archival mode it simply stops after the single field that is being processed is completed. The only difference worth noting is that during the archival mode of the pipeline, it runs on all archival data for a given field, and therefore there is no need for the archival template selection process (see Sect. \ref{pi}). The first image to be processed that passes the image quality check has an exposure time $>200s$ and is $<4'$ from the position specified by the user; this is always selected as the template. We also note that for any field processed, in both the real-time and archival modes, the template can be chosen manually by the user. By default, this does not happen since we aim to minimise user intervention (as we process several tens of fields per day). However, this can occasionally be useful, for example in the archival mode where several hundreds of images may be processed with respect to the template, because users may want to manually ensure that the template is of good quality (e.g. it has high enough exposure time, no artefacts present, and no elongated sources).

\subsection{Part III} \label{piii}

\begin{figure}
    \centering
    \includegraphics[width=.49\textwidth]{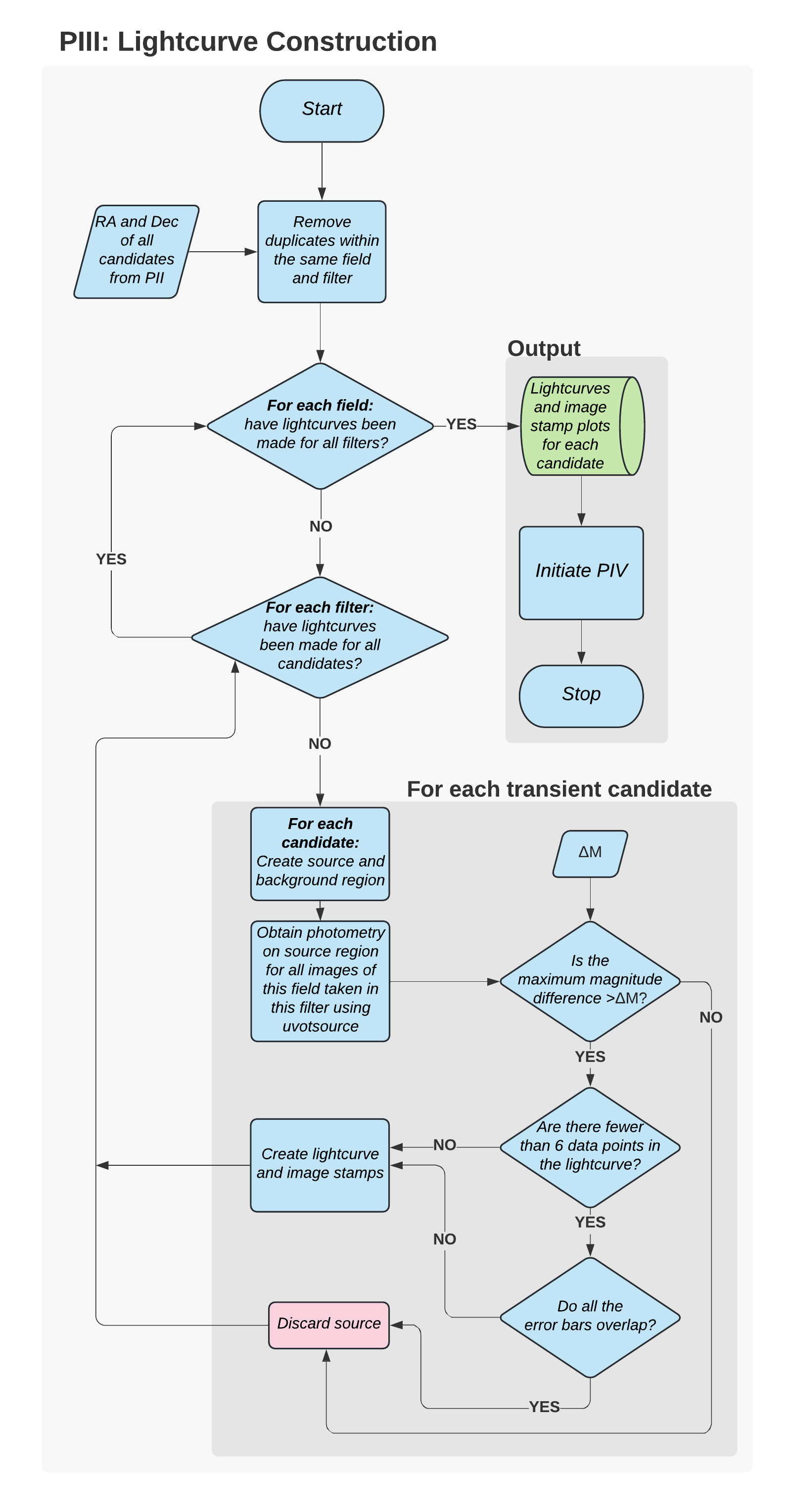}
    \caption{Workflow of PIII (discussed in detail in Sect. \ref{piii}). This part of \texttt{TUVOpipe} creates light curves and stamps for each transient candidate detected in PII and which passes the $\Delta$M check. See Fig. \ref{fig:PIII_example} for an example of the output of PIII. PIII operates in the same way for the real-time and archival modes of \texttt{TUVOpipe} since it requires only a list of candidate transients that were detected in PII. Upon completion of PIII, Part IV (PIV) of the pipeline is automatically initiated.}
    \label{fig:flow_piii}
\end{figure}

The next step in \texttt{TUVOpipe} consists of creating light curves of all transient candidates that passed all tests in PII. This is done in Part III (PIII) of the pipeline, the workflow of which is displayed in Fig. \ref{fig:flow_piii}. \ 

PIII begins by running through each field and creating a global list of all candidates per field that were output by PII and removing any duplicates. Duplicates are sources detected by PII in different difference images of the same field and the same filter, but which are close to each other and represent the same source. The radius chosen inside which sources are considered duplicates is 8"; this value was chosen empirically by examining many duplicate sources. For each source in the global list, PIII performs photometry at the source location. It does this using every science image of the field available in the user's directory (downloaded in PI), as well as the template image\footnote{For the template image, PIII also computes the on-sky distance (in arcminutes) of the source from the centre of the image. This is done to make users aware of occasions when the source is near the corners of CCD in the template image. In those cases (typically when the distance is >11'), the photometry may not be completely reliable due to the decrease in sensitivity of the detector towards the corners of the CCD. We find that this effect is not always fully accounted for in the automatic UVOT calibration.} (which was selected in PII). PIII thus outputs the (AB) magnitude of the source in each image. We note that PIII runs photometry on the uncropped, aligned, science images. This is because if the cropped images were used, some transients may be cut out of some images, therefore not allowing us to obtain photometry on those images and thus reducing the number of data points we can obtain. Despite this, for a given transient, in some images no photometric measurement is possible (consider a transient that is originally detected near the edge of a particular image). Since UVOT images of the same field do not always cover the same exact sky area (see Sect. \ref{pii_dataprep}), in a subsequent image of that field the location of that transient may fall outside the image. In these cases, no photometric measurement can be made for the transient in that image. \

\begin{figure}
    \centering
    \includegraphics[width=0.45\textwidth]{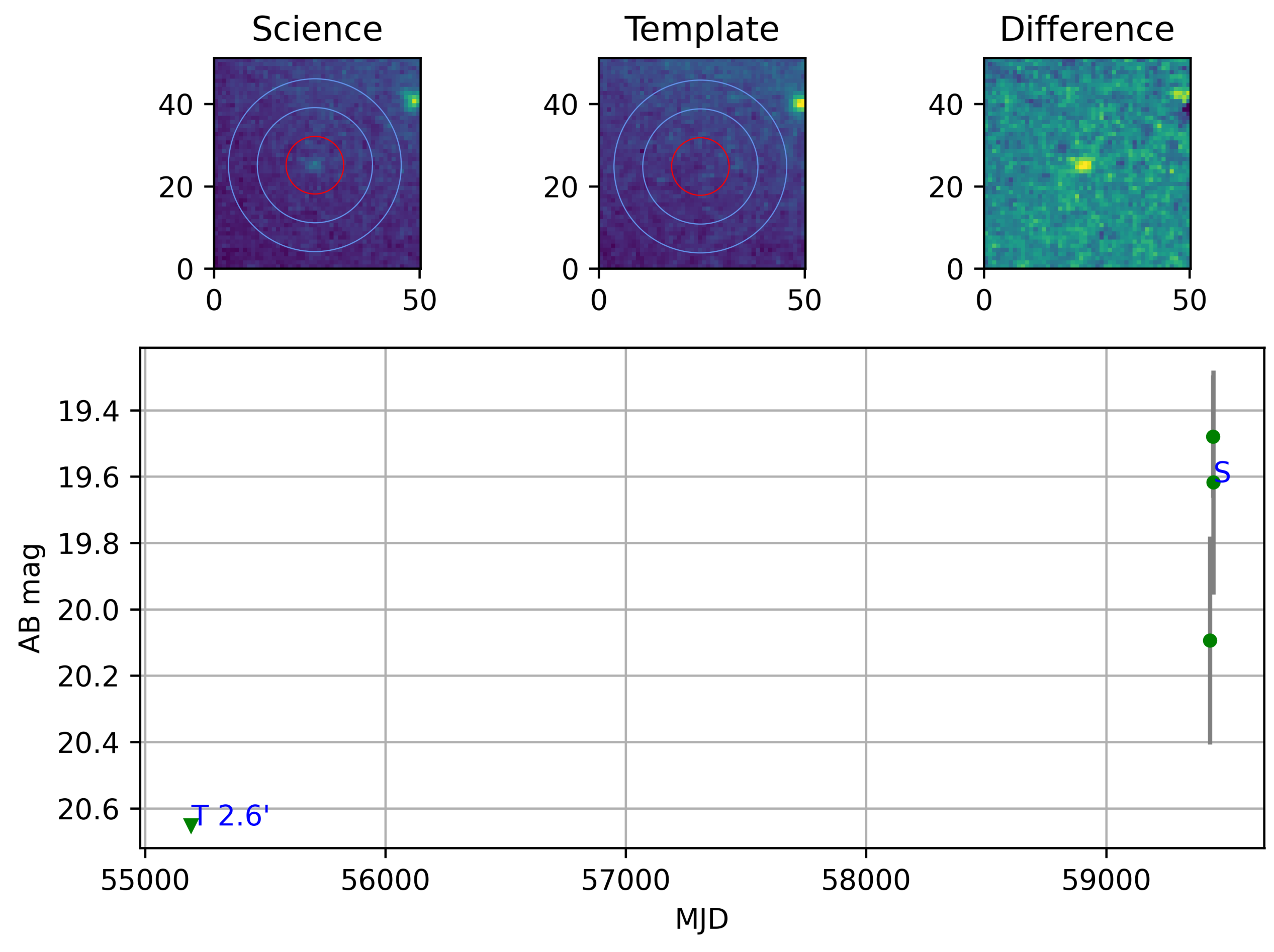}
    \caption[caption]{Example of final product of PIII for a candidate transient. Top row shows, from left to right, the (50"x50") stamps of the source in one of the science images, the template image, and the difference image. The figure at the bottom shows the light curve created from the locally available QL data of the source (plus the upper limit of the source in the template image). In the left and middle stamps in the top row, the red circle indicates the source region and the blue annulus indicates the background region used by \texttt{uvotsource}. The data points in the light curve corresponding to the template and science images are indicated with a `T' and an `S', respectively. The value alongside the template label indicates the on-sky distance (in arcminutes) of the source in the template image from the centre of the template image. Inspection of the science, template, and difference image stamps clearly show that this is a real transient source. The data point derived at the source location in the template image is an upper limit, since there was no clear detection. The transient was found at RA=00:42:46.65, Dec=+41:14:26.6 in the UVW1 filter and is a known nova in M31 (AT2021jwr).}
    \label{fig:PIII_example}
\end{figure}

Photometry is carried out using the HEASoft UVOT photometry tool \texttt{uvotsource}\footnote{\url{https://heasarc.gsfc.nasa.gov/lheasoft/ftools/headas/uvotsource.html}}. As input, \texttt{uvotsource} requires an image, a source region file, and a background region file. The regions are created automatically by PIII as each source is processed. The source region file is defined by a circle with its centre at the location of the source (as determined by the source detection method) and a radius of 7". The background region is then created as an annulus with inner radius twice the size of the source region radius (to avoid the wings of the source flux profile contaminating the background region) and outer radius three times the size of the source region radius. This ideally results in a suitable background region that is close to the target source, is significantly larger than the target source region, and is devoid of other sources (see Fig. \ref{fig:PIII_example} for an example). Unfortunately, the last criterion is occasionally not met, for example in some crowded fields where a candidate can have several nearby sources contaminating the background region. This results in incorrect magnitude measurements, generally underestimating the target source brightness\footnote{This effect may also introduce fake variability or hide true variability in the light curves of our sources. However, although it is difficult to quantify, this is likely rare because it requires the sources contaminating the background region to also be strongly variable themselves within the time frames of our light curves.}. However, the products of PIII include a stamp of the source, displaying the source and background regions (see Fig. \ref{fig:PIII_example}). From these stamps, it is always apparent when there is background contamination (see Fig. \ref{fig:PIV_output} for such an example), so the user becomes directly aware when the displayed magnitudes are not completely reliable. We note that for sources of interest, manual light curves can be created using Part V (see Sect. \ref{pv}) and a user-specified background region in order to obtain the most accurate light curve possible. We also note that the automatically produced light curves are not intended to be the most accurate possible, but simply to provide good enough indications to the user to roughly examine the variability behaviour of each candidate and determine whether or not further investigation of the transient is warranted.\ 

It is worth making a remark about how we deal with non-detections in PIII (the transient might not be active during all the images of a field, so not all images will necessarily have detections of the transient). For every \texttt{uvotsource} run, a magnitude measurement is always produced, even when there is no detection (in the cases where there is no detection and the flux in the source region is negative, the magnitude will be set to 99). However, for each \texttt{uvotsource} run an upper limit is also always given, representing the limiting magnitude of the UVOT image at the source location (based on the local background counts). Our method of determining whether there is a detection or not is thus to compare the magnitude measurement and the upper limit; if the upper limit value is lower (i.e. brighter) than the measured magnitude, then we assume there was no detection, and we take the upper limit as the data point for that image\footnote{This method was suggested to us during communication with the \textit{Swift} help desk regarding a quick and rough check on whether a source was detected.}. We later label it as an upper limit in the light curve. If the upper limit value is higher (i.e. fainter) than the measured magnitude, then we assume there was a detection and take the magnitude measurement as the data point for that image.\ 

The output of each \texttt{uvotsource} run is a table including the source (AB) magnitude, the upper limit or limiting magnitude at the source location, and the $1\sigma$ error. The error is composed of both the statistical and systematic errors on the magnitude measurement due to photometric calibration errors of UVOT (see e.g. Breeveld et al. 2011 or the instrument software guide\footnote{\url{https://swift.gsfc.nasa.gov/analysis/UVOT_swguide_v2_2.pdf}}).\

Before creating a light curve with the measured magnitudes and upper limits, PIII determines the amplitude of the variability exhibited by the light curve. The magnitude difference between the brightest and faintest data points is determined and a check is carried out: if the difference is less than 0.4 magnitudes, a light curve is not created. This is done because we are only interested in highly variable sources, and also because we noticed that sources detected by \texttt{TUVOpipe} with $<$0.4 magnitude variability very rarely pass our variability significance test (see Sect. \ref{piv_variability}).\ 

For all transient candidates that pass these tests, PIII then uses all measured magnitudes and upper limits obtained to construct light curves. The data points in the light curve correspond to the brightness of the candidate transient during all the QL data currently located in the local archive (this may include data taken $>$1 week previously, e.g. if the user's local drive contains data from previous runs of the pipeline), as well as during the one archival image chosen as the template (if it was available; see Sect. \ref{pii_dataprep}). We denote these products `PIII light curves'. Additionally, corresponding 50"x50" stamps of the science, template, and difference images are displayed, overlaid with the source and background regions used by \texttt{uvotsource}. This is done in order to contextualise the candidate for the user and to help visually distinguish between real and bogus sources (see Fig. \ref{fig:PIII_example}). The difference image stamp (and the corresponding science image stamp) shown is the first difference image in which the candidate was detected by the pipeline (not all the difference images of a given field will trigger a particular transient; e.g. if the transient was not variable in a particular image with respect to the template, or if the performance of the image subtraction or source detection was poor for a particular image). All light-curve files and stamps produced in PIII are passed to the next stage of the pipeline (see Sect. \ref{piv}), which attempts to further classify the sources.\

There are no major functional differences in PIII when running \texttt{TUVOpipe} in archival mode with respect to running it in real-time mode, except that during the archival mode only sources from one field are processed, and all archival data covering that field (within the chosen search radius) are analysed. To maximise the potential for finding historical transients, we selectively run the archival mode on fields that have been observed many times by UVOT, so light curves produced when running in archival mode typically cover long timescales and include many data points.

\subsection{Part IV}  \label{piv}

\begin{figure}
    \centering
    \includegraphics[width=.45\textwidth]{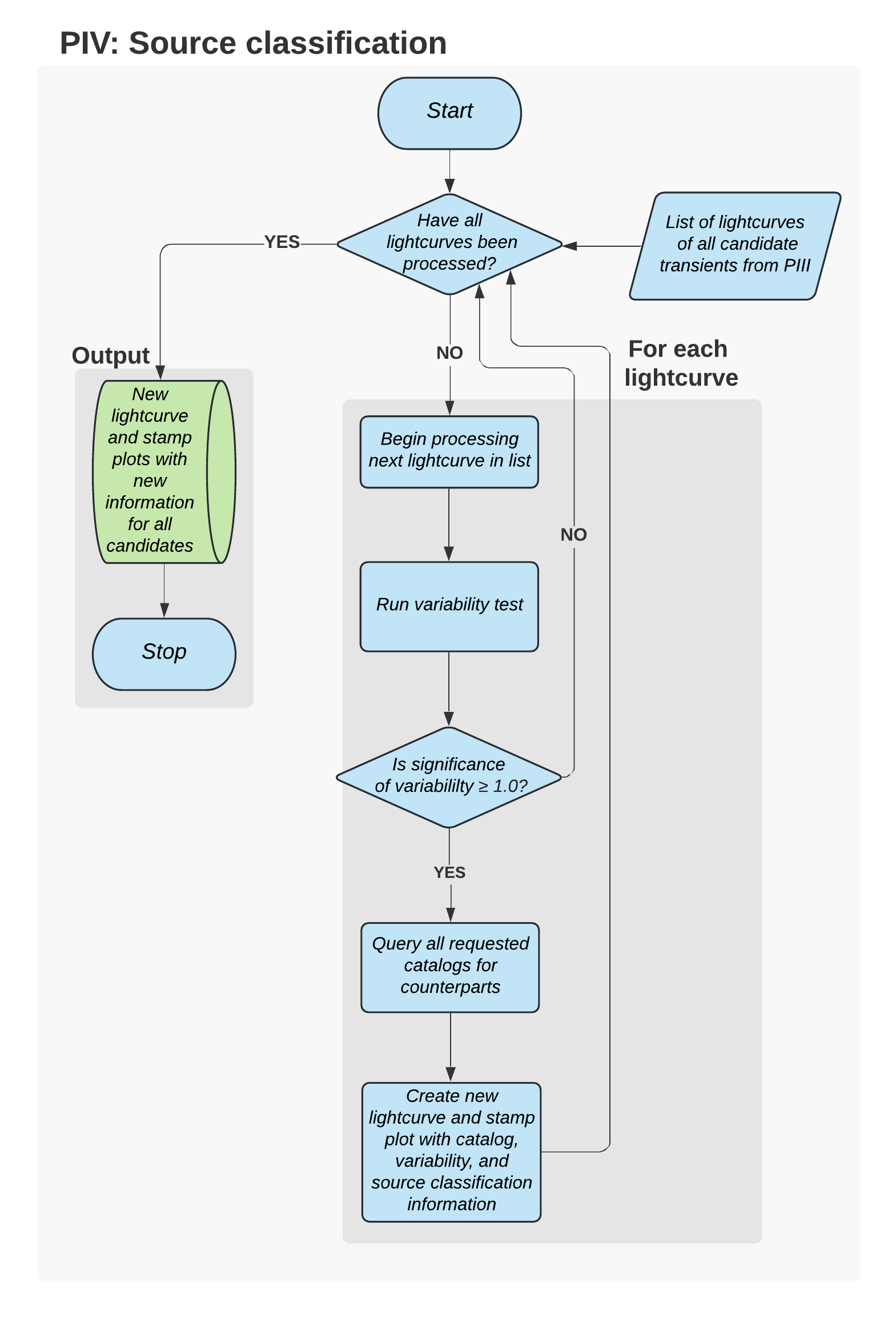}
    \caption{Workflow of PIV (discussed in detail in Sect. \ref{piv}). This is the part of \texttt{TUVOpipe} that collects information about each transient to supplement the UVOT light curve in order to obtain a preliminary classification and help determine whether further investigation is warranted. The operation of PIV is identical between the archival and real-time modes of the pipeline. The output of PIV (shown in Fig. \ref{fig:PIV_output}) is all the light curves produced in PIII supplemented with basic source information, information from external catalogue queries, and variability information.}
    \label{fig:flow_piv}
\end{figure}

\begin{figure*}
    \centering
    \includegraphics[width=0.9\textwidth]{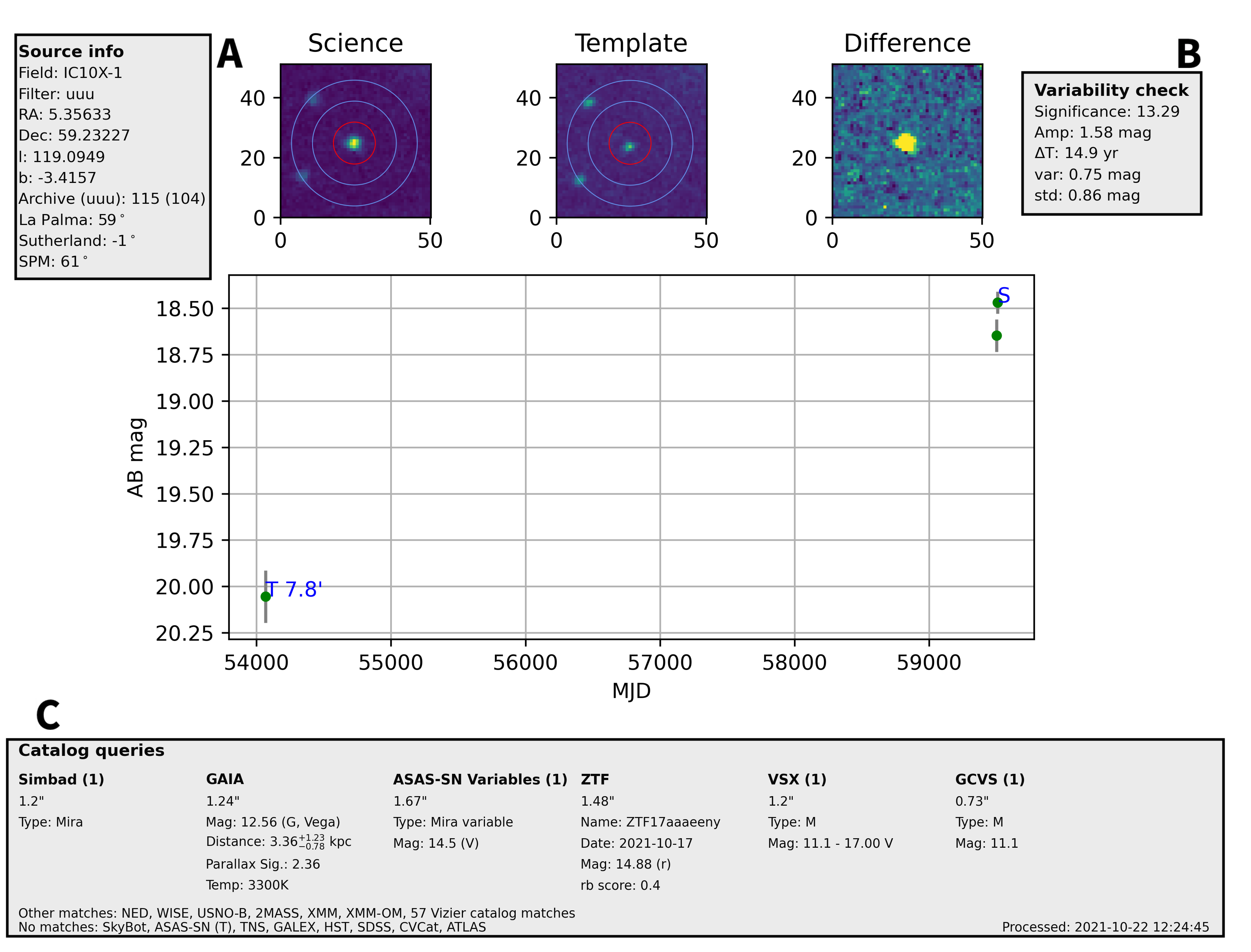}
    \caption{Output of PIV for a transient detected by \texttt{TUVOpipe}. The image stamps and light curve are produced by PIII (see Fig. \ref{fig:PIII_example} for an example). PIV supplements this with information about the UVOT observations (panel A), basic analysis of the light curve (panel B), and results from the catalogue queries performed (panel C). The information displayed in panels A, B, and C is discussed in detail in Sections \ref{piv_basic_info}, \ref{piv_variability}, and \ref{piv_queries}, respectively. In this example, with all the information provided we can deduce a fairly robust classification of this source as a Mira variable (FQ Cas), based primarily on Simbad, \textit{ASAS-SN}, VSX, and GCVS. Note that for most catalogues, magnitudes are provided in the AB system; where this is not the case (e.g. \textit{Gaia}), we specify that the magnitude stated is in the Vega system.}
    \label{fig:PIV_output}
\end{figure*}

Part IV (PIV) of \texttt{TUVOpipe} carries out further investigation of all the candidate transients for which light curves were produced in PIII. It does this by obtaining additional information both from external catalogues and from further analysis of the transient and its light curve. The workflow of PIV is displayed in Fig. \ref{fig:flow_piv}. An example of the PIV product, that is by default the final output of \texttt{TUVOpipe} with which users are presented for each detected transient, is shown in Fig. \ref{fig:PIV_output}. \ 

\subsubsection{Basic source information} \label{piv_basic_info}

For each transient candidate, PIV provides information about the \textit{Swift} observation(s) used to detect it. In the final product, PIV displays the coordinates of the source (both in RA and Dec, J2000, and in Galactic longitude and latitude), visibility information relevant for follow-up observations (maximum altitude during the night at the time of discovery) with the three ground-based observatories with which we undertake follow-up studies (see Wijnands et al., in prep., for a description of our ground-based follow-up programmes), and the number of archival observations available (total and specifically for the filter in which the transient was detected; this helps to determine whether a long-term light curve would be useful; see Sect. \ref{pv}). Figure \ref{fig:PIV_output}, panel A, shows an example of the basic source information obtained for a transient candidate.

\subsubsection{Variability test} \label{piv_variability}

Part IV then carries out a test for each source to determine a rough estimate of the level of variability we detect. The purpose of this test is for users to have an idea of the significance of the variability of each transient candidate that they are presented with, and also to automatically discard sources for which the observed variability is not significant.\ 

To do this, we fit a line of constant brightness (i.e. assuming no variability) to the light curve using a weighted least-squares approach. We determine the goodness of fit using the $\chi^{2}$ statistic, which is then converted to a significance. The significance gives an indication of the likelihood that the variability displayed represents true variability, so sources with variability significance <1.0 are discarded at this stage, and the PIV output is not created for these sources. If there are any upper limits in the light curve, then the variability significance is not computed and the transient is retained. This is because upper limits do not have errors and cannot be incorporated into this test; however, highly interesting transients may not be detected in some images, and we do not want to discard those.  We note that, due to the underlying assumptions of the method, it should only be seen as a tool to limit the number of transients picked up that do not show strong enough variability to warrant further investigation and not as a tool to determine any highly accurate statistics regarding each transient. The standard deviation and variance of the magnitudes in the light-curve data is also computed in order for users to have a quantitative indication of the scatter in the data. The following quantities derived in this analysis are then displayed alongside the light curve (see Fig. \ref{fig:PIV_output}, panel B): the significance of the variability, the maximum amplitude of the variability (`Amp', obtained by subtracting the brightest from the faintest magnitudes in the displayed data), the timescale of the light curve (`$\Delta$T'; the time between the earliest and latest data point displayed), the standard deviation, and the variance.

\subsubsection{Probing external catalogues} \label{piv_queries}

PIV automatically probes public catalogues from a variety of facilities and surveys for previous detections and potential classifications at the locations of the sources we detect. Most queries are performed using \texttt{astroquery}\footnote{\url{https://astroquery.readthedocs.io/en/latest/}}, an \texttt{astropy} tool designed to query astronomical databases using Python. If \texttt{astroquery} is not available for a particular catalogue, we attempt to query it through Vizier, for which \texttt{astroquery} allows access to all catalogues. If a catalogue is also not available in Vizier, a query of the catalogue's online database is done via Python-based web-scraping with the package \texttt{Beautiful Soup}\footnote{\url{https://beautiful-soup-4.readthedocs.io/en/latest/}}. We currently probe 22 different catalogues, including several all-sky catalogues in various wavelengths as well as transient and variable star catalogues (see Table \ref{tab:PIV_table} for a full list of all catalogues we probe, including brief descriptions and references for each). The code is set up such that when new catalogues or new versions of catalogues are published, we can easily create a new query and add the information to the PIV output.\ 

For each transient candidate, PIV queries each catalogue for any sources within 5" of the position of the candidate. The value 5" was chosen to account for errors in positions of our sources, which may be up to a few arcseconds (see Sect. \ref{uvot_data_structure}). If matches exist, relevant source information is extracted from the results of the query and displayed alongside the light curve (see Fig. \ref{fig:PIV_output}, panel C). This information can include any known designation, classification, or measured magnitudes (always in the AB system except where noted otherwise) of the source. For example, we query the full \textit{Gaia} catalogue, and we can thus immediately check for each transient whether there is a likely \textit{Gaia} counterpart, and if there is one, we know its G-band magnitude, and if available, temperature and distance values (distances are queried from an associated \textit{Gaia} catalogue that contains distances to 1.33 billion \textit{Gaia} sources, see \citealp{Bailer-Jones_2018}). Under the \textit{Gaia} information, `Parallax Sig' gives the significance of \textit{Gaia}'s parallax measurement for the source; so high numbers indicate robust distance measurements. We also query the MARS alert stream\footnote{\url{https://mars.lco.global/}} of the \textit{Zwicky Transient Facility} \citep{Bellm_2019}, which is useful in order to know whether \textit{ZTF} detected a variable source at the position of our transient, and if so, at what optical magnitude. The \textit{ZTF} query can sometimes also help to confirm the source as a real transient, given its real-bogus (`rb') score of 1.0 (the score ranges from 0 to 1, where 1 corresponds to the highest probability of it being real; see Fig. \ref{fig:PIV_output}, panel C). \

If there are multiple sources in a given catalogue within 5" of our transient, PIV extracts the information from the catalogue source with the smallest on-sky separation to the transient candidate. Since the UVOT astrometric uncertainty can be up to a few arcseconds (see Sect. \ref{uvot_data_structure}), in such cases there is a chance that PIV collects the information of a catalogued source that is not the real counterpart of the TUVO source. However, the number of sources in the catalogue within 5" of the transient candidates is given, so, in such cases, the information given in the figure is interpreted more tentatively than if only one match had been found. We further note that these automatic catalogue matches are not intended to be conclusive, but simply to add potentially useful information for a preliminary classification of the TUVO source. \

Finally, PIV assigns to each transient a name in the format TUVO-[YY][id] where [YY] is the last two digits of the year in which it was triggered by \texttt{TUVOpipe} and [id] is a unique identifier composed of letters starting with `a' and rising alphabetically, adding additional characters when necessary (this is the standard scheme for naming transient alerts, as recommended by the IAU\footnote{\url{https://www.iau.org/public/themes/naming_stars/}}). For sources we deem to be of high interest and warrant further investigation, we manually check the results obtained with the catalogue queries to ensure that the matches are correct (e.g. by astrometrically solving the UVOT images with \textit{astrometry.net}\footnote{\url{https://astrometry.net/}} and ensuring that the match obtained is still accurate).\

\subsubsection{Asteroids} \label{asteroids}

It is worth noting the regular detection of Solar System objects by \texttt{TUVOpipe}, which are triggered by the pipeline as transients due to their high proper motion. The UV to U-band properties of asteroids are so far poorly studied (e.g. only a handful of asteroids have been studied at <220nm, mostly spectroscopically, see \citealp{Becker_2020}). Therefore, although asteroids are not prime targets of TUVO (since they are not eruptive or intrinsically variable), \texttt{TUVOpipe} detections of such objects may be useful for a better understanding of asteroids, in particular when UVOT observations in multiple filters are available, allowing us to study the colour properties of these sources.\

\begin{figure}
    \centering
    \includegraphics[width=0.45\textwidth]{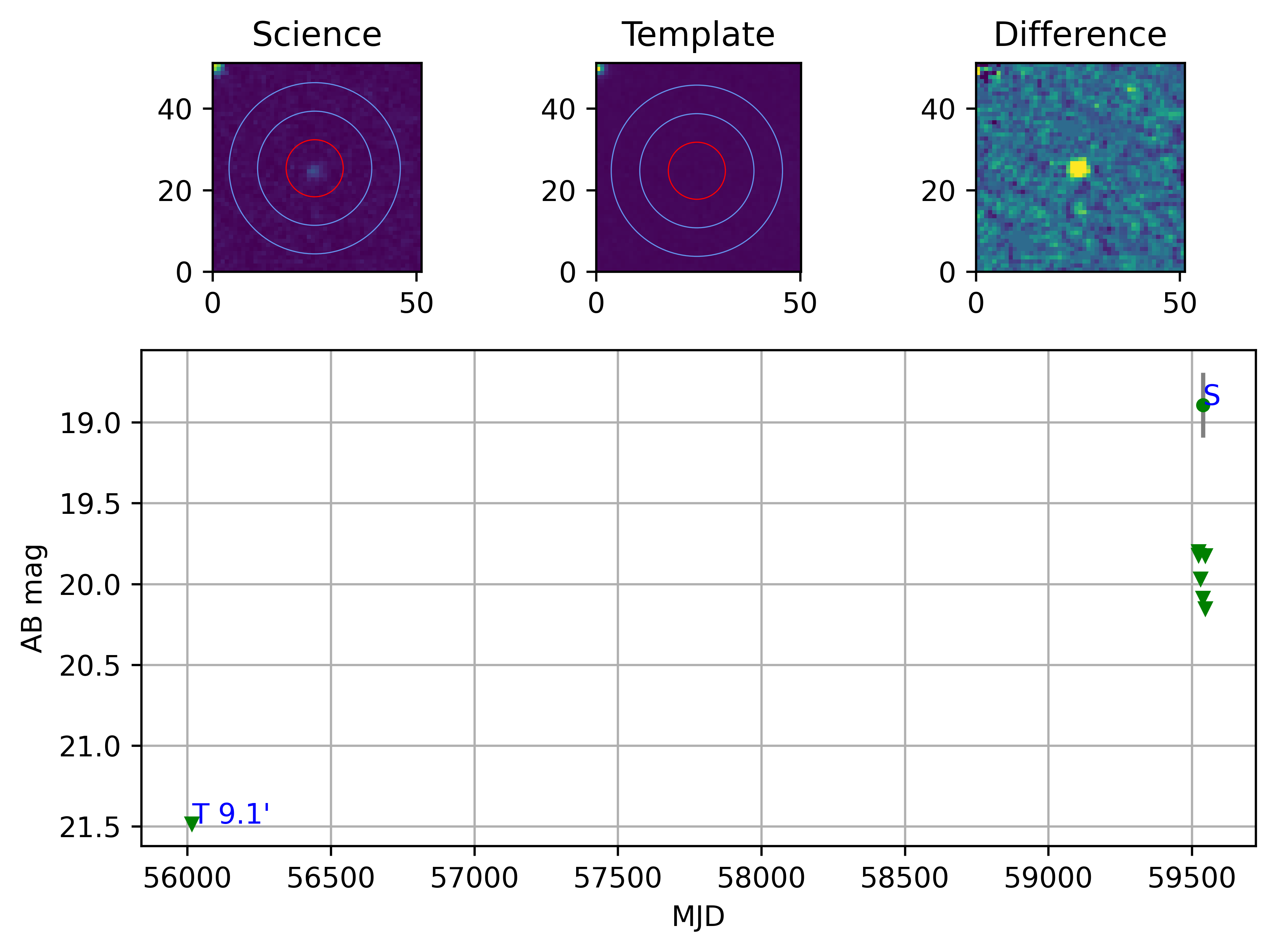}
    \caption{Example of TUVO (PIII) light curve in which an asteroid is detected in one of the images. This source was detected at position RA=09:49:16.30, Dec=+00:26:48.3 with the exposure starting on 30 November 2021 at 16:21:55 in the field of PMNJ0948+0022 in the U filter, and was matched to the known Solar System object Alleghenia 457 with the \textit{SkyBot} database.}
    \label{fig:asteroid}
\end{figure}

In Fig. \ref{fig:asteroid}, we show an example of how an asteroid detection typically appears in our pipeline: a light curve in which there is only a detection in one image (the object detected is named Alleghenia 457). To separate asteroids from astrophysical transients, we probe the \textit{SkyBot}\footnote{\url{https://vo.imcce.fr/webservices/skybot/}} \citep{skybot} database, a large catalogue of known Solar System objects. To account for the higher uncertainty in the positions of many asteroids, we query this database with a 20" radius from our transients, rather than 5" as used for all other catalogue queries in PIV. In addition, due to the high proper motion of asteroids, the time in which the detection occurred is required as input in the query (we take this from the UVOT image which triggered the detection).\ 

\begin{figure}
    \centering
    \includegraphics[width=0.5\textwidth]{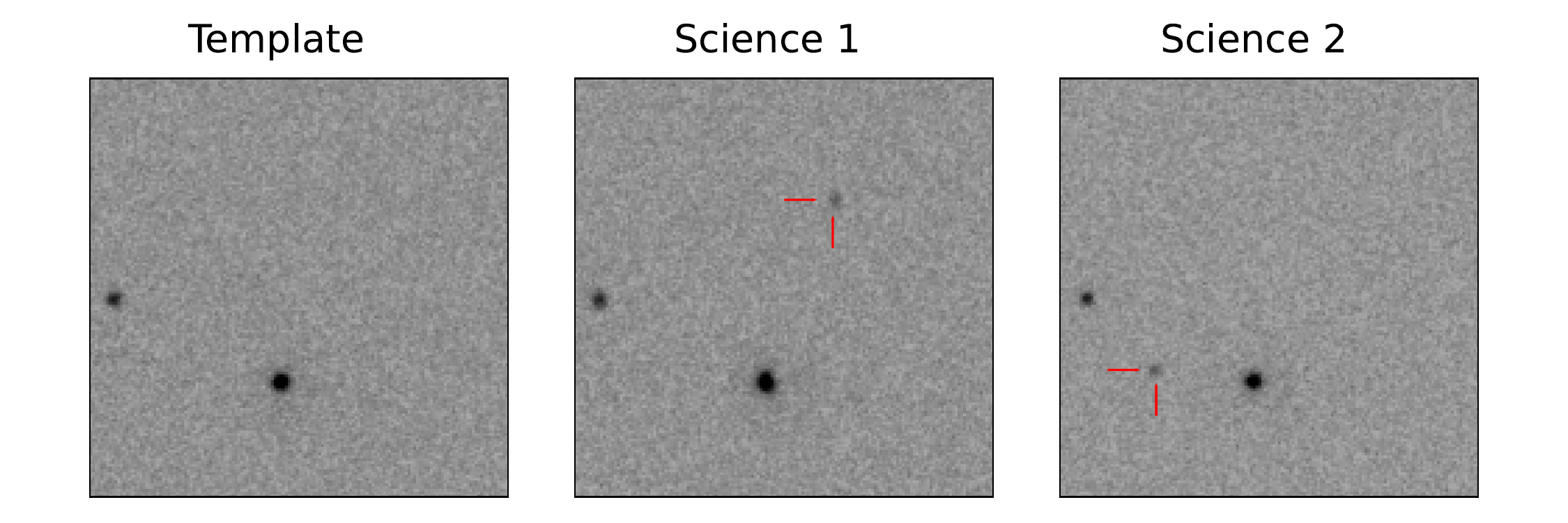}
    \caption{Example of multiple UVOT images in which our pipeline detected an asteroid. In the template image, no asteroid is observed. In the two science images, which were taken a few hours apart, a source appears at different positions, indicated by the red arrows. Our pipeline detected the source in both images as two separate transients, with both having used the left image as the template for the image subtraction. This source was matched using the \textit{SkyBot} database to the known object Alleghenia 457. All images were in the field of the known AGN PMNJ0948+0022 and taken in the U filter. The ObsIDs and extension numbers for the three images are, from left to right: 00031306073[1], 00031306074[1], 00031306074[2]. The exposures in the images began, from left to right, on 13 November 2021 at 05:54:39, 20 November 2021 at 16:21:55, and 20 November 2021 at 19:43:05. All three images are 2.5'x2.5'. North is up and east is to the left.}
    \label{fig:asteroid_stamps}
\end{figure}

For most asteroids we detect, a match is found in the database. Additionally, very bright detections present only in one image (i.e. transients with a single detection), may be indicative of asteroids. Therefore, in cases where a transient is detected in only one image and there is no match in the \textit{SkyBot} database, we inspect consecutive UVOT images by eye. If we see a source appearing at different positions in consecutive UVOT images (usually taken a few minutes to a few hours apart; potentially including different filters), we can confirm that it is a Solar System object. We invite the reader to consult Fig. \ref{fig:asteroid_stamps} for an example of an asteroid detection in consecutive UVOT images (the source shown was also matched with the \textit{SkyBot} database). \ 

So far, we have been able to identify a few tens of TUVO transient candidates as asteroids, either by matches with the \textit{SkyBot} database or by inspecting the UVOT images. We have, however, also found a handful of transients that were not in the \textit{SkyBot} database and were only seen in a single image (i.e. they do not appear as moving sources in consecutive UVOT images). These sources could be real astrophysical transients (and therefore potentially of great interest for TUVO), though they could also still be asteroids that were not listed in the \textit{SkyBot} database and do not appear in multiple UVOT images (because the time between the images was long enough that the asteroid was no longer in the FoV). We refer the reader to Wijnands et al. (in prep.) for an example and a more in-depth discussion of these types of detections in the TUVO project.

\subsubsection{Vetting transients}
\label{vetting}

All transients that are processed in PIV are vetted manually by the users. For each transient, we visually examine the output of PIV (see Fig. \ref{fig:PIV_output}) and attempt to determine first if the source is likely real (as described in Sect. \ref{pii_sextractor}, despite the various tests performed within the code, a number of `bogus' sources are not successfully filtered out and are output as viable transient candidates), and second if it is of sufficient interest to warrant further investigation. Bogus transients are often identified by significant artefacts visible in science, template, or difference image stamps. Typically $\sim$20-30$\%$ of the candidates are vetted as `real' transients, that is, sources that display real, significant variability in the UVOT data. This fraction depends in part on how well \texttt{TUVOpipe} performed on the fields processed (e.g. if the alignment does not succeed on many images of a particular field observed on a given day, then a relatively high number of bogus transients may be output from that field). This leaves us with around a few tens to $\sim$100 real transients per day.\

If a candidate is identified as real, the users proceed to determine whether it is of sufficient interest to study further based on examination of the light-curve behaviour and the results of the probed catalogues. If the transient displays one or more very bright (>1-2 mag) outbursts, it is likely to be considered for further investigation. If the source has no clear identification in the probed catalogues, or if there are disagreements between reported classifications of the source, then we also consider the transient to be potentially interesting. Transients that we consider highly interesting, of which we typically detect a handful every week, are manually passed to the final part of the pipeline (see Sect. \ref{pv}).\

\subsection{Part V}  \label{pv}

\begin{figure}[!h]
    \centering
    \includegraphics[width=.45\textwidth,trim={0 0 0 1cm},clip]{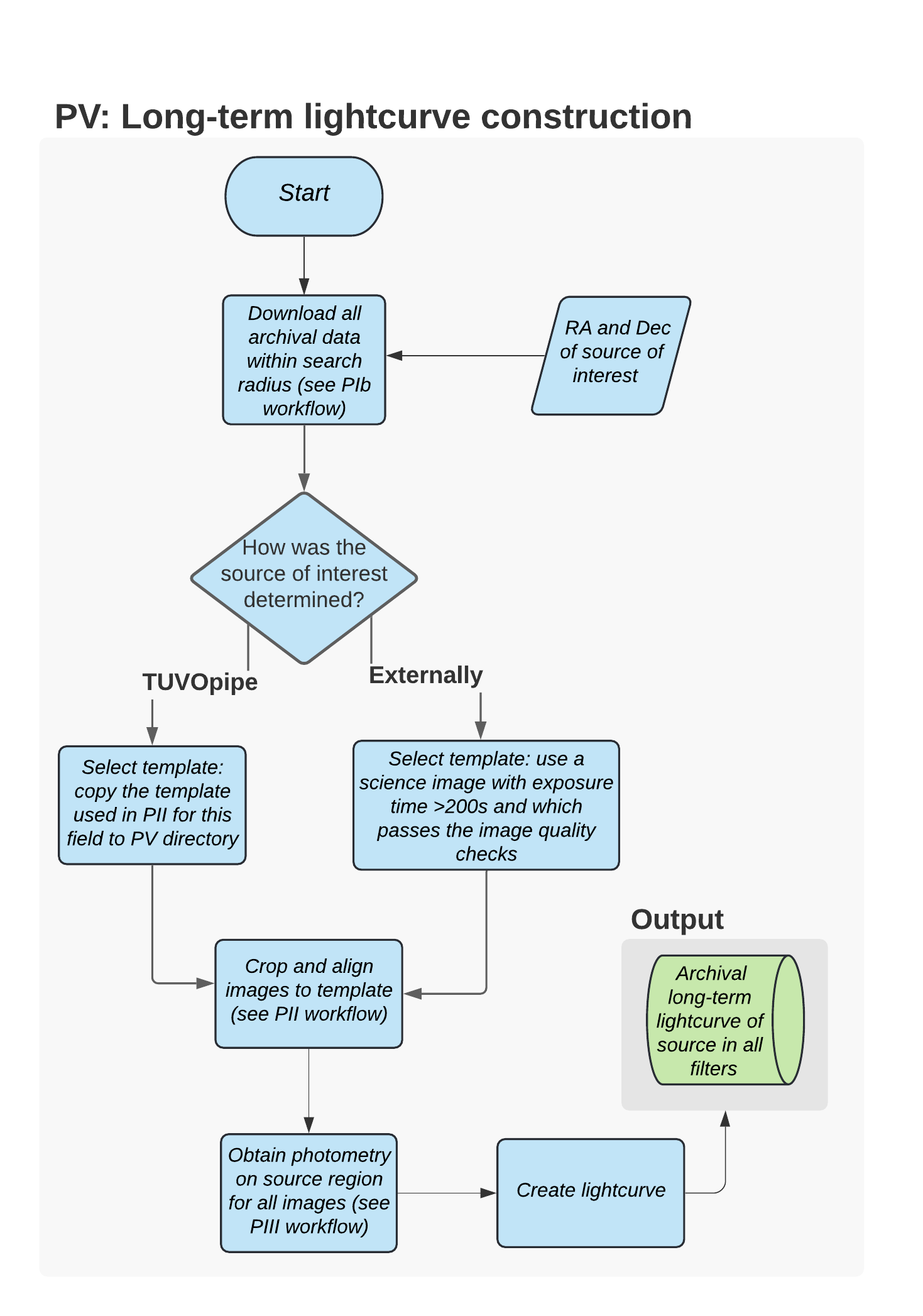}
    \caption{Workflow of PV (discussed in detail in Sect. \ref{pv}). Since most PV functionalities are identical to previous parts of the pipeline, the workflow shown here is highly simplified, and we refer the reader to Figs. \ref{fig:flow_pi}, \ref{fig:flow_pii}, \ref{fig:flow_piii}, and \ref{fig:flow_piv} for details. This part of \texttt{TUVOpipe} creates a light curve of any source of interest using all archival UVOT data in all filters. The output of PV (shown in Fig. \ref{fig:pv_example}) for a given source is the long-term, multi-band light curve created using all UVOT data.}
    \label{fig:flow_pv}
\end{figure}

\begin{figure*}[t]
    \centering
    \includegraphics[width=0.9\textwidth]{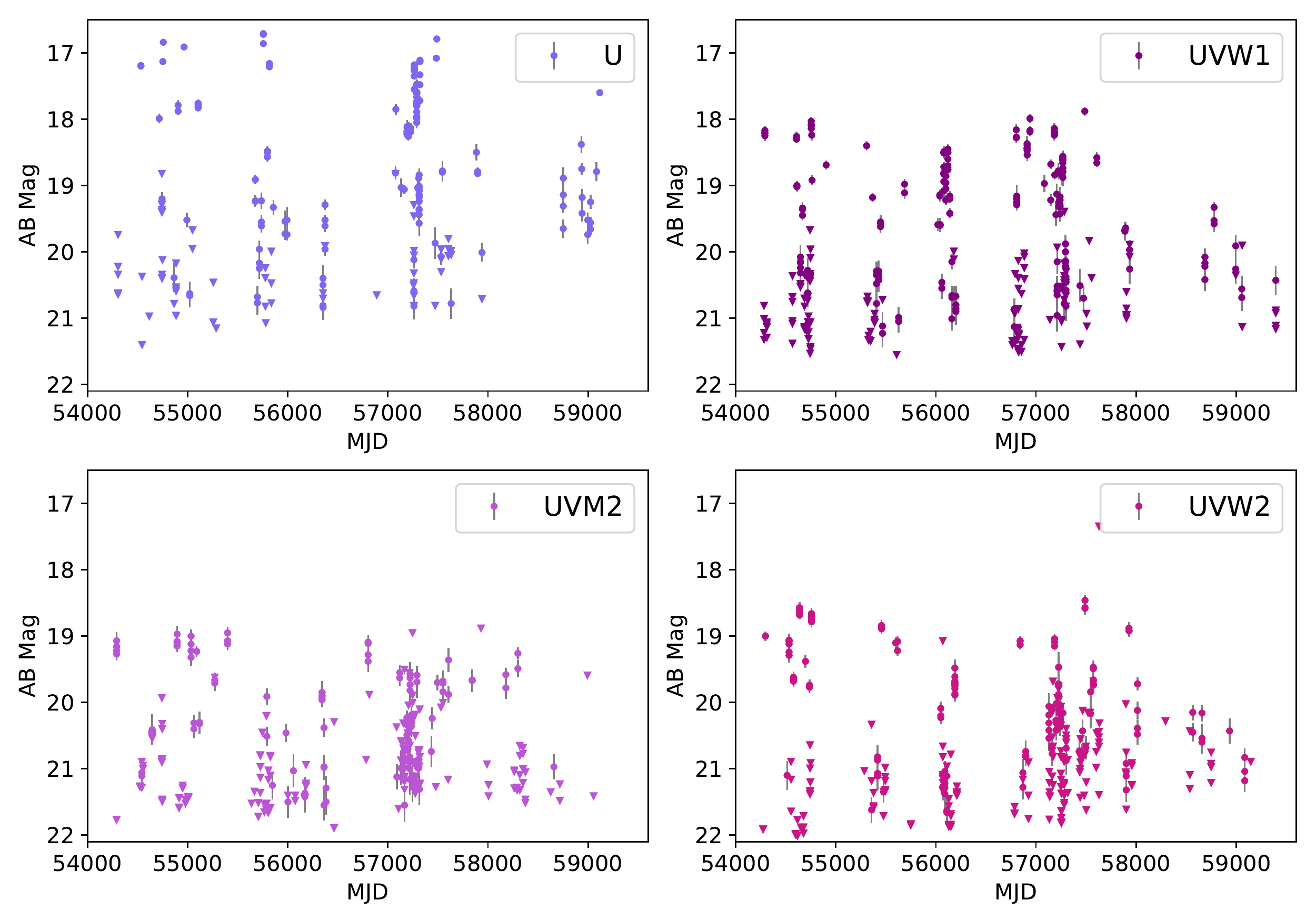}
    \caption{UVOT light curves created a variable star using PV for the highly active dwarf nova AX J1549.8-5416 (see \citealp{Zhang_2017} for a detailed study of this source). The data includes all archival UVOT data covering the position of the source, which in this case included a total of 1097 images, in the four filters U (209 data points), UVW1 (322 data point), UVM2 (264 data point), and UVW2 (302 data point), spanning 14 years. The light curves for different filters are shown in different sub-plots to avoid confusion due to the high number of data points.}
    \label{fig:pv_example}
\end{figure*}

The final part of \texttt{TUVOpipe}, Part V (PV), is used for creating long-term UVOT light curves of any sources of interest, using all archival data available (spanning up to $\sim$17 years, reflecting the operation time of \textit{Swift}). The only input required for PV is the RA and Dec of a source. Since this is a useful tool in itself, PV was built to function both as a continuation of PI-PIV (i.e. to examine transients detected by our pipeline), and also entirely independently of the rest of the pipeline (i.e. to examine any source of interest). We therefore highlight two different use cases for PV, described in this section, which are implemented depending on whether the source of interest was determined by \texttt{TUVOpipe} or externally. In both cases, the functionality and the final product of PV is the same (see Fig. \ref{fig:flow_pv} for a chart displaying the workflow of PV). We note that most of the functionalities of PV are identical to previous parts of the pipeline; so, to avoid repetition and for aesthetic purposes (i.e. to avoid a very complicated flow diagram), the displayed workflow for PV is highly simplified, and within the diagram we refer to the relevant workflows for details. An example of the output of PV is shown in Fig. \ref{fig:pv_example}: the source was a known, highly active dwarf nova that was initially triggered by our pipeline due to strong variability. There is no separation between real-time and archival modes in PV, since all archival data are always processed. \ 

In the first use of PV, it takes as input the coordinates of a source detected with \texttt{TUVOpipe} and which was found to be highly interesting when vetting the PIV product (see Sect. \ref{piv} for a description of how we determine sources of high interest). This allows further information, namely past variability or outburst behaviour (or lack thereof), to be gathered about each discovered source, aiding in classifying the source and in deciding whether further investigation (e.g. follow-up observations) is warranted.\ 

In the second mode, we input the coordinates of any source for which we are interested in the long-term UV behaviour; in other words, sources which were not necessarily transients detected by our pipeline. This mode can be used for any source which has been observed at any time by UVOT, and it is therefore very useful for studying the long-term UV variability of any source discovered with other methods and by other facilities, whether transient or persistent, previously classified or not. For many source classes (e.g. X-ray binaries, cataclysmic variables/dwarf novae, and variable stars), long-term UV variability remains largely understudied compared to other wavelengths, so PV is a very useful tool.

In both cases, PV works in the same way, making use of the various functionalities that exist in previous parts of \texttt{TUVOpipe}. It uses the input RA and Dec of the source of interest, plus a user-defined search radius as input (the latter value is by default 12'; see Sect. \ref{pi} for the reasons for selecting this radius). PV then downloads all archival UVOT data within the search radius of the source position (i.e. similarly to how PI works when operating in the archival mode). All downloaded images are then cropped and aligned (see Sect. \ref{pii_dataprep} for details of how our pipeline carries out cropping and image alignment), and PV then performs photometry at the location of the source in each image using \texttt{uvotsource} (see Sect. \ref{piii} for details regarding how \texttt{TUVOpipe} performs photometry). \texttt{Uvotsource} requires a source extraction region and a background region as input; by default, PV automatically creates these regions in the same way as in PIII (thus, see Sect. \ref{piii} for a description of how these are constructed). However, PV also has the option for the user to input manually created regions. This can improve the reliability of the photometry, because, for example, users can ensure that the background region is devoid of contaminating sources that may be present in the automatically created background region.\ 

\begin{figure}
    \centering
    \includegraphics[width=.45\textwidth]{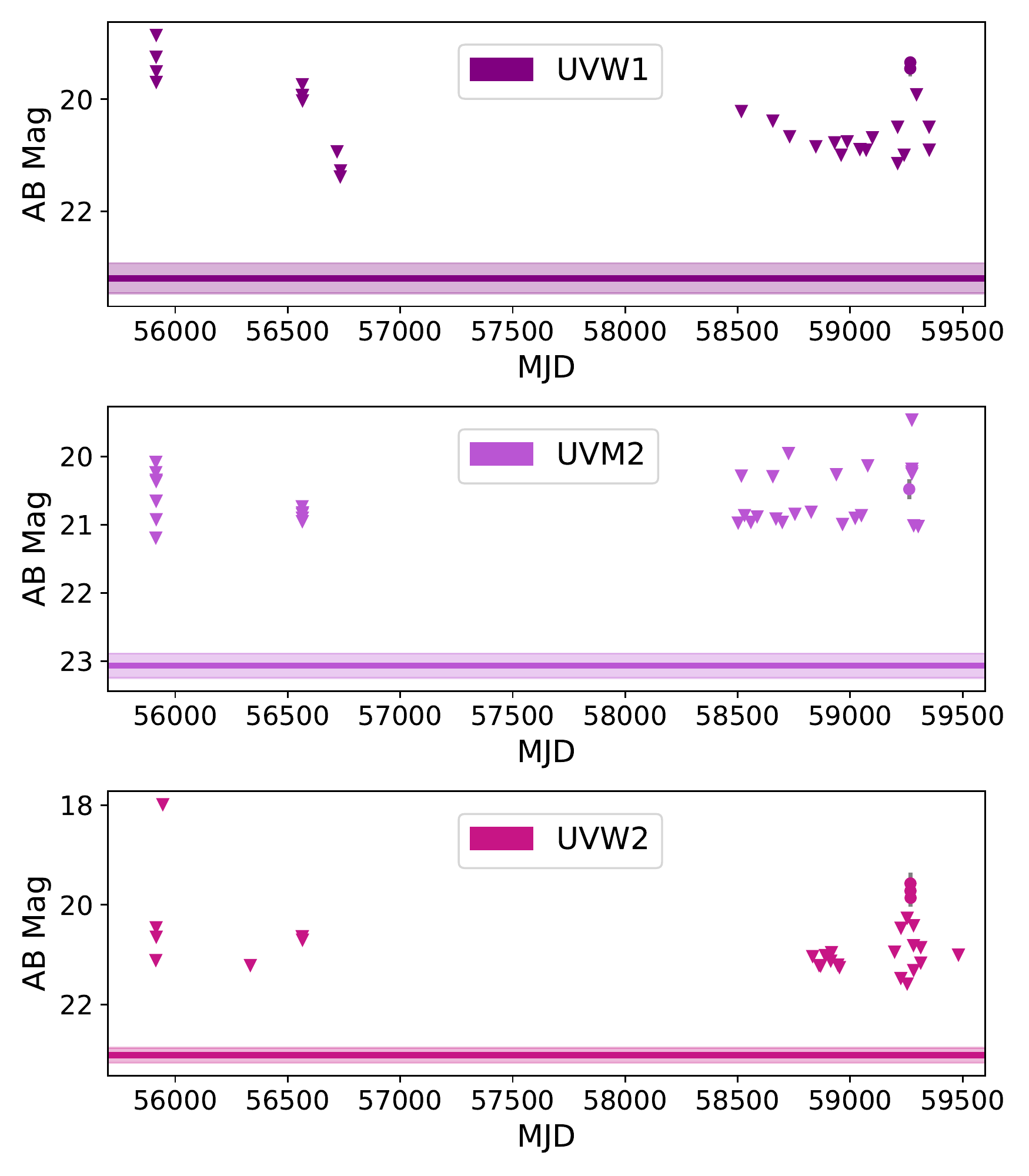}
    \caption{UVW1, UVM2, and UVW2 PV light curves for a source triggered by \texttt{TUVOpipe}, denoted as TUVO-21acq, due to an outburst detected in February 2021 (see the data points around MJD 59300). The quiescent level of the source is too faint to be detected by individual UVOT images, as seen by the upper limits displayed in the light curves. However, when all the images in which no detection was made are stacked, \texttt{uvotsource} detects a source in the stacked image due to the increased depth. This quiescent level is shown by the horizontal line, with the error shown by the shaded region. This long-term light curve shows that the recent outburst has so far been the only outburst of this source observed using the UVOT. Analysis of the outburst and further spectra obtained (see Sect. \ref{results_ql_newtrans}) allowed us to securely classify this source as a CV undergoing a DN outburst.}
    \label{fig:stacked_lc_example}
\end{figure}

Finally, using PV we attempt to obtain a better handle on the amplitude of the variability of the source of interest by stacking images in which there was no detection. Using the \textit{HEASoft} tools \texttt{fappend} and \texttt{uvotimsum}, all images in which there was no detection (see Sect. \ref{piii} for how we determine whether there is a detection with \texttt{uvotsource}) are automatically stacked, independently for each filter. PV then runs \texttt{uvotsource} on the stacked images. For each filter, this results in either a deeper upper limit (i.e. fainter than the upper limits for individual images in which there was no detection) or potentially a detection. In both cases, this feature allows us to obtain a better estimate of the amplitude of variability because we have a stronger constraint on the quiescent level (and if the stacked images result in a detection of the source, we even know the exact value). However, we note that this method assumes that at any time when there is no detection, the source is in quiescence, and that the quiescent level is constant (which is not necessarily the case). In practice, this means that although the quiescent level we determine in this way is indicative, it may not always be completely accurate, and it is useful for the user to keep this in mind when examining PV light curves. However, with PV the user can also manually select a range of observations (either by ObsID or by MJD)  to use for the summing of images. Therefore, when we detect highly interesting sources, we manually select the intervals for which to sum images, and thus we obtain deeper upper limits (or detections) as needed for the particular source. In Fig. \ref{fig:stacked_lc_example}, we show an example of PV light curves of a source which underwent an outburst that was detected by \texttt{TUVOpipe}. When in quiescence, the source is too faint to be detected in individual images in the UVW1, UVM2, and UVW2 bands, but the stacking of all non-detection images led to detections (shown as horizontal lines in the light curves). This allowed us to characterise the amplitude of the outburst with more precision. The source is a CV we discovered with \texttt{TUVOpipe} (see Sect. \ref{results_ql_newtrans} for a brief discussion and Modiano et al., in prep., for a detailed study about this source).\

There is one minor difference in operation between the two use cases of PV. As discussed in Sect. \ref{pii_dataprep}, UVOT images may be misaligned with respect to each other by up to a few arcseconds. When passing a source's coordinates to PV, we need to ensure that the coordinates will match the position of the source in all newly downloaded images to process in PV. In the mode where we run PV on a source found by \texttt{TUVOpipe}, the input coordinates are those determined in PII based on the template used there, so PV copies the template used in PII to the PV directory. It selects this as the template to which all other images in PV will be aligned. In the mode where we input any source of interest (not found by our pipeline), the coordinates are taken from external sources. In this mode, the template is selected as the first science image to be processed that has an exposure time of at least 200s and which passes the image quality checks (see Sect. \ref{pii_dataprep}). We need to ensure that this template image is correctly astrometrically solved so that the coordinates of the source of interest fall at the right location on the image. We do this by using \textit{astrometry.net} and by manually comparing the position of the source in the UVOT template image to images of the source available in the literature or in online databases. We then use the obtained UVOT position of the target as input for PV.\

\section{Results and discussion} \label{results}

\begin{figure*}[!t]
    \centering
    \includegraphics[width=.9\textwidth]{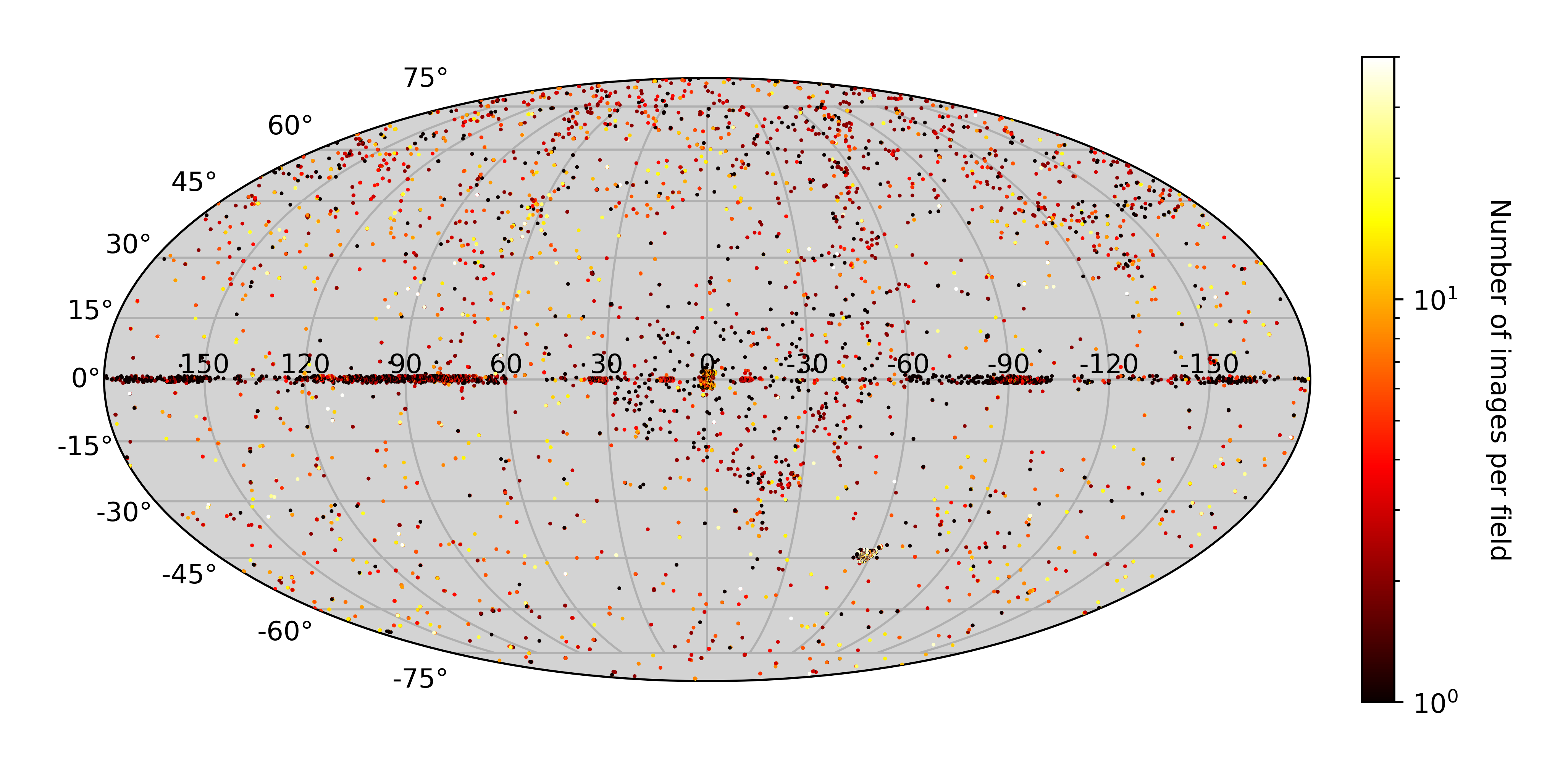}
    
    \caption{Sky map (Mollweide projection) of all data obtained using the UVOT which were processed by our pipeline (from 01 October 2020 until 23 December 2021). The colour bar indicates the number of images taken for each field (`fields' here are based on the coordinates of the centres of the images, rather than on field name). The very high number of observations at -60\degree\ and -45\degree were obtained at the position of the SMC, which \textit{Swift} has been observing extensively as part of the S-CUBED survey program \citep{Kennea_2018}. The total number of images included in the plot is 75\,183, representing a total exposure time of 22\,686 ks (see also Fig. \ref{fig:bar_charts_allobs}).}
    \label{fig:skymap}
\end{figure*}

\begin{figure*}[!b]
\centering
\includegraphics[width=.45\textwidth]{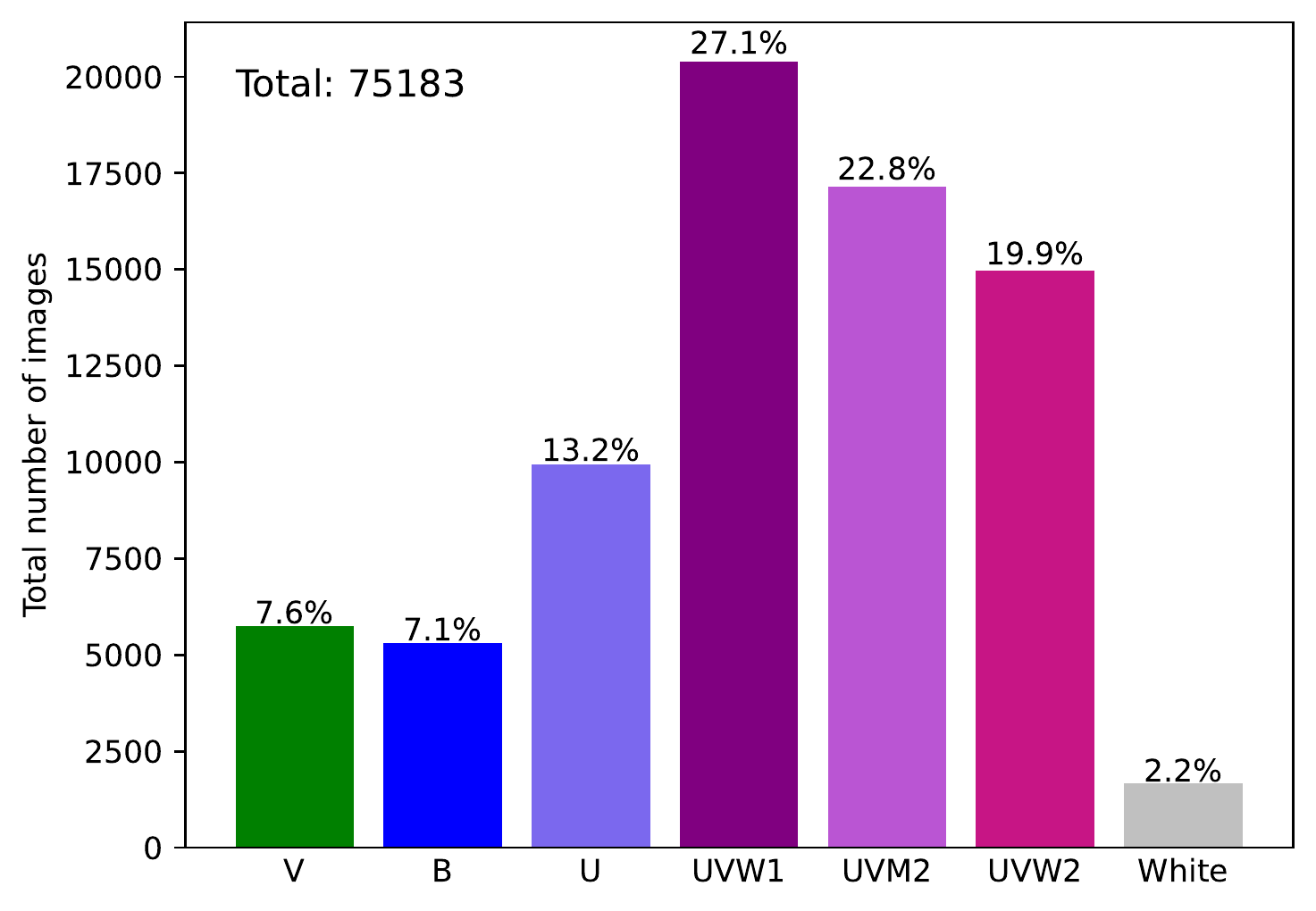} \hspace{1cm}%
\includegraphics[width=.45\textwidth]{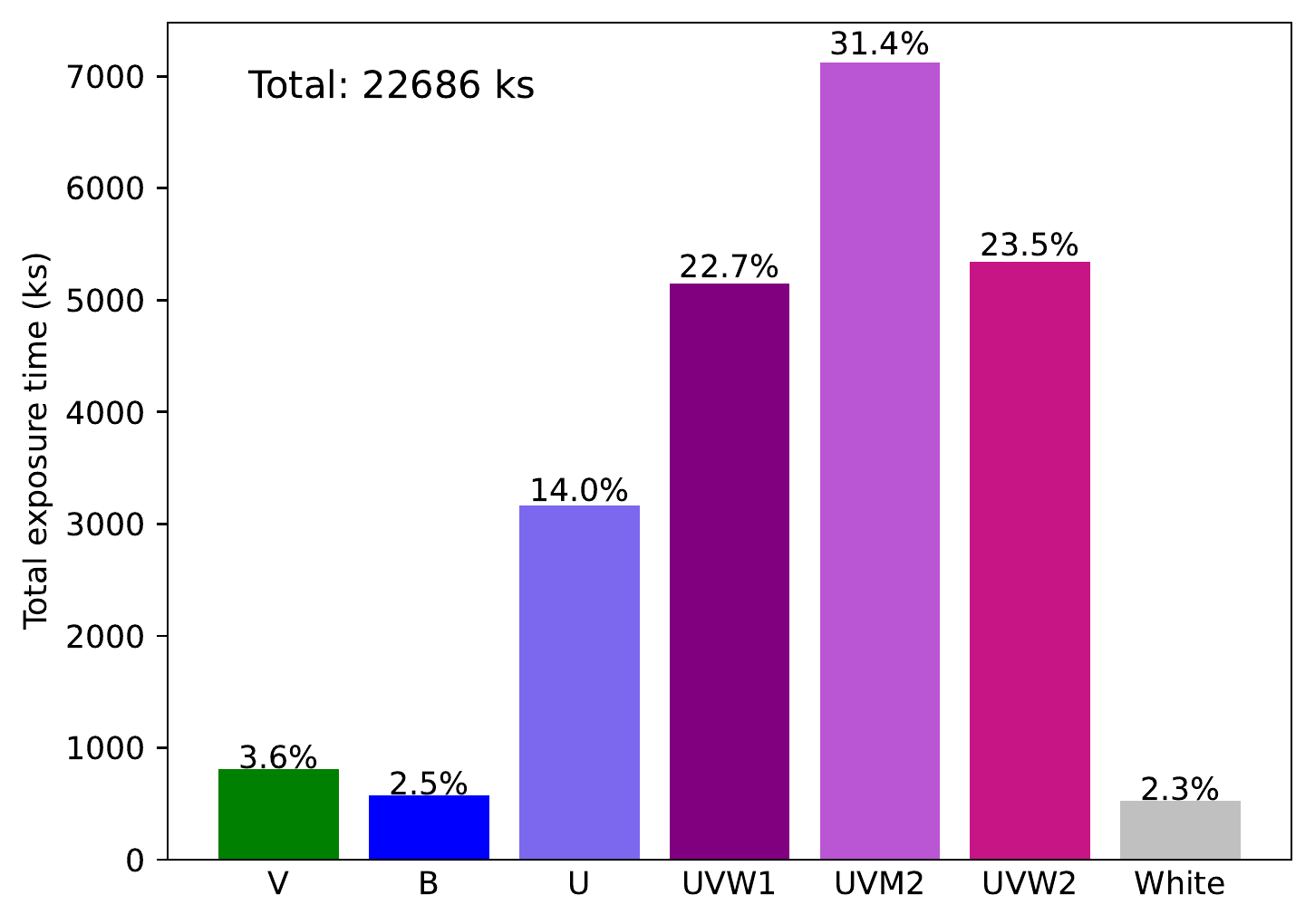}
\caption{Bar charts displaying UVOT filter usage of all data processed by our pipeline. In the left panel, the filters are separated by number of images; in the right panel, filter usage is separated by total exposure time.}
\label{fig:bar_charts_allobs}
\end{figure*}

\texttt{TUVOpipe} has been running daily since 01 October 2020 (and continues to be run daily), and we have detected several thousand real UV transients, including a few particularly interesting sources which lack clear previous identifications (see Sect. \ref{results_ql_newtrans}). In this section, we describe the results obtained since the pipeline began full operation (01 October 2020) to the time of submission (23 December 2021). \texttt{TUVOpipe} has processed all individual exposures which were on the \textit{Swift}-UVOT QL page during this period\footnote{A small fraction of images is not processed by us, including images which do not pass the elongation test (see Section \ref{pii_dataprep}), or due to errors such as repeated failed downloads or data that our software fails to read correctly (the reasons behind these failures are not always clear). In this section, we only discuss UVOT results obtained by our TUVO project; however, it is worth noting that since we process the vast majority of UVOT data (since 01 October 2020), some of the information we have obtained, such as the distribution of exposures per filter, is also a good indication of the typical characteristics of the UVOT observations.}, and typically we process all data between a few hours and a few days after the observations are obtained. In Sect. \ref{results_stats}, we discuss \texttt{TUVOpipe} statistics and present an overview of all the data processed by the pipeline. We also present breakdowns of the detected transients by filter. In Sections \ref{results_ql} and \ref{results_archival}, we discuss some specific examples of interesting transients we have detected using \texttt{TUVOpipe} in the real-time and archival modes, respectively.

\subsection{TUVOpipe statistics} \label{results_stats}

\subsubsection{Observations processed} \label{results_stats_observations}

\begin{figure}[!h]
\centering
\includegraphics[width=.45\textwidth,trim={0 1cm 1cm 2cm}]{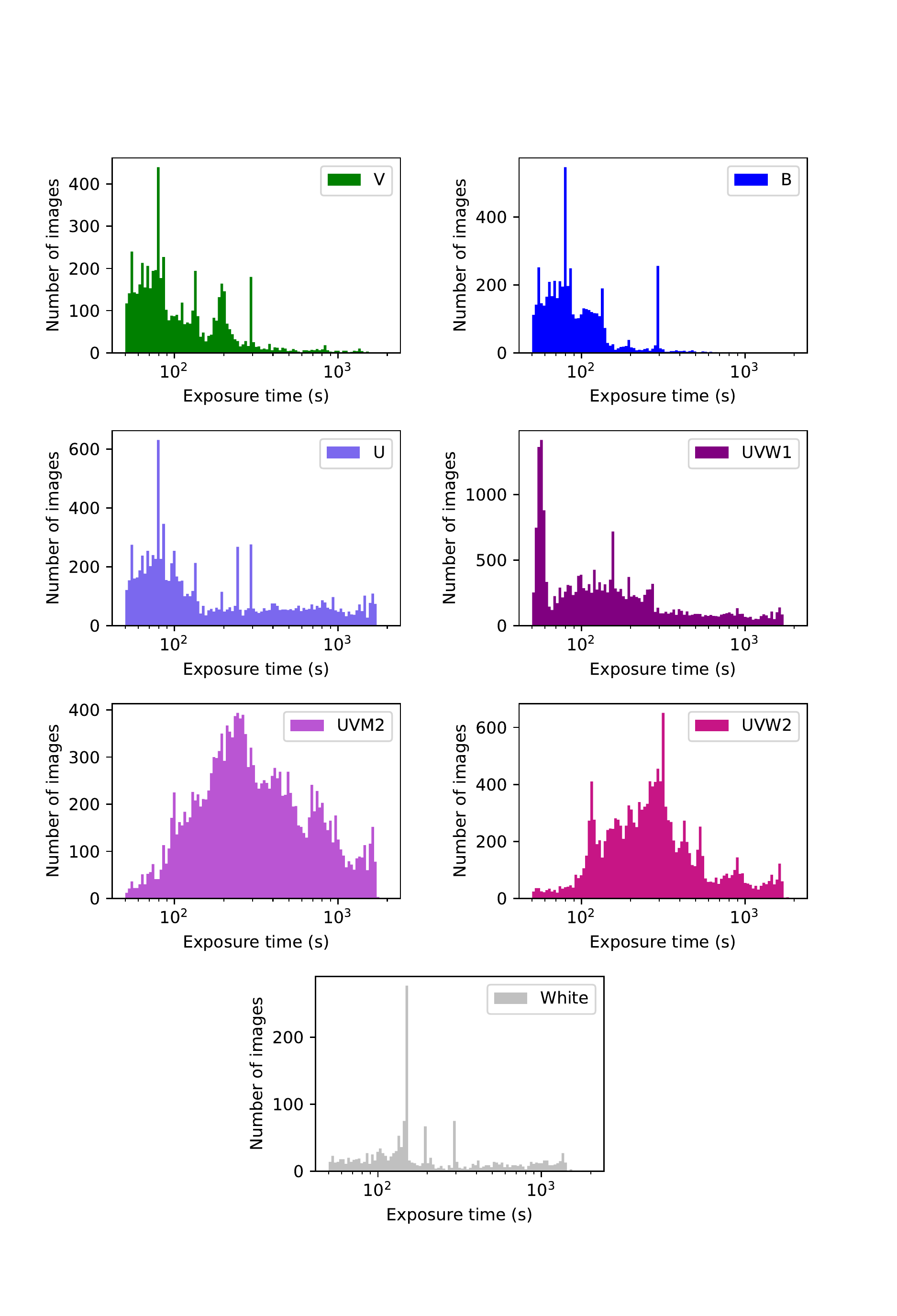}
\caption{Histograms of UVOT exposure times per filter processed by \texttt{TUVOpipe}. The x-axes are logarithmic, and the bins are evenly spaced in logarithmic space.}
\label{fig:exposures_filt}
\end{figure}

\begin{figure}[!h]
\centering
\includegraphics[width=.45\textwidth]{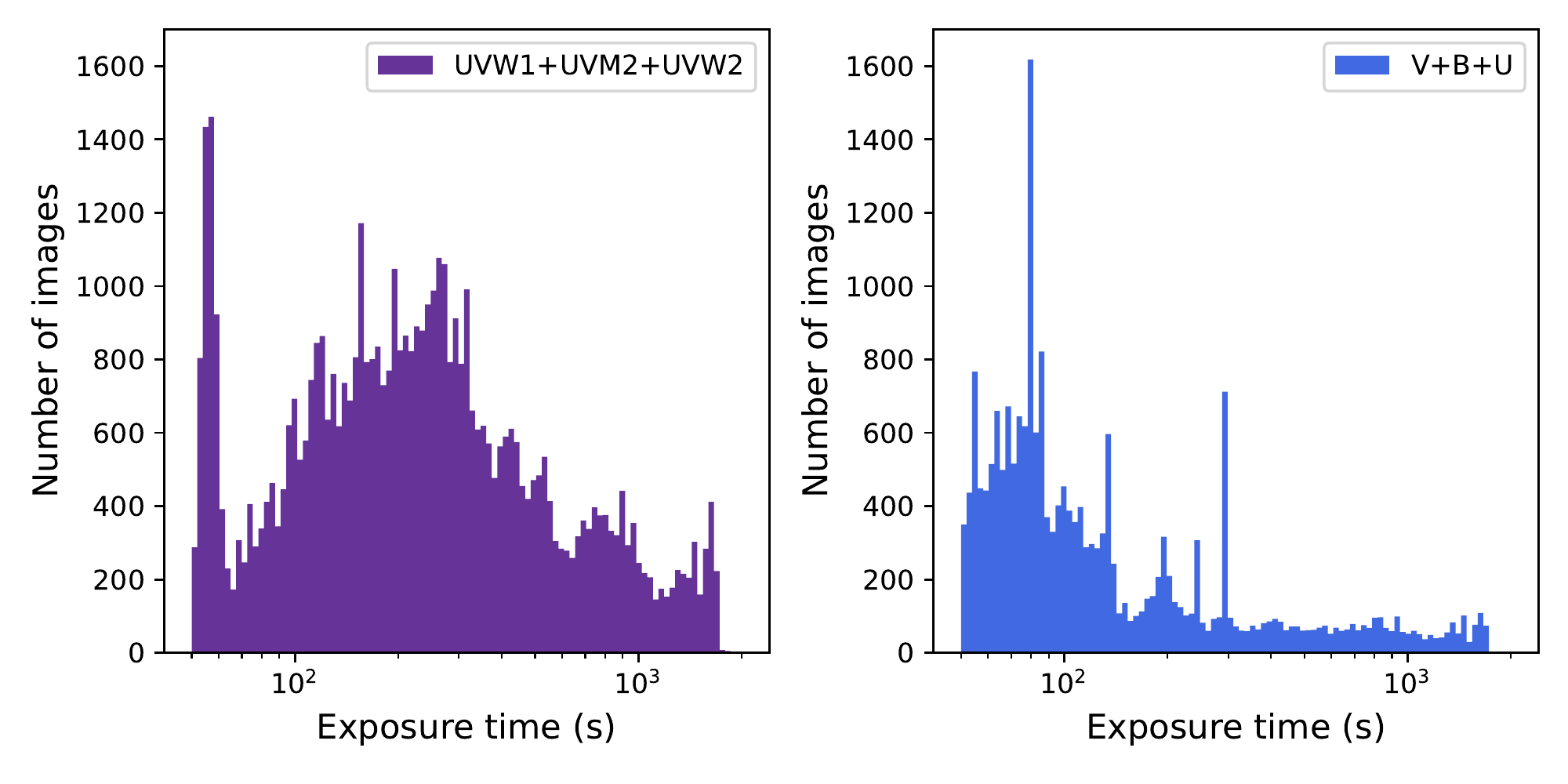}
\caption{Histograms of UVOT exposure times per wavelength regime (optical and UV) processed by \texttt{TUVOpipe}. This is the same data as shown in Fig. \ref{fig:exposures_filt}, but with the three optical filters combined and the three UV filters combined (the white filter is not included). The x-axes are logarithmic and the bins are evenly spaced in logarithmic space.}
\label{fig:exposures_combined}
\end{figure}

So far, TUVO has processed 75\,183 individual UVOT snapshots (we refer to these individual exposures simply as images), or a total of $\sim$22\,686 ks of UVOT exposure time. In Fig. \ref{fig:skymap}, we show a sky map of all the fields processed by TUVO, with the colour indicating the total number of images per field. These fields are scattered throughout the sky, with concentrations in the Galactic plane and on the SMC (-60\degree, -45\degree), the latter mainly due to the S-CUBED survey \citep{Kennea_2018}, an ongoing survey of the Small Magellanic Cloud with \textit{Swift}.\

In Fig. \ref{fig:bar_charts_allobs}, we show how the number of images (left panel) and total exposure time (right panel) are distributed between the UVOT filters. The majority of data were taken in the three UV filters ($69.8\%$ of UVOT observations by number of images and $77.6\%$ by exposure time). This reflects the primary usage of the instrument as a UV telescope in the form of the mentioned `filter of the day' strategy (see Section \ref{uvot}). Total exposure time is similar similar between these three UV filters ($\sim$22-32$\%$ each, with UVM2 being the most used filter by exposure time with $31.4\%$). Only a relatively small fraction of UVOT observations are performed with the V, B, and U filters: $27.9\%$ by number of images and $20.1\%$ by total exposure time. By number of images, the distinctions between filter usage are slightly less pronounced than by exposure time. This implies that UV exposure times are on average longer than optical exposure times (see also Figs. \ref{fig:exposures_filt} and \ref{fig:exposures_combined}, where this is clearly visible).\

Figures \ref{fig:exposures_filt} and \ref{fig:exposures_combined} show histograms of the exposure times used in UVOT observations per filter (the latter combines the UV filters and combines the optical filters). These clearly show the preference for short exposure times for the optical filters (the distribution peaks at $<$100s) and for long exposure times for the UV filters (the bulk of the distribution peaks around a few hundred seconds; the narrow peak at $\sim$50s in the UVW1 is largely due to the SMC survey that uses short exposures). As seen from the histograms, the white filter is the least used filter, similar in total usage to the V and B filters by exposure time, and significantly less by number of images (also indicating that exposures with the white filter are typically longer than those in the optical filters). \ 

\begin{figure}
    \centering
    \includegraphics[width=.45\textwidth]{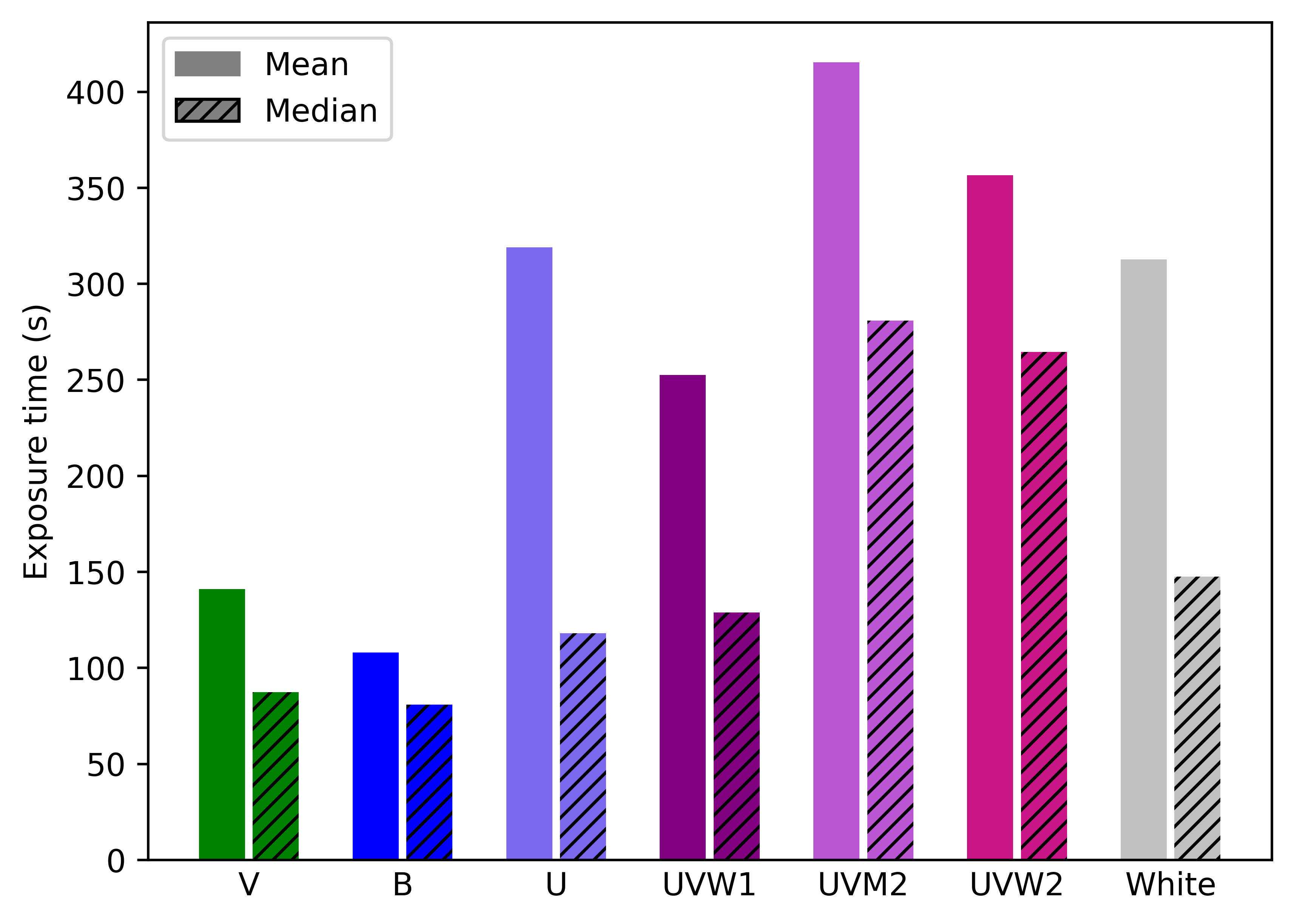}
    \caption{Mean and median exposure times per image for each UVOT filter for all data processed by \texttt{TUVOpipe}.}
    \label{fig:mean_med_exptime}
\end{figure}

In Fig. \ref{fig:mean_med_exptime}, we display the average and median exposure times per filter. The mean exposure times for the V and B filters are significantly lower than those of the U and UV filters (<150s vs. >250s), which are relatively flat (the lower average value for the UVW1 filter than for the other UV filters is again due in part to the SMC survey being carried out in that filter with short exposures). The median values are roughly flat at around 100s for the V, B, U, and UVW1 filters and jump to around 250s for the UVM2 and UVW2 bands. The higher values in the average compared to the median indicate that the latter is dominated by a few long exposures (several hundred seconds). \

\begin{figure}[!h]
\centering
\includegraphics[width=.45\textwidth]{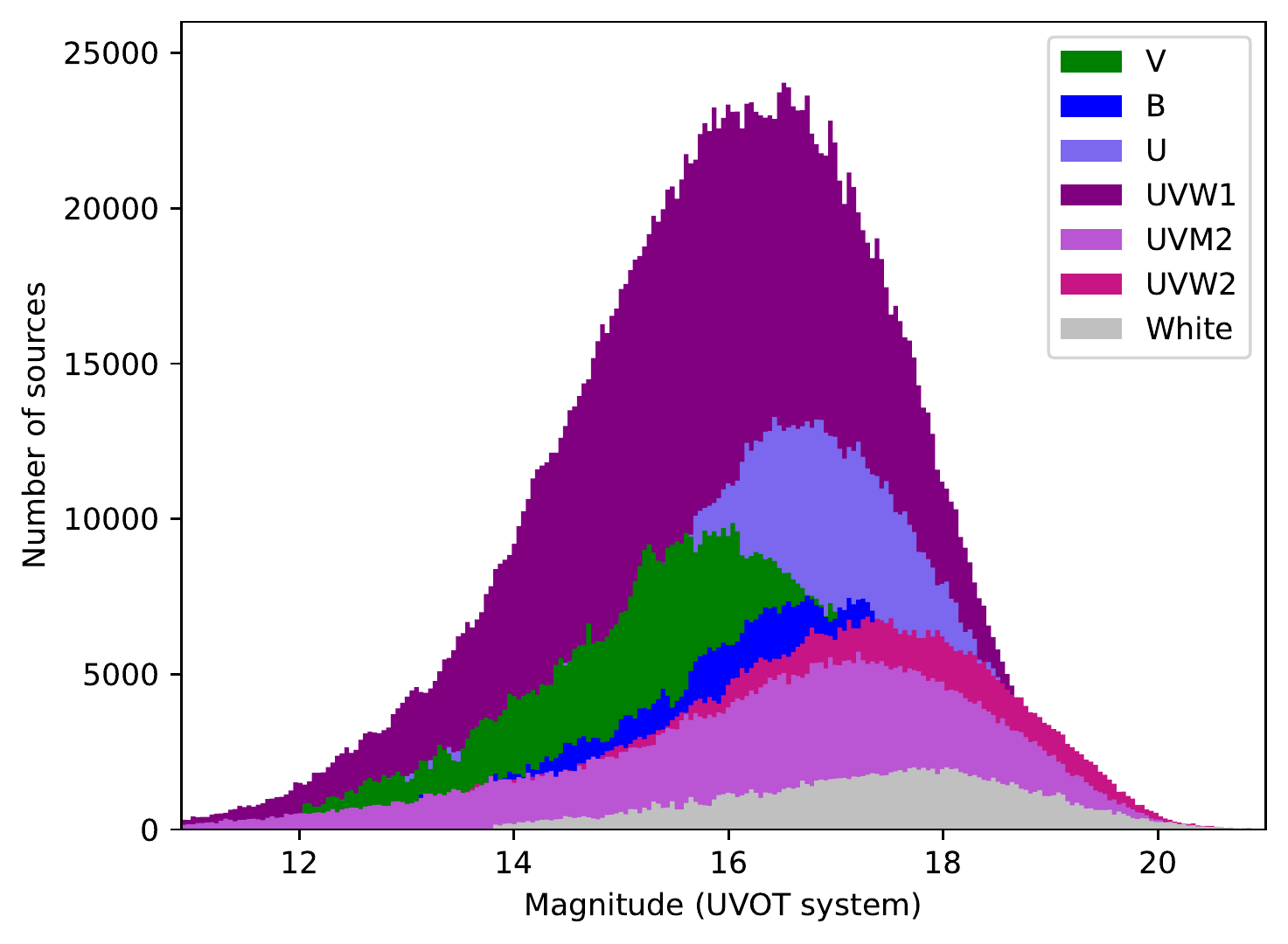}
\caption{Magnitudes of all detected sources (when using a threshold of $5\sigma$) in all science images processed using \texttt{TUVOpipe}, separated out by filter. Magnitudes are in the UVOT magnitude system (see \citealp{Breeveld_2011}). The data are not stacked, i.e. histograms are directly overplotted on each other. Each histogram contains 300 bins of equal width.}
\label{fig:mags}
\end{figure}

\begin{figure*}[!ht]
    \centering
    \includegraphics[width=.95\textwidth]{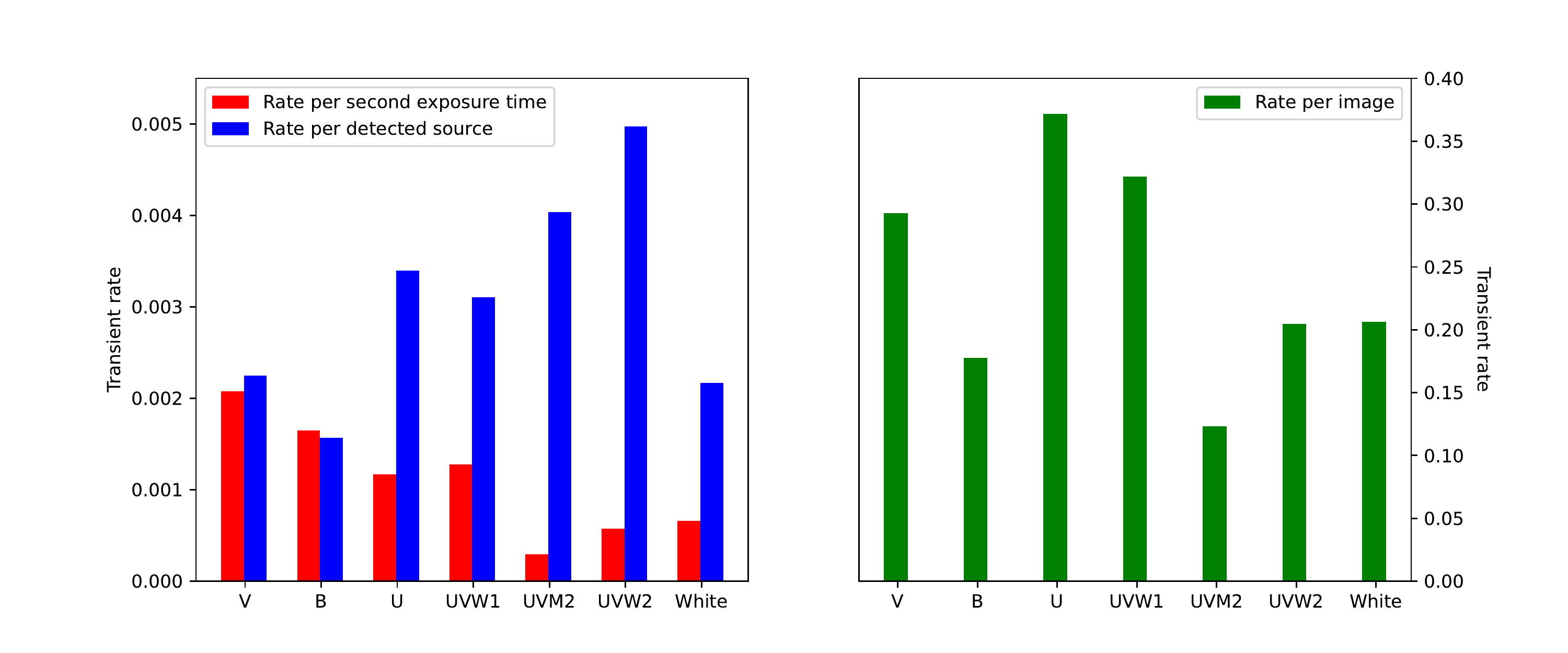}
    \caption{Transient candidate rates determined by \texttt{TUVOpipe} in each filter. The rates are computed independently per filter and are the ratios of the total number of transients detected to: the total exposure time processed by \texttt{TUVOpipe} (rate per second exposure time); the total number of sources detected at $5\sigma$ in all the science images processed (rate per detected source); and the total number of images processed (rate per image).}
    \label{fig:transient_rates}
\end{figure*}

In Fig. \ref{fig:mags}, we show the magnitude distributions of all sources detected (using a threshold of $5\sigma$) in the UVOT science images processed by \texttt{TUVOpipe} so far. Magnitudes were extracted using \texttt{uvotdetect}, which performs photometry on all detected sources in a UVOT image\footnote{We note that although \texttt{uvotsource} is typically preferred for UVOT photometry, \texttt{uvotdetect} runs significantly faster per source, and it is thus well suited to obtaining an indication of the overall statistics for a large number of sources and observations. \texttt{Uvotdetect} photometry measurements of point sources have been found to agree well with those derived from \texttt{uvotsource}. See \url{https://www.swift.ac.uk/analysis/uvot/pos.php} for details. Magnitudes shown here are in the UVOT magnitude system (see \citealp{Breeveld_2011}) since this is the only system in which magnitudes are given by \texttt{uvotdetect}.}. The histograms of the different filters are not stacked; i.e. the data overlap. The distributions peak between magnitudes 16 and 18, with a slight trend of increasing magnitudes (fainter sources) towards the UV. This may be due to a combination of the longer exposure times used in the UV bands on average resulting in greater depth and the different zero points of the filters (see Table 2 in \citealp{Breeveld_2011}).\

\subsubsection{Transient rates} \label{results_stats_transients}

In this section, we discuss candidate transients detected by \texttt{TUVOpipe}. These are transient detections that meet all automatic criteria within the pipeline and have thus made it to PIV. However, they are still `candidates', because most of the sources that pass all automatic tests are not real variable sources (typically $\sim$70\% of our transient candidates are bogus), and must be separated from real transients by eye (see Sect. \ref{vetting}). Therefore, we emphasise that the information provided in this section is mostly useful in terms of describing the output of the pipeline, and does not allow for detailed statistics about true transient rates detected by UVOT.\ 

So far, \texttt{TUVOpipe} has made a total of 28\,061 transient candidate detections; this represents 15\,154 independent candidates found (many detections are duplicate; for example, when multiple filters find the same transient or when a transient is active for many UVOT observations). Typically, between several tens and a few hundred candidates are detected by \texttt{TUVOpipe} every day, though the daily output is highly variable as it can depend on the fields observed by UVOT (e.g. crowdedness, diffuse emission, sparse fields). For example, when images from the S-CUBED SMC survey \citep{Kennea_2018} are processed, the number of candidates we detect is significantly higher than average, due to crowded fields resulting in both more real sources (having many sources in the field increases the chances of finding variable sources) and more bogus sources (due to relatively more frequent problems with alignment; see Sect. \ref{pii}).\

In Fig. \ref{fig:transient_rates}, we display the rates of TUVO transients in each filter, using three different measurements. The transient rate per second of exposure time (i.e. the ratio of the number of transients detected to seconds of exposure time) gradually drops towards the UV in the V, B, U, and UVW1 bands and then decreases significantly for the UVM2 and UVW2 filters (the UVM2 filter is around 6 times less prolific than the V filter by this metric). The transient rate per detected source (i.e. the ratio of the number of transients to the total number of sources detected in the science images) shows the fraction of sources that are variable in each filter and which our pipeline detects. Here, we see an increasing rate in the U and UV filters compared with the redder B and V optical bands. The transient rate per image (i.e. the ratio of the number of transients detected to the number of images) gives an indication of the average number of images required in each filter for \texttt{TUVOpipe} to detect a transient. This rate peaks in the U and UVW1 bands, where we detect, on average, one transient in every two-to-three images, and is lowest in the UVM2 band, where we detect one transient in every eight-to-nine images.\ 

The average rates across all filters are: 0.0011 (rate per second of exposure time), 0.0031 (rate per detected source), and 0.24 (rate per image). As mentioned, since the majority of the transient candidates discussed here are not real, we cannot make conclusive statements about any potential intrinsic astrophysical reasons for the varying rates across bands, due to the significant effect of the pipeline's performance on these rates.\

\subsection{Initial results from Quick-Look searches} \label{results_ql}

The primary goal of the TUVO project is to serendipitously discover new transients that have not been previously detected, even during quiescence. Unfortunately, no robust, confirmed astrophysical transient in this category has been discovered by the project so far, although some candidates have been detected (note, however, that these candidates were only detected in a single UVOT image, so we cannot rule out that they are uncatalogued asteroids; see Wijnands et al., in prep., for a more in-depth discussion and an example). However, transients we discover that have previously known quiescent counterparts (such as detections from  deep sky surveys), but which do not have any known classification or variable behaviour, are also considered to be very interesting by the TUVO project. Finally, detecting new outburst behaviour of known variable sources is a further aim of the project.\

In this section we describe some specific examples of different kinds of transients we have detected so far with \texttt{TUVOpipe}. For each example transient, we only provide a brief summary of the key information that can be extracted and what can be learned; in-depth discussions of each source are beyond the scope of this paper. We note that for formatting and aesthetic purposes, we omit the detailed information obtained with PIV (e.g. catalogue information, variability information, etc.) from the figures shown for each transient. Instead, we display the PIII light curves and science, template, and difference image stamps (the PIII products), and where relevant, the long-term light curves (the PV products), and we note any important information about each source in the caption(s). When displaying PV light curves, we show data for all available filters in a single plot and omit the (potential) detection in quiescence obtained by stacking the individual images in which no detection was made (see Sect. \ref{pv} and Fig. \ref{fig:stacked_lc_example} for a description and example of this feature of PV). We note that the coordinates we give for all the transients in this section are those obtained by the pipeline and therefore may have errors of up to a few arcseconds (due to the uncertainties on the astrometry of UVOT images; see Sect. \ref{pii_dataprep}).\ 

\subsubsection{New outburst discoveries} \label{results_ql_newtrans}

\begin{figure}[htb]
    \centering
    \includegraphics[width=.45\textwidth]{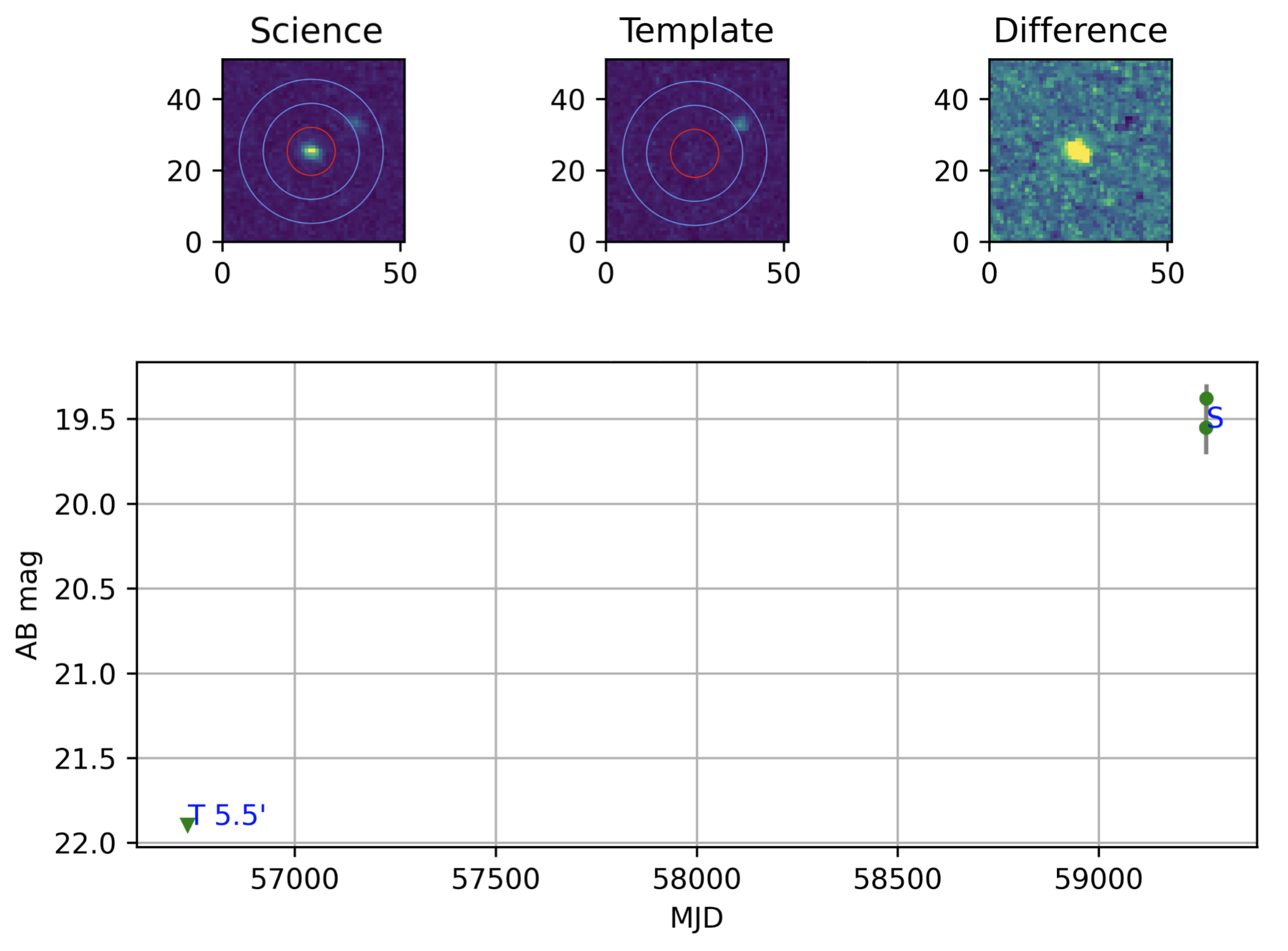}
    \caption{UVW1 light curve and different image stamps of TUVO-21acq, the transient detected by \texttt{TUVOpipe} at RA=13:05:44.74, Dec=-49:32:58.4. The source was previously detected in quiescence but had no classification. We later classified it as a CV using UVOT data and a SALT spectrum we obtained, which was taken during quiescence (see Modiano et al., in prep., for a full report about this source).} 
    \label{fig:PIV_TUVO-21acq}
\end{figure}

The source shown in Fig. \ref{fig:PIV_TUVO-21acq}, TUVO-21acq, is a TUVO discovery of a previously unknown transient, and we refer the reader to Modiano et al., in prep., for a detailed study of this source. Though there are archival detections of a source at the position of TUVO-21acq when the source was in quiescence (e.g. Gaia and XMM-OM catalogues), the nature of the source has so far not been determined. From the timescale and amplitude of the UV outburst we detected (which were, respectively, 7-21 days and >3.9 magnitudes in the UVW1 filter), we suspected that this was a DN outburst from a cataclysmic variable (CV). We created a long-term light curve using PV, which is displayed in Fig. \ref{fig:stacked_lc_example}. The PV light curves allowed us to confirm that no previous outbursts of this source have been detected by UVOT, and we also did not find any previous outbursts reported by transient searching facilities such as ZTF, ASAS-SN, or ATLAS (i.e. in any of the transient or variable star catalogues we probe with PIV; see Sect. \ref{piv_queries}). Additionally, by stacking the individual UVOT images in which no detection was made (see Section \ref{pv} for a description of this feature of PV), we detected the source during quiescence and measured its brightness at 23.20 $\pm$ 0.27, 23.07 $\pm$ 0.18, and 23.02 $\pm$ 0.15, in the UVW1, UVM2, and UVW2 bands, respectively. This allowed us to better characterise the outburst amplitude. After our discovery of this source, we obtained a spectrum with the Southern African Large Telescope (SALT; \citealp{Buckley2006}) during quiescence, which we used to securely classify it as a CV.\

\begin{figure}
    \centering
    \includegraphics[width=.45\textwidth]{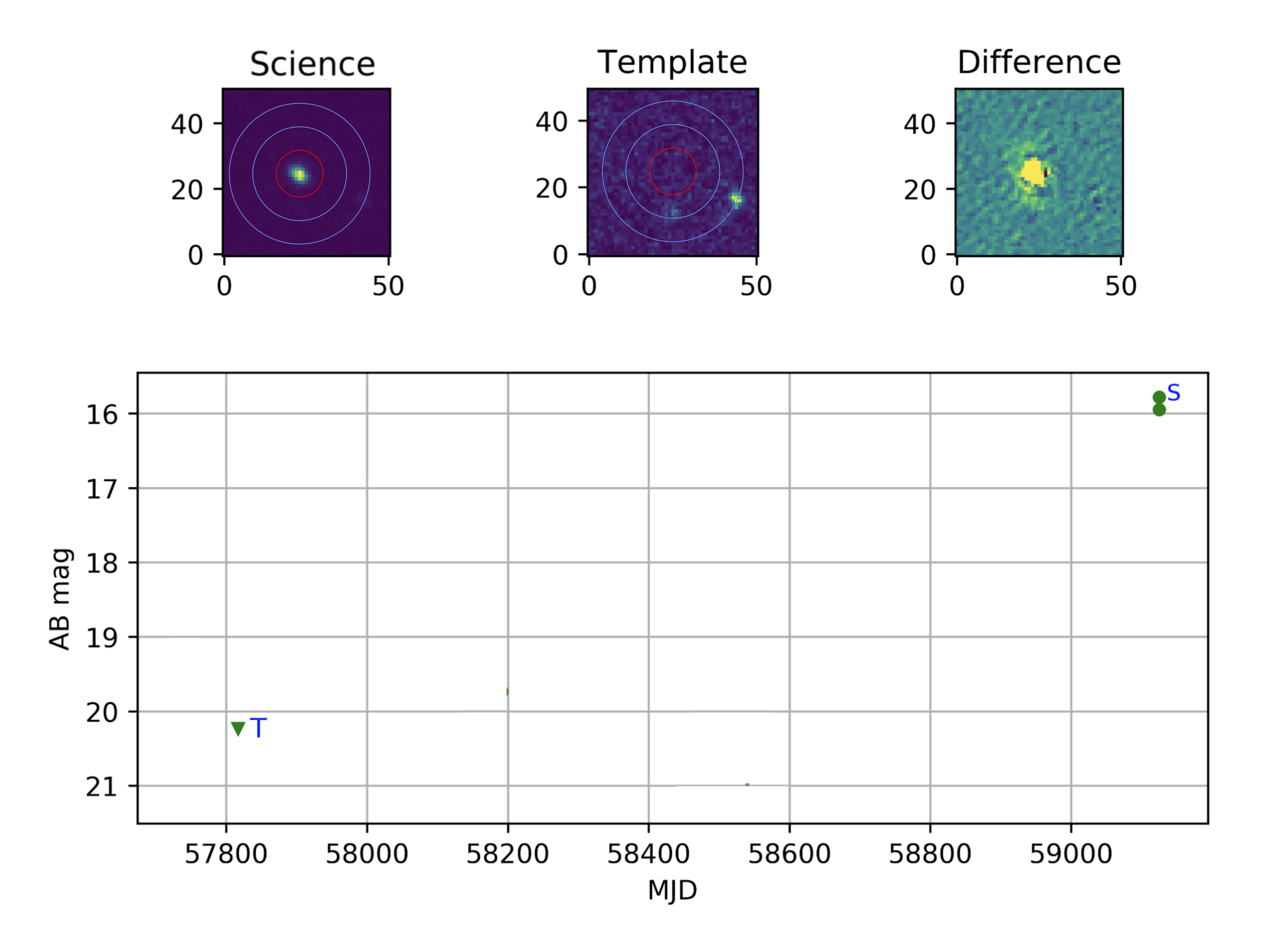}
    \caption{UVW1 light curve and the different image stamps of the transient detected using \texttt{TUVOpipe} at RA=14:28:33.52, Dec=-46:11:26.5. This source was associated with a known CV (ASASSN-18eh), and it was not the target of the \textit{Swift} observations.} 
    \label{fig:new_outburst_CV}
\end{figure}

The source shown in Fig. \ref{fig:new_outburst_CV}, for which we refer the reader to \cite{Verberne_2020} for a detailed discussion, is a TUVO discovery of the first reported UV outburst of the known CV ASASSN-18eh\footnote{\url{http://www.astronomy.ohio-state.edu/asassn/transients.html}}. Using \texttt{TUVOpipe}, we detected this transient in October 2020 and found the UV brightness in the UVW1, UVM2, and UVW2 to increase by at least 6.0 magnitudes compared to quiescent levels. This is consistent with relatively high-amplitude DN brightenings, which are typically in the two-to-six magnitude range in the optical (though they may be stronger in the UV; see e.g. \citealp{Giovannelli_2008} and \citealp{Neustroev_2017}). Due to solar constraints, we could not obtain further observations, so we only have a lower limit on the outburst duration of just over one day. The only previously known outburst of this source was reported by ASAS-SN and occurred in February 2018, indicating a maximum recurrence time of around 2.7 years. This is consistent with known DN recurrence times of days to decades (although several outbursts may of course have been missed, so the recurrence time is likely lower than our upper limit). For further examples of new outbursts we detected with \texttt{TUVOpipe}, including a bright UV transient that was previously seen in quiescence but with no previous identification (\citealt{ATEL_Modiano}), we invite the reader to consult Wijnands et al. (in prep). \

\subsubsection{Recovered known transients} \label{results_ql_knowntrans}

\begin{figure}
    \centering
    \includegraphics[width=0.45\textwidth]{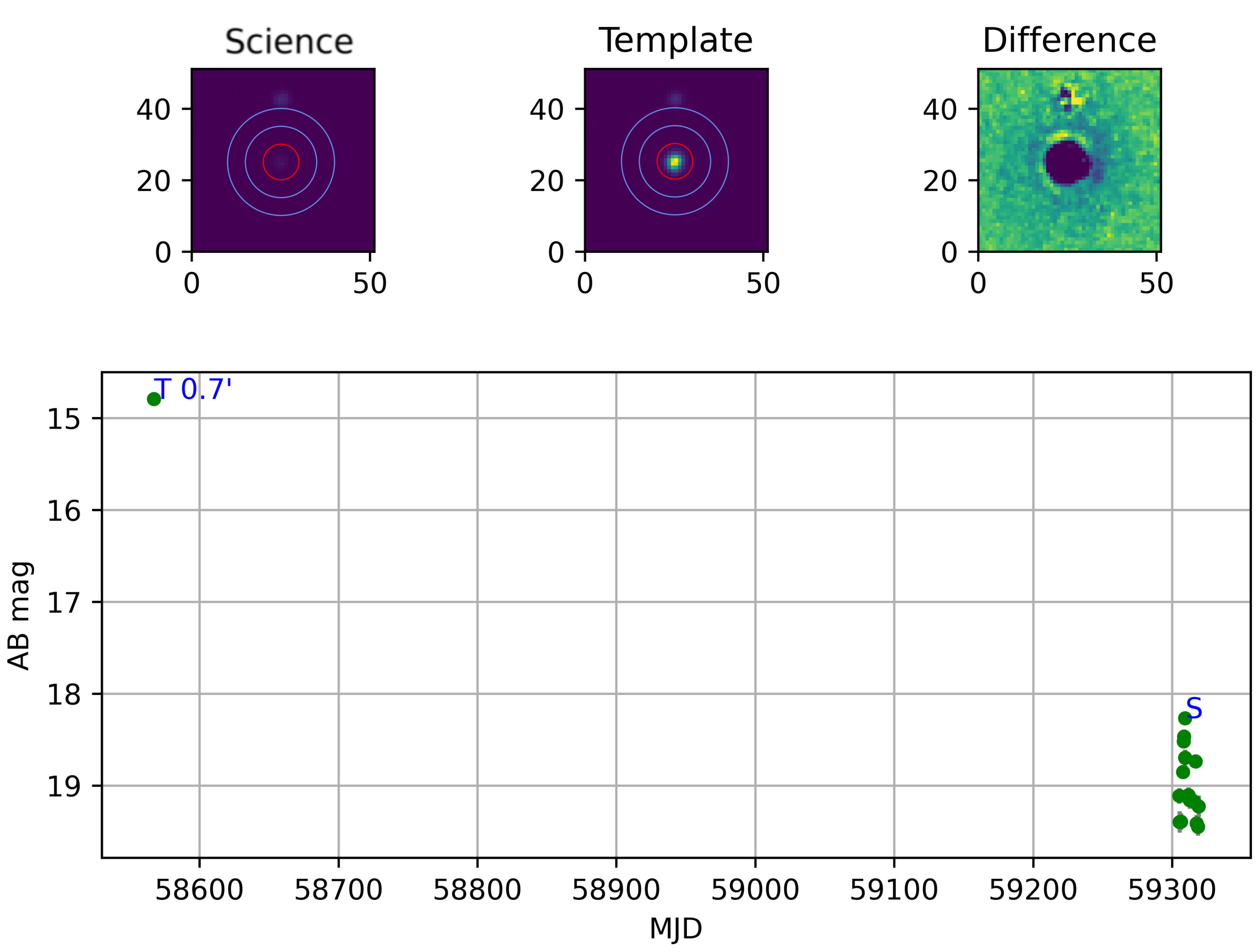}
    \caption{UVW1 light curve and the different image stamps of the transient detected using \texttt{TUVOpipe} at RA=18:20:21.94, Dec=+07:11:07.3. This source was identified as the known source MAXIJ1820+070, a black hole X-ray binary, and was the target of the \textit{Swift} observations. In Fig. \ref{fig:PV_maxij1820}, we show the long-term (PV) light curve of this source.}
    \label{fig:piv_maxi}
\end{figure}

\begin{figure}
    \centering
    \includegraphics[trim={0 0 0.3cm 0},clip,width=0.45\textwidth]{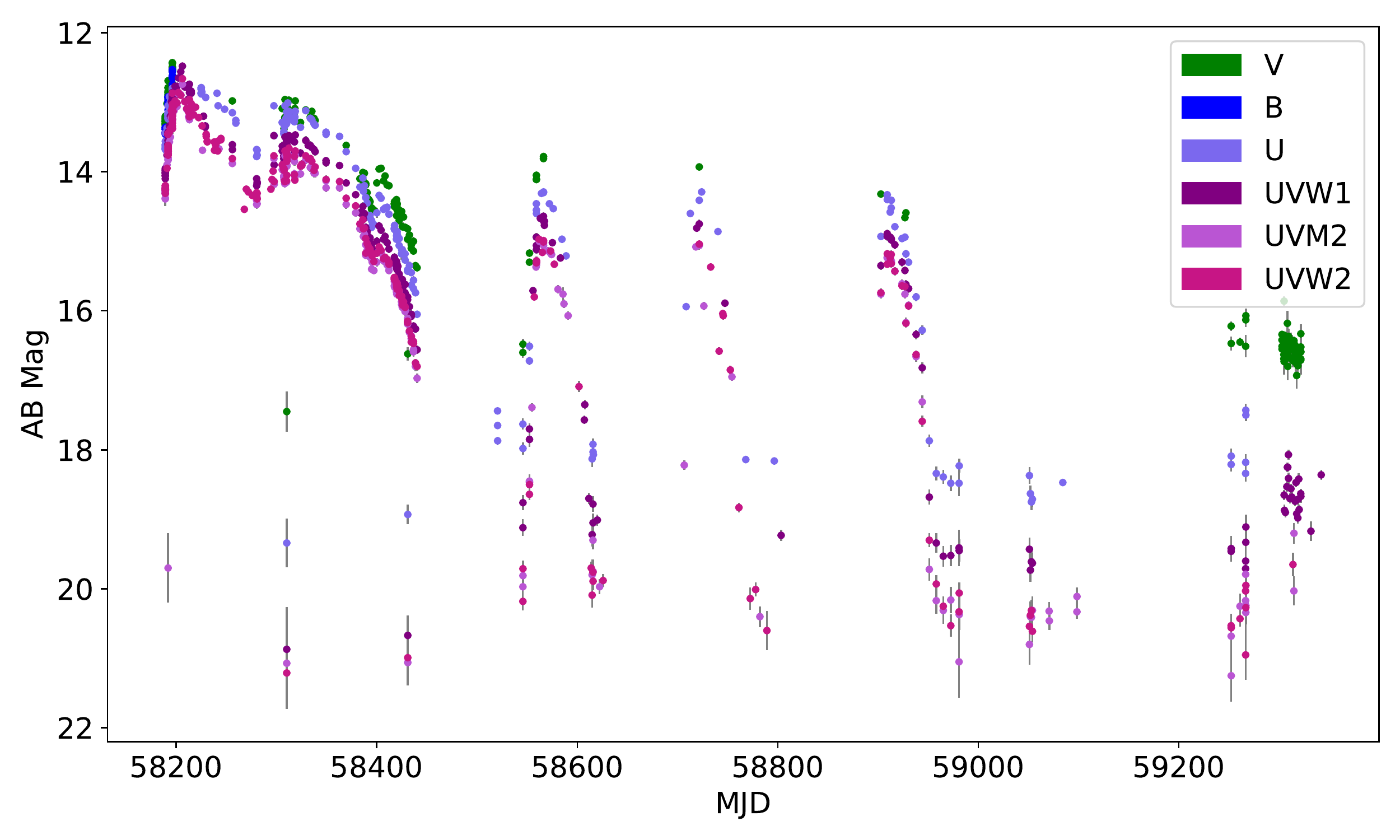}
    \caption{Long-term light curves created using PV of MAXIJ1820+070, a transient that was detected by our pipeline (see Fig. \ref{fig:piv_maxi}). The few data points obtained from images taken in the white filter are not shown. This source was targeted very often by \textit{Swift} due to the outburst and several reflares in recent years.}
    \label{fig:PV_maxij1820}
\end{figure}

\begin{figure}
    \centering
    \includegraphics[width=.45\textwidth]{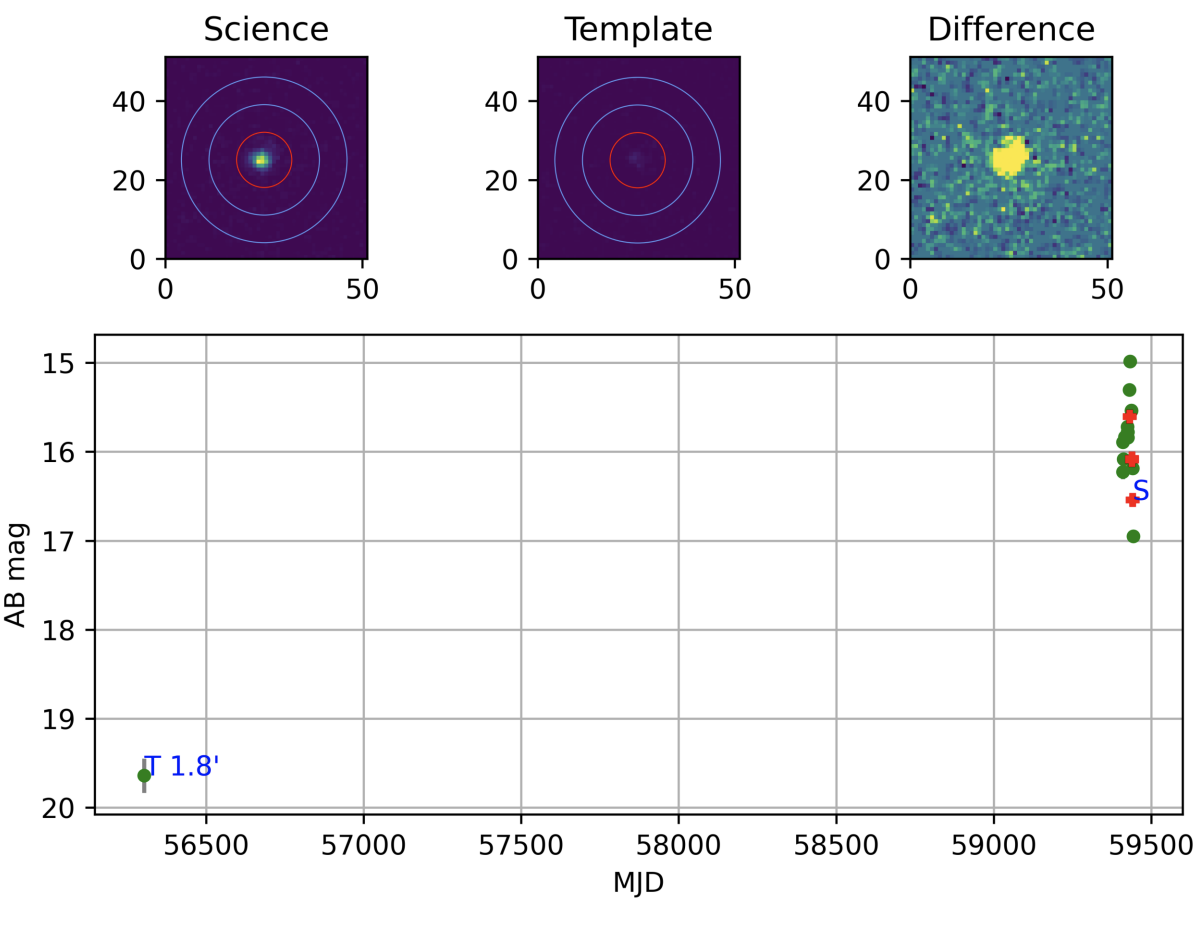}
    \caption{UVM2 light curve and the different image stamps of the transient detected using \texttt{TUVOpipe} at RA=22:02:43.26, Dec=+42:16:40.0. This source is the prototype BL Lacertae object, a highly variable blazar that is frequently observed by \textit{Swift} and is therefore often detected as a transient by our pipeline. It was the target of the \textit{Swift} observations. The red points indicate data points obtained on images that may not have been successfully aligned in the pipeline (see Sect. \ref{pii_dataprep}). In Fig. \ref{fig:bllac_pv}, the long-term (PV) light curve of this source is displayed.} 
    \label{fig:bllac_piv}
\end{figure}

\begin{figure}
    \centering
    \includegraphics[width=.45\textwidth]{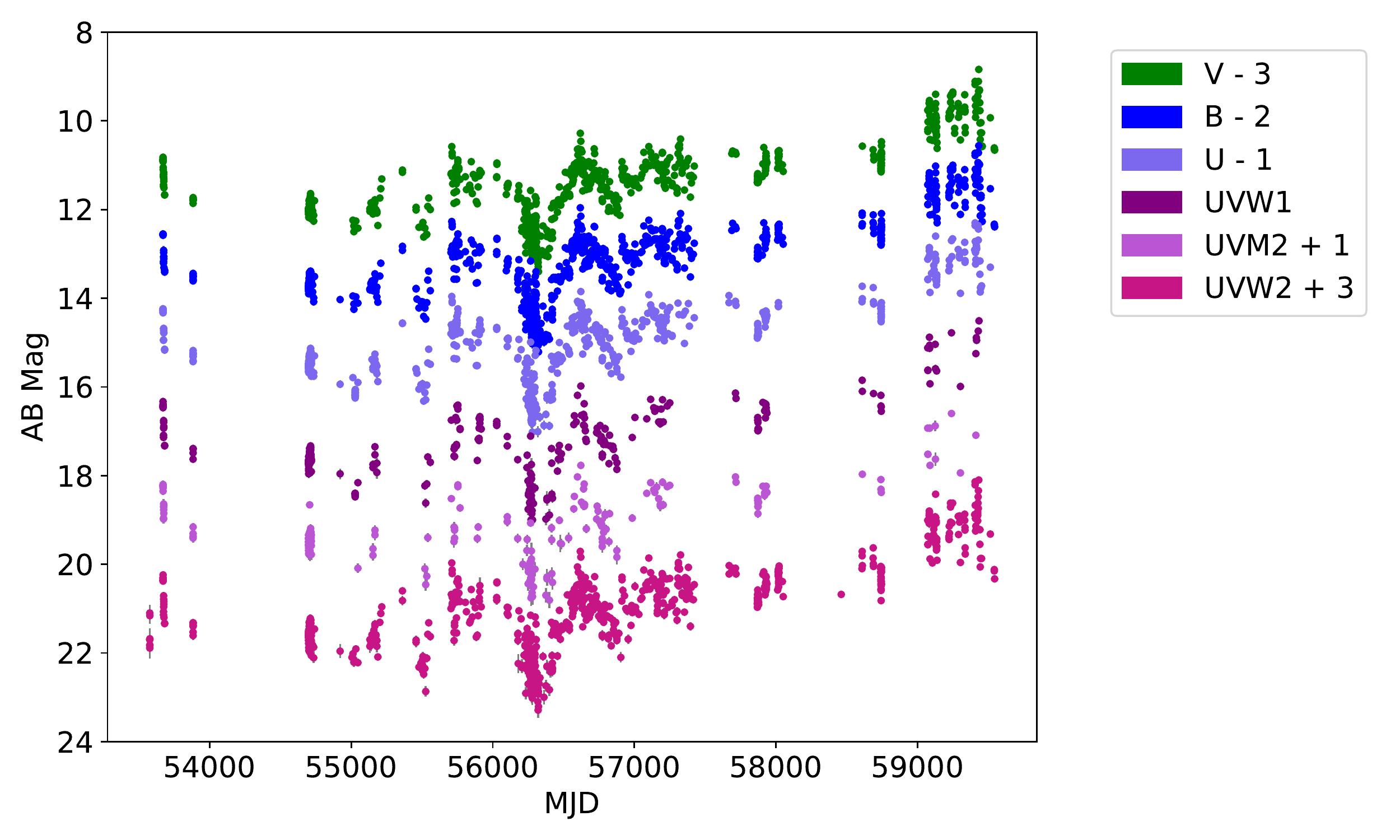}
    \caption{Long-term light curves created using PV of the flaring blazar BL Lacertae with all six primary UVOT filters. This transient was detected with \texttt{TUVOpipe} (see Fig. \ref{fig:bllac_piv}).} 
    \label{fig:bllac_pv}
\end{figure}

\begin{figure}
    \centering
    \includegraphics[width=.45\textwidth]{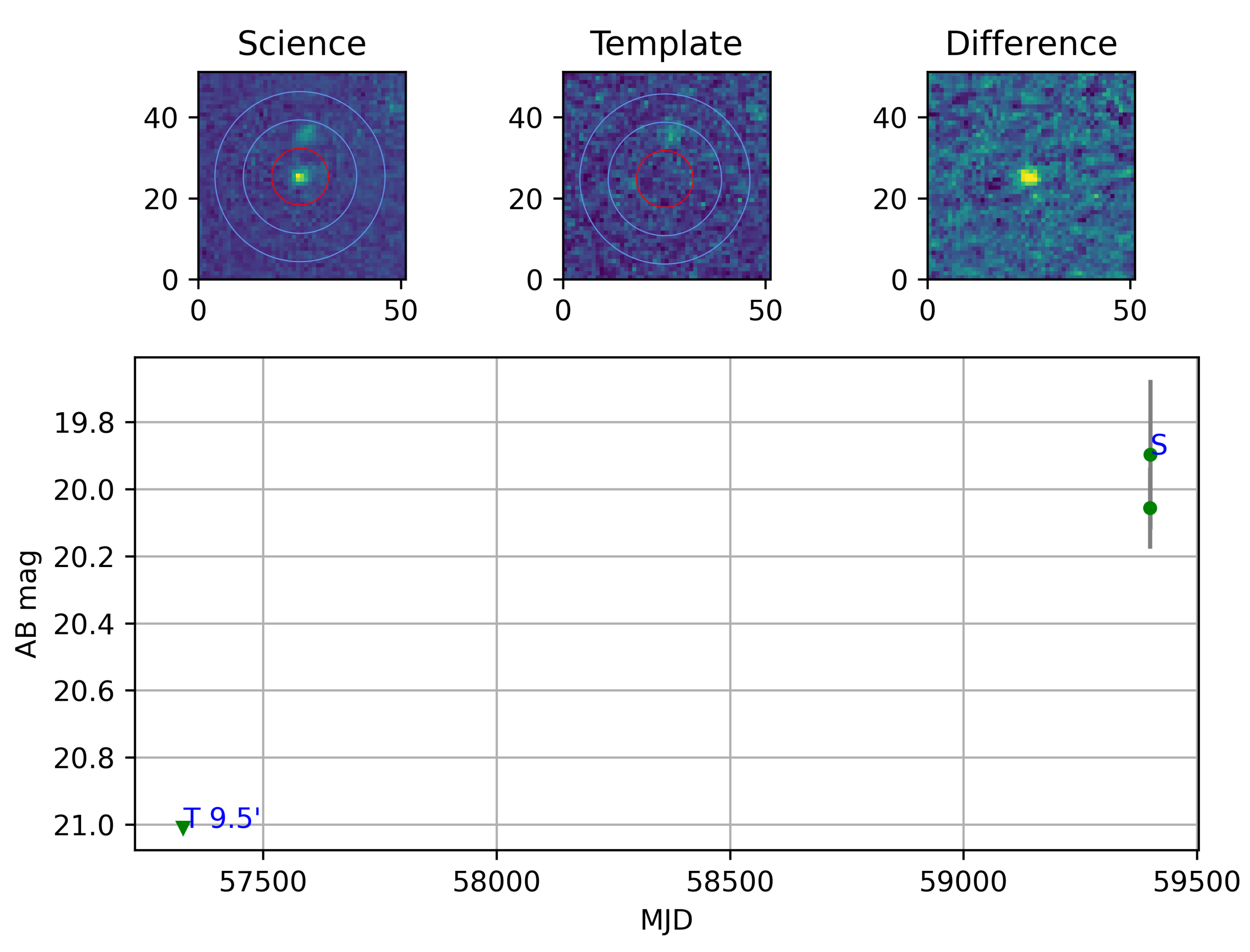}
    \caption{UVW2 light curve and science, template, and difference image stamps of a transient detected using \texttt{TUVOpipe} at RA= 00:43:05.81, Dec=+41:15:47.5. This source was identified as a known nova in M31 (AT2021pap; \citealp{AT2021pap}), and it was not the target of the \textit{Swift} observations.} 
    \label{fig:known_nova_m31}
\end{figure}

Using \texttt{TUVOpipe,} we recover known transients daily, and in particular sources that are the target of the \textit{Swift} observations and are undergoing known outbursts (UVOT is used extensively to follow up on transients discovered both by \textit{Swift}'s XRT and BAT and by other facilities), but also serendipitous detections of outbursts that happened to be in the FoV of a UVOT observation but are being monitored by other facilities. These sources are usually not of prime interest since they are not new TUVO discoveries. However, aside from demonstrating the effectiveness of the pipeline, the recovery of these known transients is still very useful, because they may have exhibited past UV outbursts (or any other potentially interesting UV behaviour) that were observed by UVOT but have not yet been studied or reported previously. \texttt{TUVOpipe} therefore still provides valuable information in these cases. Here, we describe some examples of different types of such known transients detected by \texttt{TUVOpipe} (see Figs.~\ref{fig:piv_maxi} to \ref{fig:known_nova_m31}). For additional examples, we suggest consulting Wijnands et al. (in prep.).\

\textit{\emph{MAXIJ1820+070}} is a black hole XRB source, which has been monitored extensively by \textit{Swift} in recent years (see \citealp{Sai_2021} for an optical and UV monitoring study of this source using only a part of the UVOT data currently available). It was the target of the observations in which our pipeline detected it as a transient. In Fig. \ref{fig:piv_maxi}, we show the stamps and light curve from the \texttt{TUVOpipe} detection, and in Fig. \ref{fig:PV_maxij1820} we show the long-term PV light curve in optical and UV bands. The primary outburst and several subsequent reflares are clearly visible in the PV light curve, most of which have so far not been studied (\citealp{Sai_2021} described only the main outburst and the first reflare). Similar reflaring behaviour has been observed for other neutron star and black hole X-ray transients (e.g. in the black hole XRB MAXIJ1535-571, see \citealp{Russell_2019}; or in the neutron star system SAX J1808.4-658, see \citealp{Wijnands_2001} and \citealp{Campana_2008}). The physical processes responsible for these outbursts and the subsequent reflares are still poorly understood (for an in-depth discussion about the possible mechanisms responsible for these reflares and an overview of systems that show this kind of behaviour, including accreting white dwarfs, see \citealp{Patruno_2016}). In addition, only a limited number of systems in these reflaring states have been observed in the optical and UV. Therefore, our extended optical/UV light curve of this source will be helpful in further constraining the mechanisms responsible for these reflares (although this is beyond the scope of our paper; see \citealp{Patruno_2016} for an example of the benefits of obtaining quasi-simultaneous UV, optical, and X-ray data for these types of sources).\

\textit{} The prototype BL Lac object, BL Lacertae, has been heavily monitored by \textit{Swift} over the past, $\sim$13, years. Due to its highly variable and flaring properties, it is regularly picked up by our pipeline whenever new observations are obtained (see Fig. \ref{fig:bllac_piv} for the PIII light curve of one such \texttt{TUVOpipe} detection in the UVM2 filter and for observations in which BL Lac was the target). To investigate the long-term variability of this system across the UV-optical range, we also created a PV light curve, which is displayed in Fig. \ref{fig:bllac_pv}. From the PV light curve, it can be seen that this source was observed extensively in all six UV and optical filters, with observations obtained up to several times per day. The long-term light curve thus shows significant variability behaviour both on long timescales (e.g. the rise over the last few years resulting in the brightest known state of this source in the optical; see also \citealp{Marchini_2021}) and on short (hours to days) timescales (e.g. the variability shown in the inset axes in Fig. \ref{fig:bllac_pv}). Long-term, quasi-simultaneous, multi-wavelength photometry can help to elucidate some of the still unanswered questions about BL Lac and similar objects. For example, in a monitoring study of this object, \citet{Stalin_2005} find that when examining short-timescale (<1 day) variability, the source was seen to become bluer when it brightened. This effect was not significantly detected when looking at longer timescales (days to months). A tentative comparison of the colours (we examined UVW2-V) of BL Lac using the UVOT data averaged over one day indeed shows an increase in blueness when the source intensity increases. However, we notice a similar effect for longer term brightenings (e.g. weeks to months). A complete, detailed understanding of the physical mechanisms causing the variability in BL Lac (and other blazars) at all timescales may therefore require the characterisation of this property, though a detailed study of this source is beyond the aim of this paper.\

Due to the rate of occurrence and typical brightness of novae in M31, and the relatively frequent \textit{Swift} observations of this galaxy, \texttt{TUVOpipe} regularly detects novae in M31 serendipitously. The observations are sometimes specifically targeting active novae in M31 for follow-up, but additional novae active at the same time and which are located within the UVOT FoV are also often detected by our pipeline. In Fig. \ref{fig:known_nova_m31}, we show an example of one of these detections. Although there have been several studies of novae in M31 in the optical (see e.g. \citealp{Shafter_2011} and \citealp{Cao_2012}), they have not been studied extensively in the UV. Studying nearby (i.e. Galactic) novae may provide better constraints on the outburst physics than more distant objects; nonetheless, the advantage of M31 is that we can find a relatively large number of novae at a known distance. This can help to perform studies of the general properties of the UV emission of novae. To study UV light curves of a sample of novae in detail (i.e. with more frequent observations than the typical non-uniform monitoring of M31 with \textit{Swift}), we obtained additional UVOT observations for five of the novae we detected, every two weeks \citep{ATEL_novae}. We also refer the reader to Wijnands et al. (in prep.) for a description of archival UVOT studies of novae in M31 in the TUVO project.\

\subsubsection{Variable stars} \label{results_ql_vars}

Due to their prevalence in the Galaxy, the majority of transients detected by \texttt{TUVOpipe} are variable stars of many different classes (including RR Lyrae, Cepheids, Mira's and other long-period variables, delta-Scuti variables, and eclipsing binaries). Most of these are previously known sources (as determined from catalogue matches in PIV; see Sect. \ref{piv_queries}), though some sources exhibit clearly variable behaviour as seen in their \texttt{TUVOpipe} light curves, but they do not have any known classification as variable sources. For many classes of variable stars, the UV emission is not well understood, and UV observations may help to explain the underlying physics of the variability (see e.g. \citealp{Sagiv_2014} for an overview of the scientific potential of UV studies of several types of variable stars; see also \citealp{Siegel_2015} for a UV study of RR Lyrae in globular clusters). Here, we briefly discuss some examples of periodic variables that were detected by our pipeline. For additional examples of some irregular variables we find, we suggest Wijnands et al. (in prep.). \

\begin{figure}
    \centering
    \includegraphics[width=.45\textwidth]{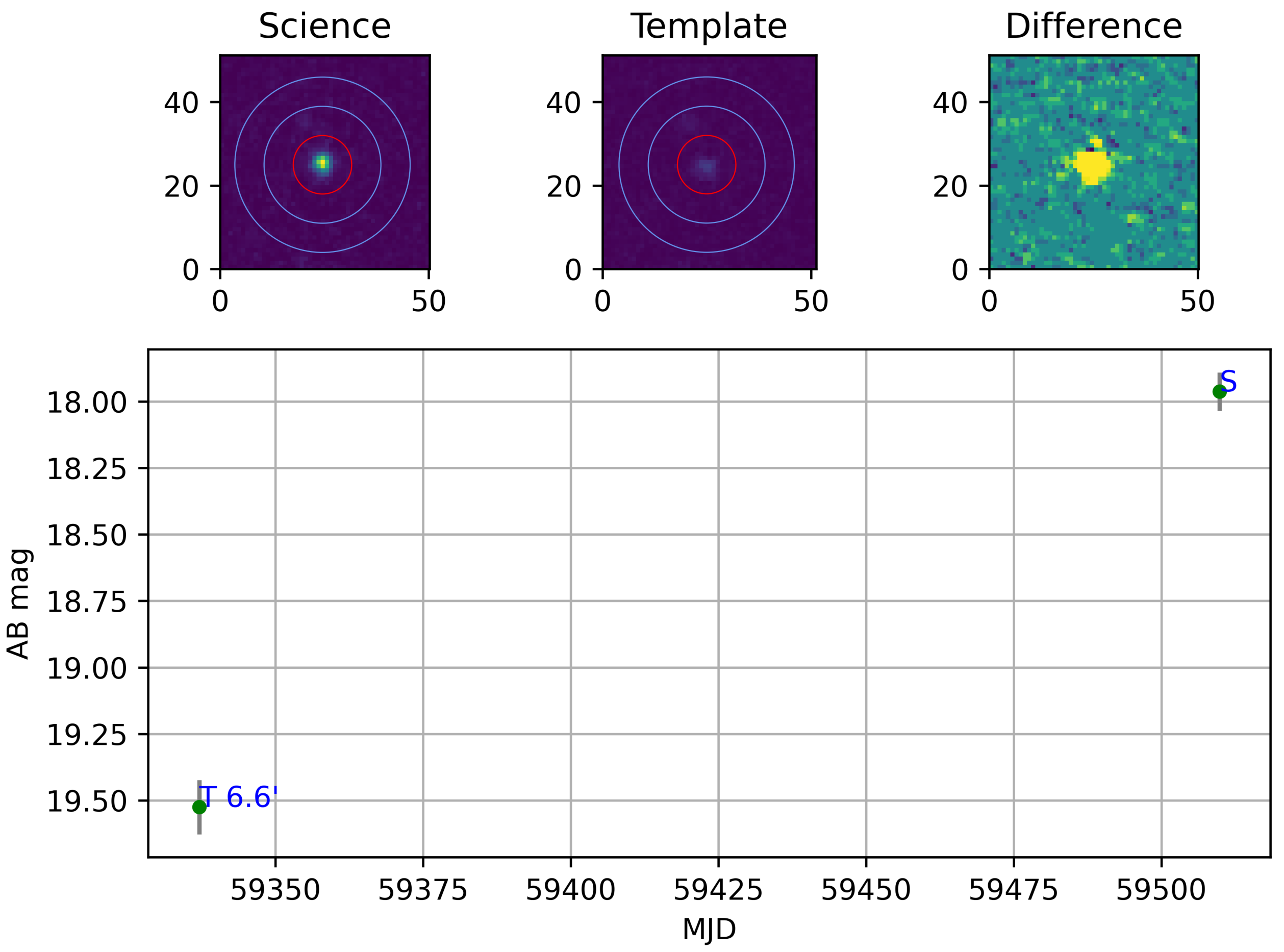}
    \caption{UVW2 light curve and the different image stamps of the transient discovered by \texttt{TUVOpipe} at RA=18:37:35.25, Dec=-33:49:59.6. This source could be securely associated with the known RR Lyrae variable ASASSN-V J183735.24-334960.0, and was not the target of the \textit{Swift} observations.}
    \label{fig:variable_rrl}
\end{figure}

A particularly interesting case regards the class of pulsating variables known as RR Lyrae stars (see \citealp{Modiano_2020} for a discussion of an RR Lyrae star we detected with \texttt{TUVOpipe} in the globular cluster 47 Tuc), for which the strong UV variability can be up to several magnitudes \citep{Siegel_2015}. This is much larger than the amplitude of their pulsations in the optical, which is typically <2 mags (see e.g. \citealp{Horace_2011, Hoffman_2021}). Near-UV (NUV) light curves are highly sensitive to underlying atmospheric models, temperatures, and surface gravities \citep{Siegel_2015}, so extensive UV observations of RR Lyrae can provide much better constraints on these parameters compared to what is possible using only optical bands (in particular when complementing the NUV with optical data; see \citealp{Siegel_2015}). Our lack of a detailed, complete understanding of the pulsations in RR Lyrae is in part due to a deficiency of UV observations. The frequency of RR Lyrae we detect with \texttt{TUVOpipe} indicates that the UVOT archive likely contains a wealth of long-term UV data of these (and other) variable stars, therefore making it a valuable data set for the study of RR Lyrae. In Fig. \ref{fig:variable_rrl}, we show an example of the known RR Lyrae star ASASSN-V J183735.24-334960.0, recovered serendipitously by \texttt{TUVOpipe} (i.e. it was not a target of the \textit{Swift} observations). We note the amplitude of variability of this RR Lyrae exhibited in the UVW2 filter is of at least 1.5 magnitudes, while the \textit{ASAS-SN} light curve\footnote{\url{https://asas-sn.osu.edu/variables/6687137d-360f-57ed-b386-dfafc94bc581}} shows a V-band amplitude of only 0.4 magnitudes.\

\begin{figure}
    \centering
    \includegraphics[width=.45\textwidth]{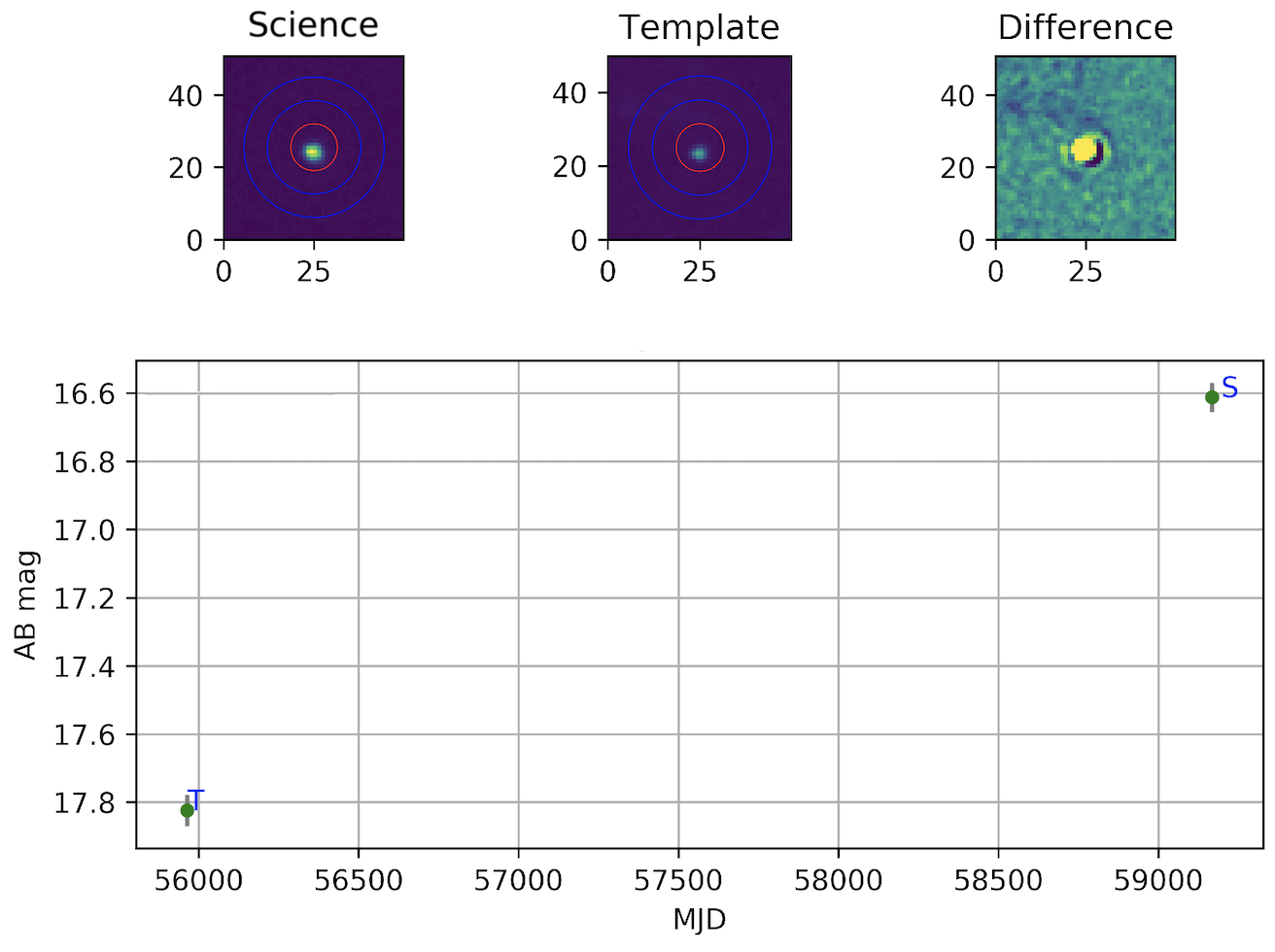}
    \caption{U-band light curve and different image stamps of a transient of a transient triggered by \texttt{TUVOpipe} at RA=06:32:47.14, Dec=+05:41:02.5. We identified the source as the known variable source AT J098.1962+05.6837 and confirmed its nature by creating a long-term light curve with PV (see Fig. \ref{fig:pv_eclipse} and \ref{fig:folded_eclipse}). It was not the target of the \textit{Swift} observations.}
    \label{fig:piii_eclipse}
\end{figure}

\begin{figure}
    \centering
    \includegraphics[width=.45\textwidth]{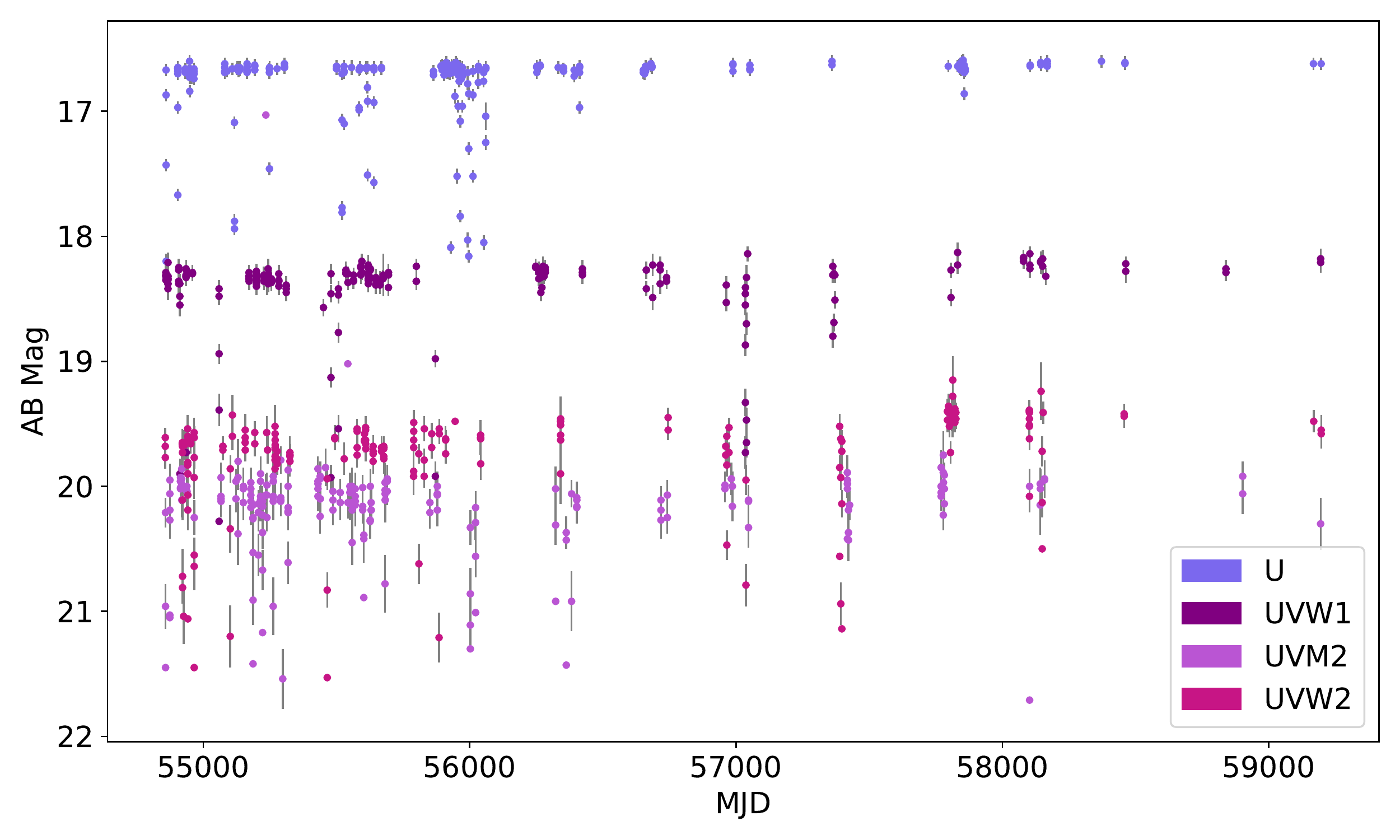}
    \caption{Long-term, multi-band PV light curve (created using all available UVOT data) of the variable source AT J098.1962+05.6837 (see Fig. \ref{fig:piii_eclipse} for the original \texttt{TUVOpipe} detection).}
    \label{fig:pv_eclipse}
\end{figure}

\begin{figure}
    \centering
    \includegraphics[width=.45\textwidth]{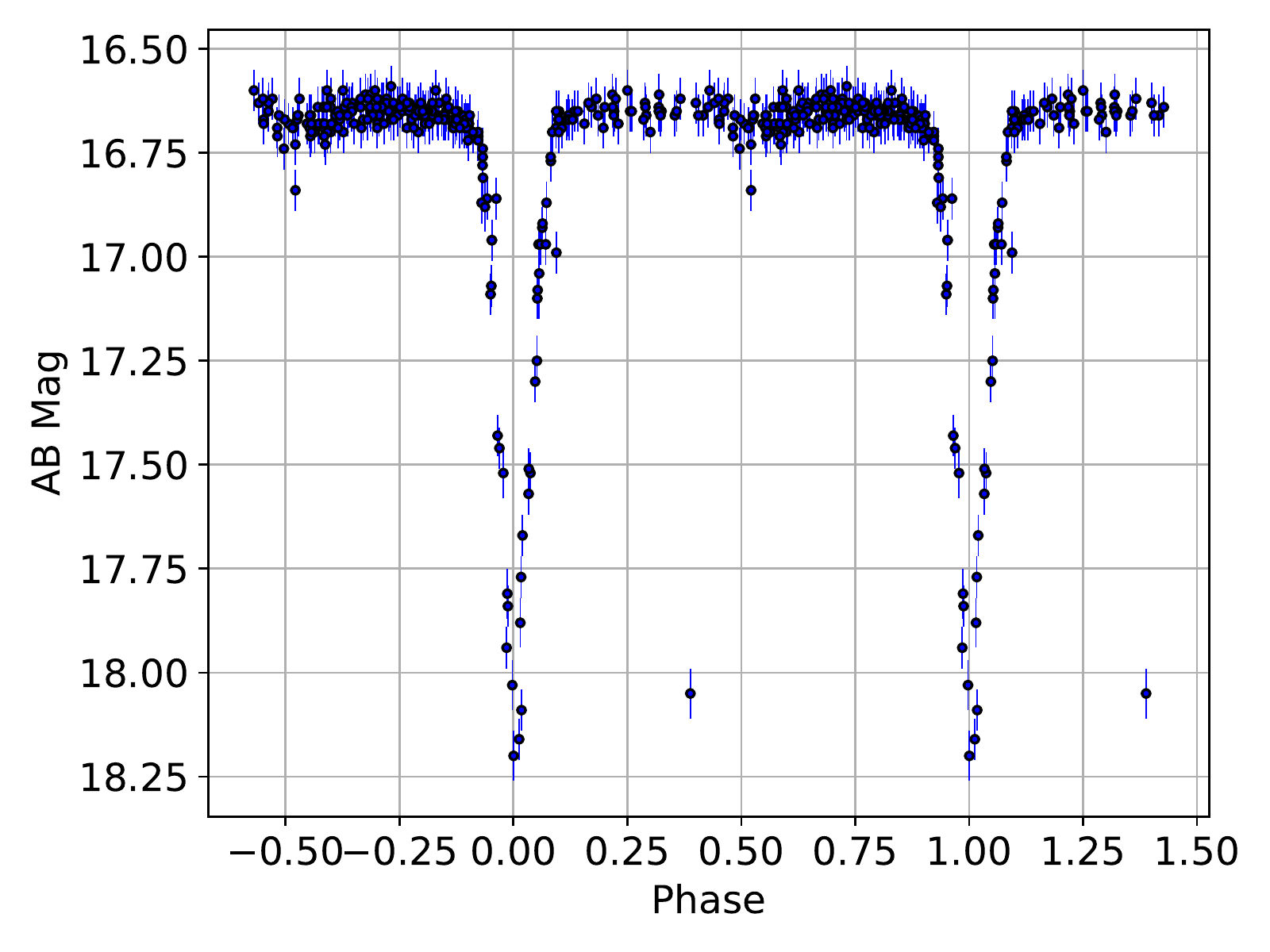}
    \caption{Light curve in the U filter of the eclipsing binary shown in Figs. \ref{fig:piii_eclipse} and \ref{fig:pv_eclipse} (AT J098.1962+05.6837), folded with a period of 1.9899 days and repeated to show two entire cycles. The times have been offset such that the deepest point occurs at phases 0 and 1.}
    \label{fig:folded_eclipse}
\end{figure}

\interfootnotelinepenalty=10000
With the real-time pipeline, we also frequently detect eclipsing binaries, due to their significant changes in brightness. Often, when we trigger these sources, PV light curves reveal potentially eclipsing behaviour. In these cases, we can attempt to verify the nature of the variable source by creating Lomb-Scargle periodograms for the light curves\footnote{We note that in these cases, to obtain the best possible period we apply barycentric corrections to the data before determining the period, using tools available within \texttt{astropy}. By default, we do not automatically barycentre-correct the data in \texttt{TUVOpipe}; we only do this when we find a variable for which we are interested in accurately measuring its period.} and inspecting, by eye, the folded light curves that we create using the best periods found. We note that this is done independently of the rest of pipeline; it is not part of the automatic software. An example of an eclipsing binary we found using \texttt{TUVOpipe} is shown in Fig \ref{fig:piii_eclipse}. The strong variability (by over 1 optical magnitude) led us to create a long-term light curve; the PV light curves of all the used UVOT filters are shown in Fig. \ref{fig:pv_eclipse}. The source is a known source that appears in many sky survey catalogues, including Gaia, where its G-band magnitude is listed as 14.1 and its inferred temperature is $\sim$5000K (see \citealp{Andrae_2018} for how the listed temperature we extract from the Gaia catalogue is inferred from the Gaia data). This G-band magnitude is brighter than the U-UV bands seen in the UVOT light curve, indicating (along with the temperature) that it is a relatively red source. The source is also present in the ATLAS catalogue of variable stars \citet{Heinze_2018}, where it is denoted as AT J098.1962+05.6837. The ATLAS catalogue gives a period of 1.985 days. However, no classification of the source was so far known, so the cause of the found periodicity is not clear. Our method gave a best period of 1.9899 days, so we used this period to fold the UVOT U-band light curve, which is shown in Fig. \ref{fig:folded_eclipse}. The light curve clearly reveals an eclipse in which the source becomes dimmer by around 1.5 magnitudes, and the shape of the light curve confirms it as an eclipsing binary. There may also be additional very weak minima seen halfway between the deep eclipses (see around phase -0.5 and 0.5 in the plot shown). The fact that any potential secondary minima are very weak may suggest that this is an Algol-type system (these are known as EA binaries; see \citealp{Carmo_2020} for a review of these systems and example light curves; the inferred Gaia temperature is also consistent with this being an EA system).\

We also note that the proximity of the found period to exactly two days may suggest an artificial effect due to the Earth's rotation. However, the period is consistent within data obtained from Earth (ATLAS) and from a satellite (\textit{Swift}), the latter of which in itself is unlikely to introduce artefacts in the periodicity due to Earth's rotation (moreover, we have not found the same period in any other source, suggesting this is not a systematic effect in the UVOT data). Additionally, the shape of the folded light curve clearly resembles that of an eclipsing binary. Therefore, we are relatively confident that the found period is real and the classification as an eclipsing binary is correct.

\subsection{Initial results from archival searches} \label{results_archival}

As discussed in Sect. \ref{tuvopipe}, \texttt{TUVOpipe} can also be run in archival mode, whereby all archival UVOT data of any user-identified field is searched in order to look for historical transients. Archival searches are not of the highest priority for the TUVO project, since transients discovered in archival data are usually no longer active at the time of discovery and therefore cannot be investigated further with follow-up observations. However, these transients can still be highly interesting, so the archival mode of \texttt{TUVOpipe} is also of prime importance for the TUVO project.\ 

We determined fields of interest based both on fields containing what we call `confined stellar populations' (fields that harbour many stars or galaxies that fit within one or few UVOT FoVs, such as globular clusters and nearby galaxies; see Wijnands et al. (in prep.) and on the number of UVOT observations. In both cases, the selection of the fields has the aim of maximising the potential for the discovery of historical transients. Similarly to the real-time mode, the vast majority of transients we discover in the archival mode are known variable stars; however, using this mode we also discovered several previously unknown transient sources. Here, we briefly discuss initial results from two types of fields we selected on which to run the archival pipeline, where we targeted fields that have been observed many times by \textit{Swift}. For results regarding archival pipeline runs on confined stellar populations, we direct the reader to \citet{Modiano_2020} and Wijnands et al. (in prep.).\

\begin{figure}
    \centering
    \includegraphics[width=.45\textwidth]{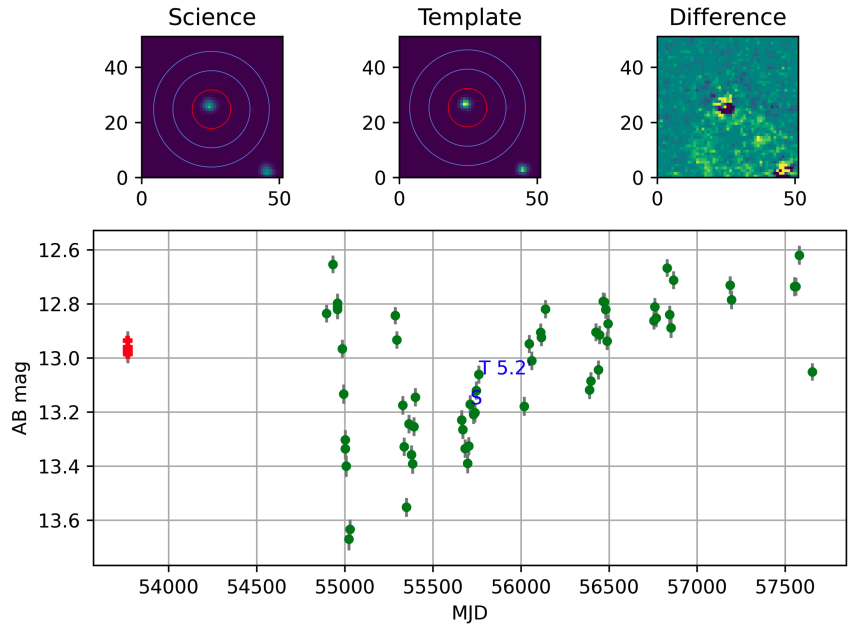}
    \caption{V-band light curve and the different image stamps of an archival transient discovered using \texttt{TUVOpipe} at RA=16:25:29.82, Dec=-29:51:35.6. This source was identified as the long-period variable candidate ASASSN-V J162529.87-295134.3, and it was recovered serendipitously (i.e. it was not the target of the \textit{Swift} observations) with \texttt{TUVOpipe} running in archival mode in the field of the blazar PKS 1622-297. The red points indicate data points obtained on images that may not have been successfully aligned in the pipeline (see Sect. \ref{pii_dataprep}). We note that the diffuse structure seen in the difference image stamp is likely the result of artefacts in the science image caused by very bright stars in the FoV.}
    \label{fig:archival_transient_1}
\end{figure}

Some fields that are very frequently observed with \textit{Swift} are part of monitoring programmes designed to regularly observe flaring sources detected by Fermi-Lat (mostly blazars although also several Galactic transients), denoted as `sources of interest'\footnote{\url{https://www.swift.psu.edu/monitoring/}} \citep{Stroh_2013}. \textit{Swift} observations of each of these fields amount to between a few tens and several hundred kiloseconds of exposure time. We ran the archival pipeline on all 23 fields in the original list\footnote{Besides the 23 primary fields in the list of Fermi sources of interest, there is also a secondary list of additional sources, which also have extensive \textit{Swift} monitoring. We continue to run the archival mode of \texttt{TUVOpipe} on the large number of these additional fields.}, and we mostly detected known variable stars, typically a few tens in each field (see Fig. \ref{fig:archival_transient_1} for an example of a long-period variable we detected with an archival \texttt{TUVOpipe} run on the field of the blazar PKS 1622-297). \ 

\begin{figure}
    \centering
    \includegraphics[width=.45\textwidth]{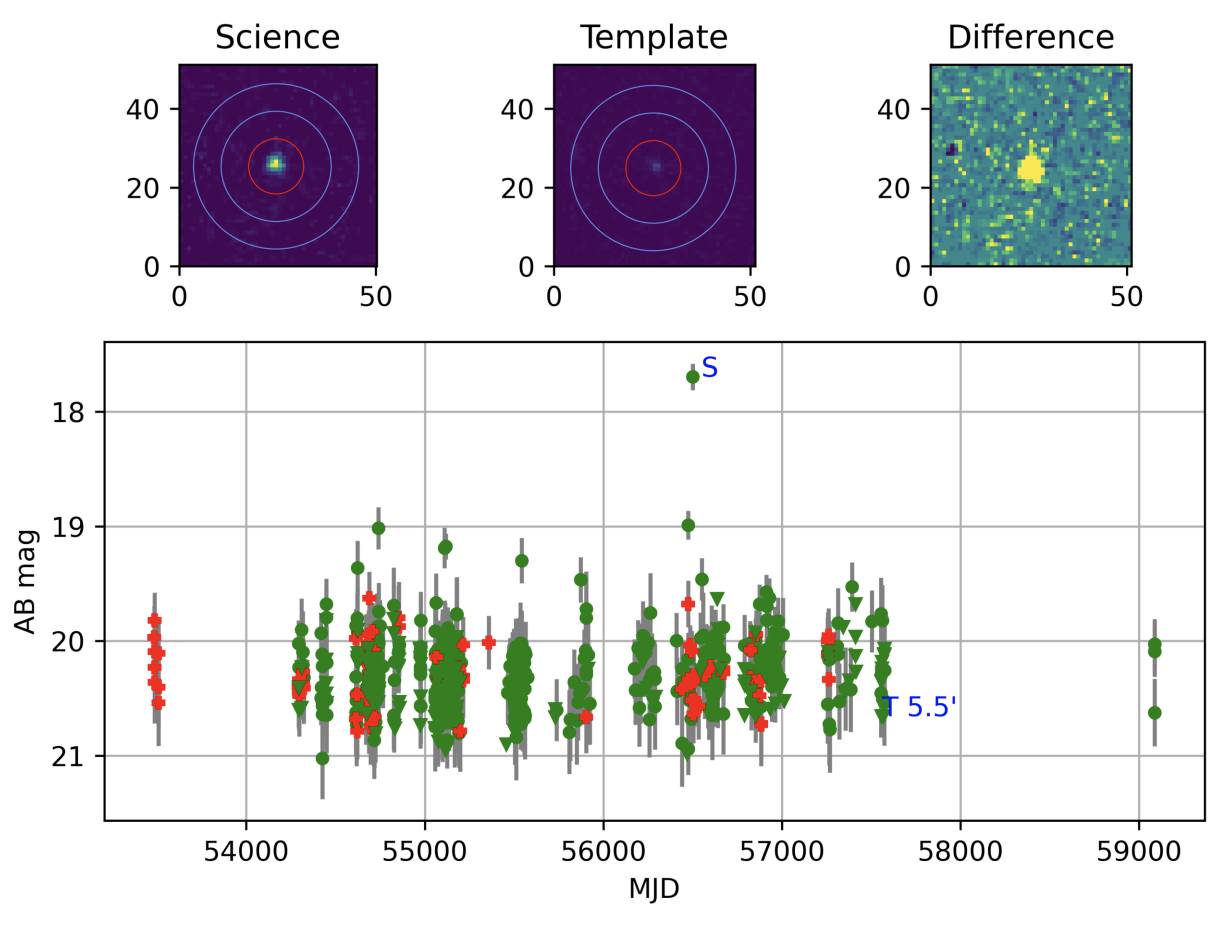}
    \caption{Archival UVW1 PIII light curve of the M-dwarf Sp2254+1606 detected with \texttt{TUVOpipe} at RA=22:54:11.09, Dec=+16:06:55.5. The light curve shows the source possibly exhibiting several flares. The target of the observations was the known blazar 3C 454.3; this M-dwarf was detected with \texttt{TUVOpipe} serendipitously in the archival mode. The red points indicate data points obtained on images that may not have been successfully aligned in the pipeline (see Sect. \ref{pii_dataprep}). See Fig. \ref{fig:flare_pv} for the PV light curves in all filters for this source.}
    \label{fig:flare_piv}
\end{figure}

\begin{figure*}[t]
    \centering
    \includegraphics[width=.95\textwidth]{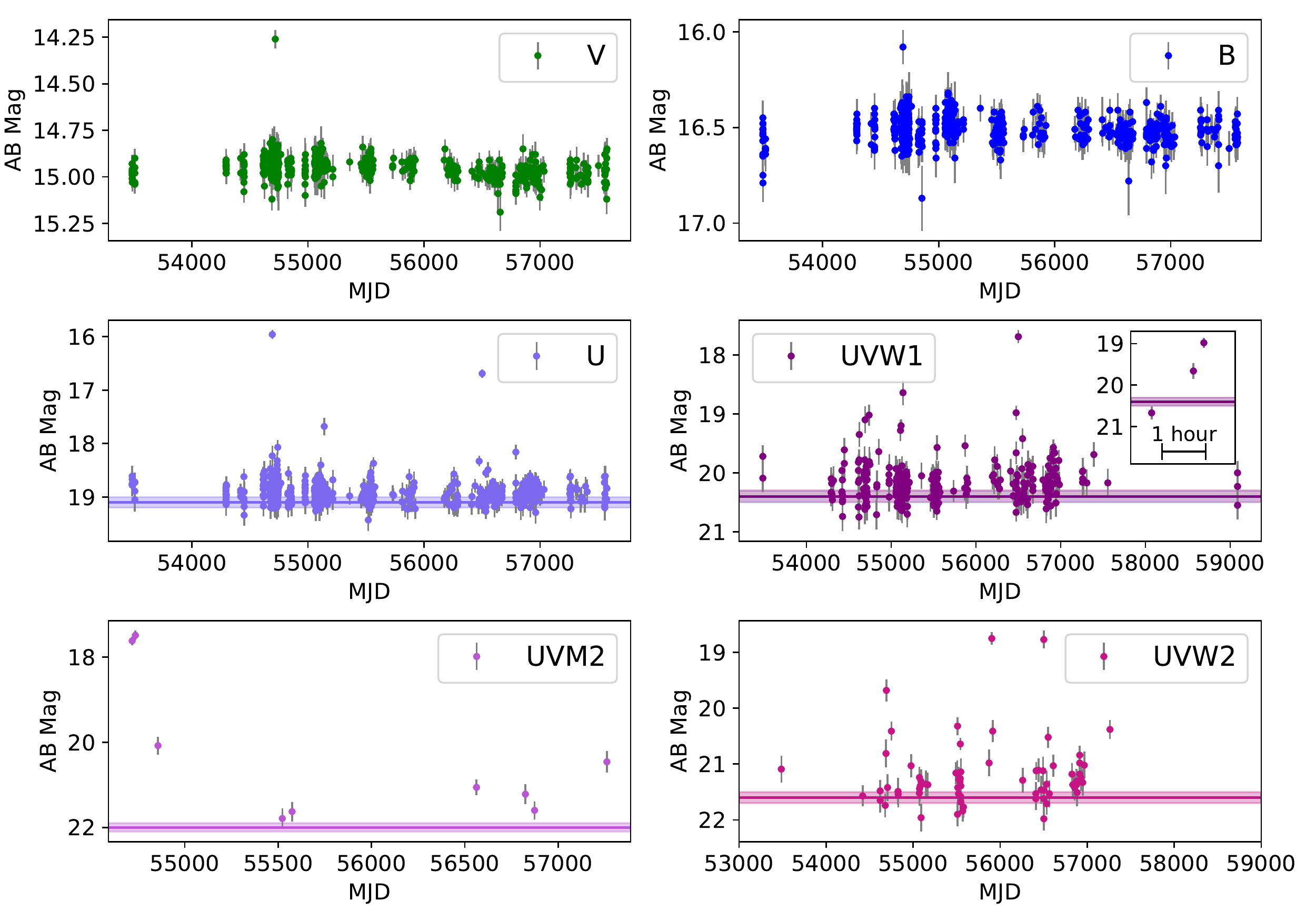}
    \caption{PV light curves, in all available filters created, of the M-dwarf and possible flaring star Sp2254+1606, detected serendipitously with \texttt{TUVOpipe} running in archival mode. For all filters, there were similar numbers of observations, but for figure clarity, upper limits derived from images in which the source was not detected are omitted. Therefore, the lower numbers of displayed data points in the UV light curves compared with the optical light curves simply indicate that the persistent source was not detected in many individual UV images. The horizontal lines represent the brightness (and the shaded regions the corresponding error) of the source detected in the stacked images, which were created by combining all individual images in which there was no detection (this value is therefore not shown for V and B filters, because the source was always detected in those filters). The inset plot in the UVW1 light curve shows a zoomed-in view of a light curve of the only flare for which there were multiple detections in the same filter, showing the timescale of the event more clearly. See Fig. \ref{fig:flare_piv} for the original (PIII) \texttt{TUVOpipe} detection.}
    \label{fig:flare_pv}
\end{figure*}

In this search, when processing the field of the blazar 3C 454.3 in the archival mode, we detected a possible M-dwarf flare shown in Fig. \ref{fig:flare_piv} (UVW1 PIII light curve with image stamps) and Fig. \ref{fig:flare_pv} (PV light curves in six filters). The PV light curves show that this is a very red source (as seen by the quiescent brightness across the optical-UV range) that undergoes occasional brightenings in all the UVOT bands.\

Information extracted in PIV showed that matches to this source appeared in several M-dwarf catalogues (e.g. SPECULOOS, \citealp{Sebastian_2021}, where this source is denoted as Sp2254+1606). In addition, using the PIV information obtained for match in the Gaia catalogue, we noticed the very red colour of the source (its Gaia magnitude in the G band is 13.6, compared with $\sim$20.5 in UVW1) and its inferred Gaia temperature of 4200K. These characteristics confirm it as a cool, red source, consistent with an M-dwarf. These stars are known to exhibit flares that can be highly energetic, brightening by several magnitudes in the optical and UV (see e.g. \citealp{Welsh_2007}; also note the flare spectrum may peak in the UV, see \citealp{Paudel_2021}), and lasting from seconds to hours. Relatively few photometric studies of M-dwarf flares have been performed in the UV (for an example, see \citealt{Welsh_2007}). \

Interestingly, despite similar numbers of data points between all filters, fewer flares are detected in the optical bands compared with the UV filters (e.g. only one flare each in V and B, while at least a few to several events were seen in U, UVW1, UVM2, and UVW2). Additionally, the increase in brightness for the flares detected in V and B bands is relatively small ($\sim$0.75 and $\sim$0.5 in V and B, respectively) compared with brightenings of up to several magnitudes in some of the U-band and UV flares (for example, the earliest flare seen in the UVM2 causes an increase in brightness by 4 magnitudes). These characteristics might result from a combination of flares intrinsically causing stronger brightenings in the UV than in the optical, along with the fact that the UV emission from M-dwarfs in quiescence is very low compared to the optical. The latter point may suggest that while in the optical only the strongest flares are bright enough to be detected above the optical emission of the star, the lower stellar emission in the UV allows for even weaker flare events to be detected, hence the more numerous detections in the UV filters. This may therefore make the case that the UV is the optimal band in which to search for flares.\

For only one of the detected flares, UVOT data in multiple filters had been obtained. This event (at MJD $\sim$57000) was seen in the B, U, UVW1, and UVW2 filters (the V-band flare seen around this time does not correspond to the same event; it occurs a few weeks later). The optical and UV (quasi-)simultaneous data may indicate that the flare causes a more significant brightening in the U and UV bands than in the optical, with the peak brightening observed in the U band: we measure the amplitude with respect to the quiescent level of around 0.43, 3.1, 1.3, and 1.9 magnitudes in the B, U, UVW1, and UVW2 filters, respectively. However, we note the caveat that the images were taken a few minutes apart, and since flare timescales can range from seconds to hours, the differences in the brightenings observed between filters may also depend in part on the stage of the outburst that happened to be observed with each filter. \

The flares we detected have relatively short durations, based on the cadence of the \textit{Swift} observations. For some of the events, the data points in quiescence immediately prior and subsequent to flares are separated from the flare by down to around 1.5 hours, which gives an upper limit for the flare duration (for other events this upper limit is much weaker, up to several days). Additionally, one of the UVW1 brightenings does have coverage during a possible rise stage of the flare event (see the inset axes in Fig. \ref{fig:flare_pv}). This shows a rise from 20.94 $\pm$ 0.23 to 18.99 $\pm$ 0.13 magnitudes in just over an hour. Overall, the constraints we obtained on both the timescales and outburst amplitudes of the observed brightenings are consistent with the known properties of flare stars, so we suggest that this source is very likely a flare star.\

Due to a few past and ongoing surveys targeting fields in the Galactic centre with \textit{Swift} \citep{Degenaar_2010,Degenaar_2013,Degenaar_2015}, there are several thousand archival UVOT images available covering the innermost region around the Galactic centre. We are currently undertaking two projects in which we make use of this large data set. However, due to the very long times associated with processing all the data, they are currently works in progress, which will be described in detail in separate publications. Here, we briefly describe our plans and preliminary results.\ 

\begin{figure}
    \centering
    \includegraphics[width=.45\textwidth]{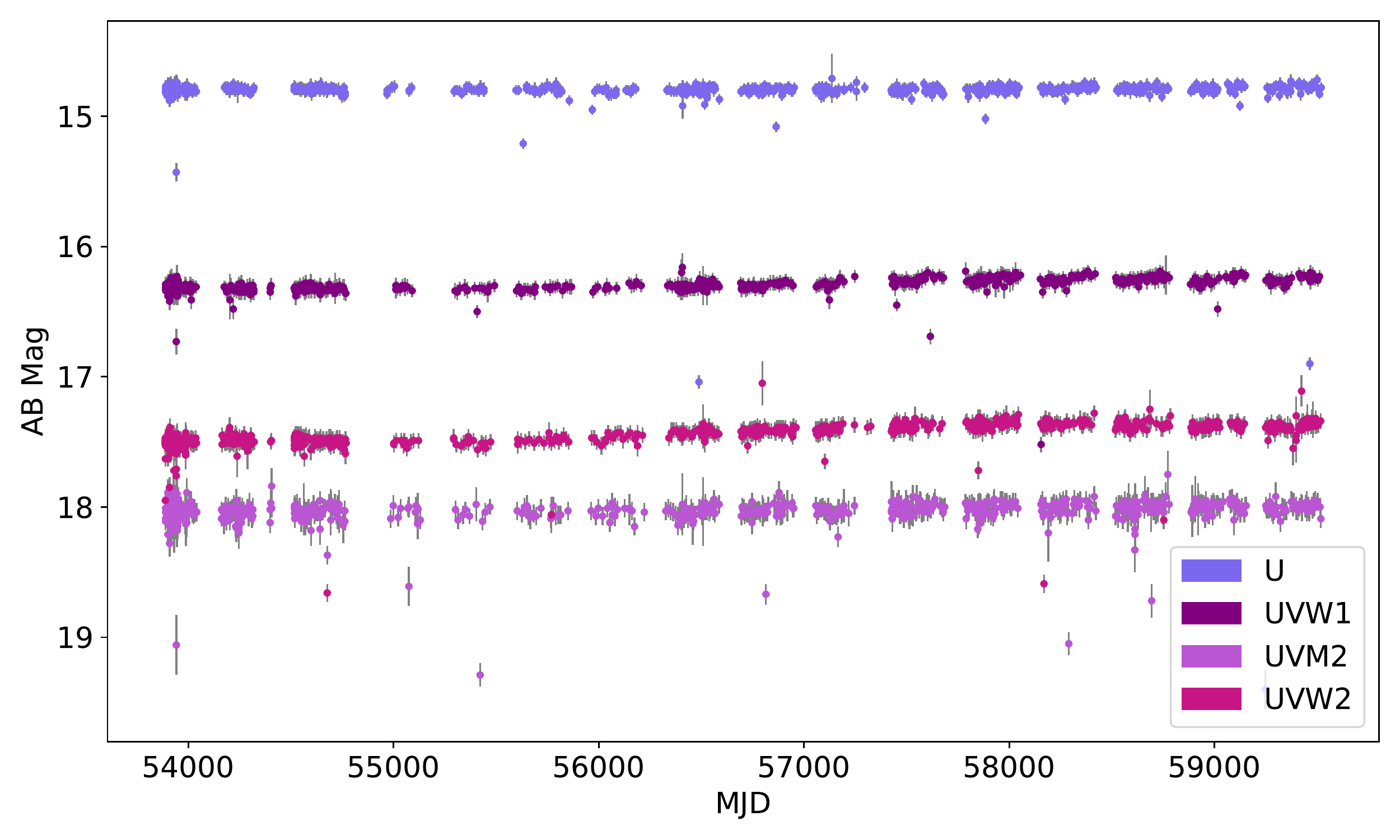}
    \caption{Long-term light curve created with PV of a persistent source in the Galactic centre, at RA=17:45:44.00, Dec=-29:07:01.1 in the U, UVW1, UVM2, and UVW2 filters (we do not show the V and B data points here because only a handful of observations of this field were performed with those filters). The few data points straying from the stable levels are likely the result of artefacts in the science images or imperfect alignments.}
    \label{fig:galcen_pv}
\end{figure}

One of the projects consists of examining persistent sources. Using PV, we can create long-term light curves of all the persistent sources that are detected in these fields (i.e. by using one of the images as a reference image in which to detect the sources). This method does not allow for the discovery of transients with no quiescent counterparts detected in the reference image. However, the PV light curves could still reveal interesting variable behaviour of the persistent sources, since we can use them to characterise the long-term UV behaviour of many sources with relatively high sampling cadences and spanning over a decade. As a preliminary study, we focused on one Galactic centre field (around the inner 12'), creating PV light curves of all the 459 sources detected at 5-$\sigma$ in one U-band image. The selected image for source detection was chosen based on a relatively high exposure time of a few hundred seconds as well as inspecting it by eye to ensure there were no major artefacts in the image (ObsID 00095660216; extension 1). In this preliminary search, we did not find any significantly variable sources (see Fig. \ref{fig:galcen_pv} for an example of one of the PV light curves). \ 

\begin{figure}
    \centering
    \includegraphics[width=.45\textwidth]{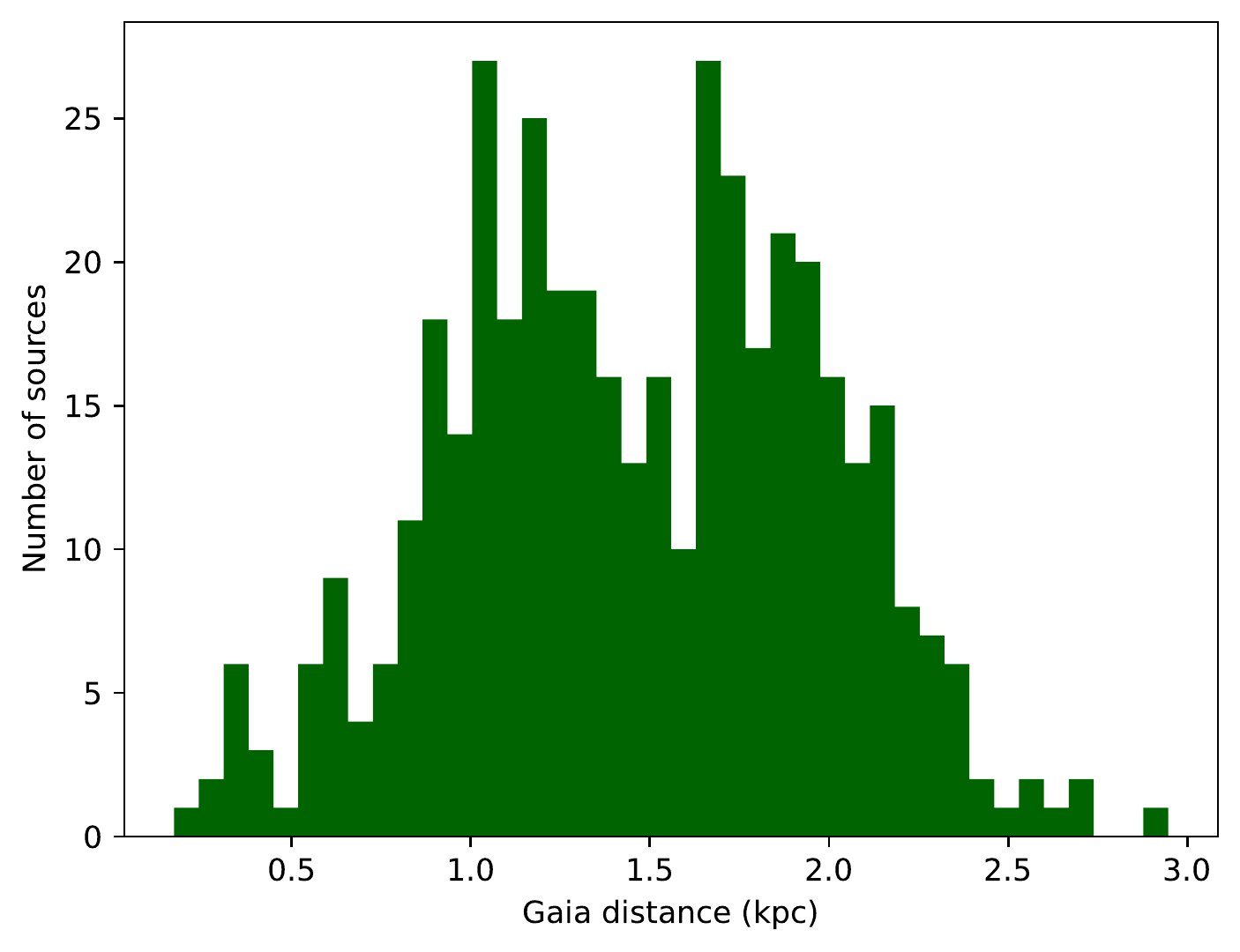}
    \caption{Distribution of distances for sources detected at 5-$\sigma$ in the U-band in a Galactic centre field. Distances were obtained by querying the Gaia catalogue for counterparts to the UVOT sources we detect (and retaining only those with at least a 5.0 significance in the parallax, to ensure the distances we retain are reliable). The data is grouped in 40 equally sized bins.}
    \label{fig:gc_distances}
\end{figure}

To determine the distance towards the Galactic centre, which we are probing with this study, we queried the Gaia catalogue for counterparts to all the sources. For each source, if there was match in the Gaia catalogue (see Sect. \ref{piv_queries} for how we determine if a catalogue source is a match with a source we detect) in which the error on the parallax was small (significance of at least 5.0), we retained the associated distance measurement from Gaia. We created a histogram of the results in order to examine the distribution of distances of the sources that the UVOT can detect towards the Galactic centre (shown in Fig. \ref{fig:gc_distances}). From the distribution, we can see that when looking towards the Galactic centre with the UVOT at 5-$\sigma$ in the U-band, we can see up to only around 2.5 kpc. We also notice a clear bimodality in the distribution, with peaks around 1.1 and 1.8 kpc.\ 

We emphasise that this is still a preliminary study. We are currently expanding it to sources detected in all filters (i.e. not just the U-band), down to 3-$\sigma$, and to a larger region around the Galactic centre; this may affect the preliminary results mentioned here. In the second planned project, we will run \texttt{TUVOpipe} in the archival mode to look for transients in all the UVOT observations available of these Galactic centre fields. It is also worth noting that the long-term light curves we already created show exceptionally stable photometry over the course of the observations (spanning over a decade), indicating the excellent long-term calibration of the UVOT (see Fig. \ref{fig:galcen_pv}).

\section{Summary and concluding remarks} \label{conclusion}

In this paper we introduce \texttt{TUVOpipe}, a specialised pipeline we constructed in order to automatically discover serendipitous transients in the UV wavelength range in data obtained with UVOT aboard \textit{Swift}. UVOT is the primary focus of the TUVO project (see Wijnands et al., in prep.), a recently launched investigation in which we aim to explore transients in the relatively understudied UV regime. Every day, the pipeline is run to search for variability in all recent ($<$1 day old) images obtained with UVOT. Between a few hours and one day after the observations are performed, transient candidates are displayed to users. This includes, for each transient: light curves, image stamps, parameters derived from basic analysis of the light curves, and information regarding any known counterparts or previous classifications from astronomical catalogues. \

The first year of running the pipeline has yielded promising results. Having processed over 75\,000 images, a few tens of thousands of candidate UV transients have been detected serendipitously, currently at a rate of a few tens of real transients per day. Most of these are known variable sources or previously discovered transients, though a fraction of these represent sources with no clear previous identifications, indicating that \texttt{TUVOpipe} can indeed discover new transients in the UV. This `high-interest' category included the detection of a new outburst from a known candidate CV \citep{Verberne_2020}, the discovery of a strong UV outburst of a previously unknown CV (Modiano et al., in prep.), and the discovery of a new unidentified transient \citep{ATEL_Modiano}. The pipeline also successfully recovers many previously known transients such as DNe, XRB outbursts, novae, SNe, and AGN.\ 

We are considering further upgrades to \texttt{TUVOpipe}. One enhancement to be integrated in PV is a tool we are currently developing to run Lomb-Scargle methods on the PV light curves to attempt to automatically find periods. We also plan to develop this aspect of the pipeline further, with additional period-searching tools as well as further statistical methods to characterise variability in the light curves. We are also investigating the option of probing additional information from online databases: we will automatically collect light curves of the counterparts of our sources determined in PIV (e.g. from \textit{ZTF}, \textit{ASAS-SN}, \textit{Gaia}, \textit{ATLAS}) to display alongside our PIV and PV products. Finally, we are currently investigating the possibility of automatically querying XRT light curves\footnote{\url{https://www.swift.ac.uk/user_objects/API/pyDocs/README.md}} and images; this would allow us to immediately verify whether there is any X-ray detection at the position of our transients (currently this is done manually). Some of these further improvements may be implemented in a Part VI of \texttt{TUVOpipe}.\

We note that \texttt{TUVOpipe} is designed to detect UV transient sources that are of the highest interest to the TUVO project (relatively high amplitude transients; see Wijnands et al., in prep., for details), and is not intended to produce a comprehensive picture of all true variable and transient sources present in the UVOT data. The complexity of the data analysis, in particular the image subtraction and source detection stages, inevitably introduces imperfections. This means that \texttt{TUVOpipe} cannot securely detect all real variability exhibited by all sources detected using the UVOT, nor can it automatically remove all false positives. Therefore, we would like to stress that the pipeline is not constructed in order to be complete, so it cannot be used for comprehensive population and statistical studies of UV transients. Rather, it is a method used to maximise the potential for efficiently discovering as many interesting UV transients as possible in real-time and archival UVOT data.\ 

After discovering UV transients, the next, and crucial, stage for our TUVO project consists of ground- and space-based photometric and spectroscopic follow-up. These observations target the transients that we find highly interesting based on the output of PIV and PV (see Wijnands et al., in prep., for details on how we are currently pursuing this strategy with several facilities, and see Modiano et al. (in prep.) for an example).\\

\begin{acknowledgements}
DM is partly supported by the Netherlands Research School for Astronomy (NOVA). We acknowledge the use of public data from the Swift data archive. This research has made use of the VizieR catalogue access tool, CDS, Strasbourg, France (DOI : 10.26093/cds/vizier). This research has made use of NASA's Astrophysics Data System Bibliographic Services. We thank Aaron Tohuvavohu for pointing out a previous error in our calculation of the number of UVOT images processed, which led us to fix a bug in our code and obtain the correct numbers.
\end{acknowledgements}

\bibliographystyle{aa} 
\bibliography{mybibliography} 

\newpage
\onecolumn
\begin{appendix}

\section{Appendix}

\begin{table}[!htbp]
    \tiny
    \centering
    \begin{tabular}{P{5.5cm}|P{5.5cm}|P{5.5cm}}
    \textbf{Database or facility} & \textbf{Probed catalog description} & \textbf{Catalog reference} \\
    \hline \hline
        Simbad\footnote[1] & Database of astronomical objects published in astronomy journal papers & \citet{Simbad} \\ \hline
        Gaia\footnote[2] & Catalog of all Gaia sources (DR 2); Catalog of distances to 1.33 billion Gaia stars & \citet{Gaia_cat}; \citet{Bailer-Jones_2018}  \\ \hline
        NASA/IPAC Extragalactic Database (NED)\footnote[3] & Database of multi-wavelength data for extragalactic objects &  \citet{NED_ref} \\ \hline
        Sloan Digital Sky Survey (SDSS)\footnote[4] & Catalog of all Sloan Digital Sky Survey sources (DR 16) & \cite{SDSS_2020}  \\ \hline
        Hubble Space Telescope (HST)\footnote[5] & Large UV-optical-infrared high-resolution space telescope source catalog & \citet{hubble_cat}  \\ \hline
        Galaxy Evolution Explorer (GALEX)\footnote[6] & Database of all GALEX UV sources & \citet{GALEX_cat}  \\ \hline
        XMM-Newton\footnote[7] & X-ray mission point-source catalog (4XMM-DR9) & \citet{xmm_cat}  \\ \hline
        XMM-Newton Optical Monitor (XMM-OM)\footnote[8] & Optical monitor aboard XMM-Newton source catalog (SUSS4.1) & \citet{xmm_om_cat}  \\ \hline
        Zwicky Transient Facility (ZTF) MARS alert stream\footnote[9] & Ground-based optical transient searching facility alert stream & \citet{ZTF_alert}  \\ \hline
        Transient Name Server (TNS)\footnote[10] & Transient database &  \\ \hline
        All Sky Automated Survey for SuperNovae (ASAS-SN; transients) \footnote[11] & Ground-based optical transient searching facility - transient catalog & \citet{Shappee_2014}  \\ \hline
        All Sky Automated Survey for SuperNovae (ASAS-SN; variables)\footnote[12] & Ground-based optical transient searching facility - variables catalog & \citet{Shappee_2014}  \\ \hline
        The Panoramic Survey Telescope and Rapid Response System (PanSTARRS)\footnote[13] & Ground-based optical transient searching facility catalog & \citet{PanSTARRS_cat}  \\
        \hline
         Wide-field Infrared Survey Explorer (WISE)\footnote[14] & Infrared space survey mission catalog & \citet{WISE_cat}  \\
        \hline
        USNO-B\footnote[15] & All-sky optical catalog & \citet{usno-b_ref}  \\
        \hline
        The International Variable Star Index (VSX)\footnote[16] & Variable star database & \citet{vsx_ref}  \\
        \hline
        The Two Micron All Sky Survey (2MASS)\footnote[17] & All-sky infrared catalog & \citet{2mass_cat}  \\
        \hline
        General Catalog of Variable Stars (GCVS)\footnote[18] & Variable star database & \citet{gcvs_ref}  \\
        \hline
        A Catalog And Atlas of Cataclysmic Variables\footnote[19] & Cataclysmic variable database & \citet{cvcat_ref}
        \\
        \hline
        Asteroid Terrestrial-impact Last Alert System (ATLAS)\footnote[20] & Ground-based optical asteroid and transient searching facility catalog & \citet{atlas_ref}\\
        \hline
        SkyBot\footnote[21] & Solar system object database & \citet{skybot}  \\
        \hline
        Vizier\footnote[22] & Meta-compilation of astronomical catalogs & \citet{vizier_ref}
        
    \end{tabular}
    \caption{All catalogs which are probed automatically by \texttt{TUVOpipe} in PIV$^{a}$. \\ \\
    \begin{footnotesize}
    $^{a}$ The first column shows the database or facility which produces the catalog. The second column describes the catalog we probe. The references in the third column are to the relevant catalog/data release papers. Footnotes point to the web pages where the catalogs we probe are hosted. \\ \\ 
    $^{1}$\url{http://simbad.u-strasbg.fr/simbad/}\\
    $^{2}$\url{https://gea.esac.esa.int/archive/}\\
    $^{3}$\url{https://ned.ipac.caltech.edu/}\\
    $^{4}$\url{http://skyserver.sdss.org/dr16/en/home.aspx}\\
    $^{5}$\url{https://hla.stsci.edu/}\\
    $^{6}$\url{https://galex.stsci.edu/GR6/}\\
    $^{7}$\url{http://xmm-catalog.irap.omp.eu/}\\
    $^{8}$\url{https://www.cosmos.esa.int/web/xmm-newton/xsa}\\
    $^{9}$\url{https://mars.lco.global/?page=1}\\
    $^{10}$\url{https://www.wis-tns.org/}\\
    $^{11}$\url{http://www.astronomy.ohio-state.edu/asassn/transients.html}\\
    $^{12}$\url{https://asas-sn.osu.edu/variables}\\
    $^{13}$\url{https://catalogs.mast.stsci.edu/panstarrs/}\\
    $^{14}$\url{https://wise2.ipac.caltech.edu/docs/release/allsky/}\\
    $^{15}$\url{http://tdc-www.harvard.edu/catalogs/ub1.html}\\
    $^{16}$\url{https://www.aavso.org/vsx/}\\
    $^{17}$\url{https://irsa.ipac.caltech.edu/Missions/2mass.html}\\
    $^{18}$\url{http://www.sai.msu.su/gcvs/gcvs/}\\
    $^{19}$\url{https://archive.stsci.edu/prepds/cvcat/}\\
    $^{20}$\url{https://atlas.fallingstar.com/home.php}\\
    $^{21}$\url{https://vo.imcce.fr/webservices/skybot/}\\
    $^{22}$\url{https://vizier.u-strasbg.fr/viz-bin/VizieR}
    \end{footnotesize}
    }
    \label{tab:PIV_table}
\end{table}

\end{appendix}

\end{document}